\documentclass[fleqn,usenatbib]{mnras}

\usepackage[T1]{fontenc}
\usepackage{ae,aecompl}

\usepackage{graphicx}	
\usepackage{amsmath}	
\usepackage{amssymb}	
\usepackage{bm}		
\usepackage{pdflscape}	
\usepackage{color}
\usepackage{hyperref}
\usepackage{lscape}
\usepackage{float}
\usepackage{newtxtext,newtxmath}

\title[Flux density variability of 286 pulsars]{Flux density variability of 286 radio pulsars from a decade of monitoring}

\author[H. Kumamoto et al.]
{H. Kumamoto$^{1,2}$\thanks{E-mail: hiroki\_kumamoto@kumadai.jp},
S. Dai$^{2}$\thanks{E-mail: shi.dai@csiro.au},
S. Johnston$^{2}$,
M. Kerr$^{3}$,
R. M. Shannon$^{4,5}$,
\newauthor
P. Weltevrede$^{6}$,
C. Sobey$^{7}$,
R. N. Manchester$^{2}$,
G. Hobbs$^{2}$,
and K. Takahashi$^{1,8}$. 
\\
$^{1}$Kumamoto University, Graduate School of Science and Technology, 2-39-1 Kurokami, Chuo-ku, Kumamoto, 860-8555, Japan \\
$^{2}$CSIRO Astronomy and Space Science, Australia Telescope National Facility, PO Box 76, Epping, NSW 1710, Australia \\
$^{3}$Space Science Division, Naval Research Laboratory, Washington, DC 20375 5352, USA \\
$^{4}$Centre for Astrophysics and Supercomputing, Swinburne University of Technology, Mail H30, PO Box 218, VIC 3122, Australia\\
$^{5}$OzGrav: The ARC Centre of Excellence for Gravitational-wave Discovery, Hawthorn, VIC 3122, Australia\\
$^{6}$University of Manchester, Jodrell Bank Centre of Astrophysics, Alan Turing Building, Manchester, M13 9PL, United Kingdom \\
$^{7}$CSIRO Astronomy and Space Science, PO Box 1130 Bentley, WA 6102, Australia \\
$^{8}$Kumamoto University, International Research Organization for Advanced Science and Technology, 2-39-1 Kurokami, Chuo-ku, \\
Kumamoto, 860-8555, Japan. \\
}

\date{Accepted XXX. Received YYY; in original form ZZZ}

\pubyear{2020}

\begin{document}
\label{firstpage}
\pagerange{\pageref{firstpage}--\pageref{lastpage}}
\maketitle

\begin{abstract}
The Parkes telescope has been monitoring 286 radio pulsars approximately monthly since 2007 at an observing frequency of 1.4\,GHz. The wide dispersion measure (DM) range of the pulsar sample and the uniformity of the observing procedure make the data-set extremely valuable for studies of flux density variability and the interstellar medium. Here, we present flux density measurements and modulation indices of these pulsars over this period. We derive the structure function from the light curves and discuss the contributions to it from measurement noise, intrinsic variability and interstellar scintillation. Despite a large scatter, we show that the modulation index is inversely correlated with DM, and can be generally described by a power-law with an index of $\sim-0.7$ covering DMs from $\sim10$ to 1000\,cm$^{-3}$\,pc. We present refractive timescales and/or lower limits for a group of 42 pulsars. These often have values significantly different from theoretical expectations, indicating the complex nature of the interstellar medium along individual lines of sight. In particular, local structures and non-Kolmogorov density fluctuations are likely playing important roles in the observed flux density variation of many of these pulsars.
\end{abstract}



\section{Introduction}

Radio pulsars show flux variability on a wide variety of timescales. The cause of the variability is both intrinsic to the mechanism which produces the radio emission and extrinsic due to propagation of the emission through the magnetised, ionised interstellar medium (ISM). Single pulses from radio pulsars vary widely in both shape and flux density~\citep[e.g.,][]{wes06,bjb+12} but when integrated over many hundreds of rotations produce a profile which is (for the most part) stable in shape and flux density. This comes with caveats as pulsars can also switch off (null) for time spans of milliseconds to years and can sometimes show different emission states with distinct flux densities and profile shapes~\citep[e.g.,][]{wmj07}. It also appears to be the case that older pulsars are weaker than younger pulsars~\citep[e.g.,][]{fk06}. If this is correct then the flux density must slowly decrease over their lifetime.

The radio emission from pulsars is affected by propagation through the ISM. As pulsars are highly compact sources, when observed at frequencies near 1.4\,GHz all but the very closest pulsars are in the strong scintillation regime. In the strong scintillation regime, two different branches of scintillation are present, diffractive scintillation (DISS) and refractive scintillation (RISS) \citep[e.g.,][]{rcb84}.
DISS is characterised by flux density variations caused by interference patterns with a diffractive length scale $r_{\rm{diff}}$. Such interference patterns are produced by multipath propagation of signals with random phases. RISS is a result of large-scale inhomogeneities in the scattering screen of length $r_{\rm{ref}}$, the size of the projected scatter-broadened image. Since $r_{\rm{ref}}$ is defined as $r_{\rm{ref}}=r_{\rm{F}}^{2}/r_{\rm{diff}}$, where $r_{\rm{F}}$ is the Fresnel scale, $r_{\rm{ref}}$ is much larger than $r_{\rm{diff}}$. For a given transverse velocity between the scattering screen and the pulsar, RISS has a much longer time scale than DISS and a lower modulation index.

DISS has been known in pulsars since soon after discovery~\citep{ric69} and a large number of papers have been published on the DISS parameters of pulsars \citep[e.g.,][]{cor86,bgr98,jnk98,wmj+05,kcs+13,rch+19}. Extensive observations of DISS have allowed a theoretical framework to be developed over the past decade, which enables an estimate to be made for DISS parameters given basic pulsar information such as their dispersion measure and galactic location~\citep[e.g.,][]{cr98}. However, our current knowledge of the distribution and nature of turbulence in the Galaxy is limited, and therefore such estimates are only accurate to an order of magnitude at best. For the pulsars under consideration here at observing frequencies of 1.4\,GHz, the DISS scintillation bandwidth is of order MHz or less, the diffractive timescale less than 60 minutes or so. 

Observations of RISS and our understanding of these observations are poorer than those of DISS. The first observational evidence of RISS in pulsars was published by~\citet{Sieber}, who showed a correlation between long-term flux density variations of a pulsar and its dispersion measure (DM). Since then several surveys of RISS with pulsar data sets have been carried out~\citep[e.g.,][]{Kaspi, grc93, lrc94, brg99, Stinebring}. Typically these surveys contained small numbers of pulsars and/or relatively short data spans, and are therefore limited to small numbers of line of sight through our Galaxy and could not measure RISS parameters of high DM pulsars.
While the underlying physics of RISS is relatively well understood~\citep[e.g.,][]{rcb84}, RISS studies have been difficult to interpret because of the aforementioned uncertainty about how inhomogeneities in the ionised ISM are distributed in the Galaxy. In particular, discussions have been focused on the spectrum of phase fluctuations and whether it follows a simple Kolmogorov model across a wide scale range \citep[e.g.,][]{Armstrong}. While some previous observations show that RISS measurements are in good agreement with the Kolmogorov prediction~\citep[e.g.,][]{Kaspi,Stinebring}, others reported enhanced levels of refractive scintillation. It has been suggested that the spatial spectrum could have a steeper component and is non-turbulent~\citep{bn85,gn85}, or that the inner scale of turbulence is much larger than the diffractive scale~\citep{cfr+87}.

In addition to DISS and RISS, so-called `extreme scattering events (ESEs)' have been observed in pulsars via variations in DM and scintillation strength~\citep{cks+15} and/or large modulations in light-curves~\citep{kcw+18}. Such events are also observed in extragalactic sources~\citep{bst+16} and are generally believed to be produced by local ISM structures with enhanced density or asymmetric structures~\citep[e.g.,][]{ww98,pk12,wtb+17}.

In this work we present flux density measurements of 286 pulsars at 1.4\,GHz using the Parkes radio telescope. Monthly observations of these pulsars were carried out for four to eleven years under a monitoring program of young and energetic pulsars~\citep{wjm+10}. This represents the largest sample of pulsar and the longest time span for studies of long-term flux density variation of pulsars.
Long-term monitoring of the same sample of pulsars have previously been used to study dispersion measure variations~\citep{pkj+13} and ESEs in PSRs J1057$-$5226 and J1740$-$3015~\citep{kcw+18}.
We describe the data set, observational system, and data reduction in Section~\ref{section;obs}.
In Section~\ref{section;str} and \ref{section;flu}, we provide details in calculating the structure function and modulation index. In \ref{section;res} we present our main results. In Section~\ref{section;dis}, we compare our results with theoretical predictions and discuss their implications. Finally, we summarise our results in Section~\ref{section;con}.

\section{Observations and data reduction}
\label{section;obs}
In this paper, we use observations of pulsars taken with the Parkes radio telescope under project P574 \citep{wjm+10}. The project has carried out monthly observations of a sample of about 300 pulsars since early 2007. For our purposes, we only include data taken at an observing frequency around $1.4$\,GHz with either the multibeam receiver~\citep{Staveley-Smith} or the H-OH receiver~\citep{Thomas}. Data were recorded with the Parkes Digital Filter Bank (PDFB) system, providing 256\,MHz bandwidth centred at 1369\,MHz with 1024 frequency channels for the multibeam receiver and 512\,MHz bandwidth centred at 1465\,MHz with 1024 frequency channels for the H-OH receiver. The data are folded modulo the topocentric spin period to form sub-integrations of typical duration of $\sim30$~s and typical observations are 180~s in length. In addition to the pulsar data, we record data from a square-wave signal injected directly into the feed by a noise diode. Observations of the noise-diode are taken once per hour and are typically 80~s duration. Finally we also make observations of the flux calibrator Hydra-A with the noise-diode operating.

Data processing follows the route described in~\citet{Johnston2}. The critical step of flux calibration proceeds as follows. First the observations of Hydra-A are used to measure the flux density of the noise-diode. Then, the nearest noise-diode observation in time to the pulsar observation is used to convert the pulsar data into units of Jy. Observations without valid calibration data were identified and removed from our analysis. We estimate the stability of the noise-diode to be better than 3\% over the decade of this observing program, which agrees with the estimate by \citet{krh+20}. A detailed study of flux calibration with the Parkes telescope has also been published by \citet{xwh+19}. The {\sc psrchive} routine {\sc psrflux} was used to measure the flux density, $S$ and its error, $e$, as follows:
\begin{equation}
    S = (\sum_{i}^{n_{\rm on}} I_{\rm i} ) / n_{\rm tot}
\end{equation}
and 
\begin{equation}
    e = \sigma \sqrt{(n_{\rm on})} / n_{\rm tot},
\end{equation}
where $n_{\rm on}$ is the number of phase bins across the `on' part of the profile, $n_{\rm tot}$ is the total number of phase bins per period and $\sigma$ is the rms of the `off-pulse' profile. The location and width of the profile is determined by using a standard template which is cross-correlated with the data.
For data taken using the H-OH receiver, we ensured that the flux densities were measured in the same frequency band as those using the multibeam receiver. Strong radio-frequency interference (RFI) can corrupt the data and result in outliers in the flux density time series. We therefore iteratively identified outliers in the time series for each pulsar, and then performed removal of the RFI via the routine {\sc pazi}.

\begin{figure*}
\begin{center}
\includegraphics[width=8cm]{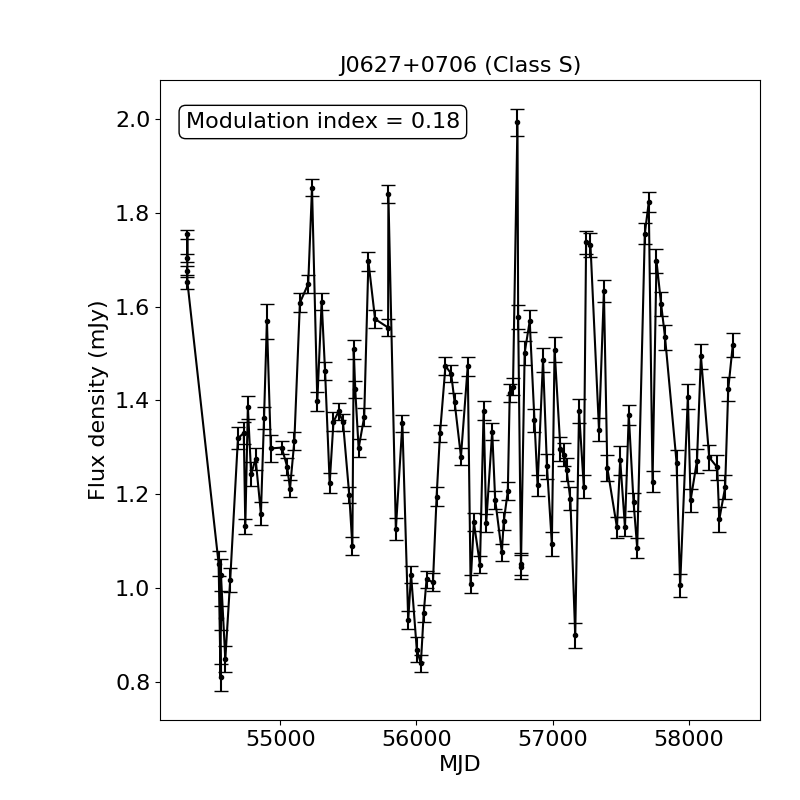}
\includegraphics[width=8cm]{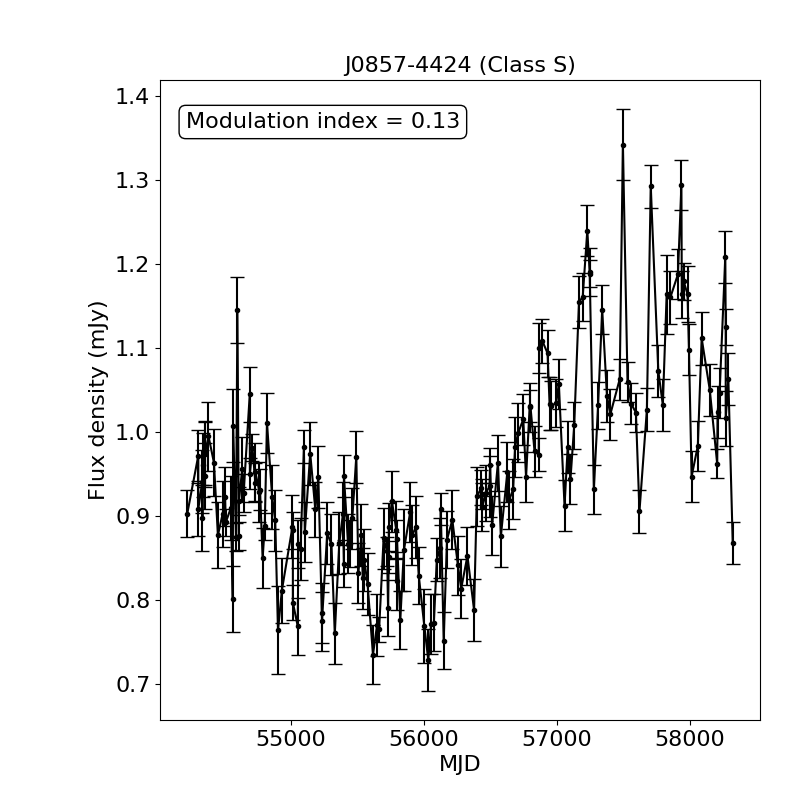}
\includegraphics[width=8cm]{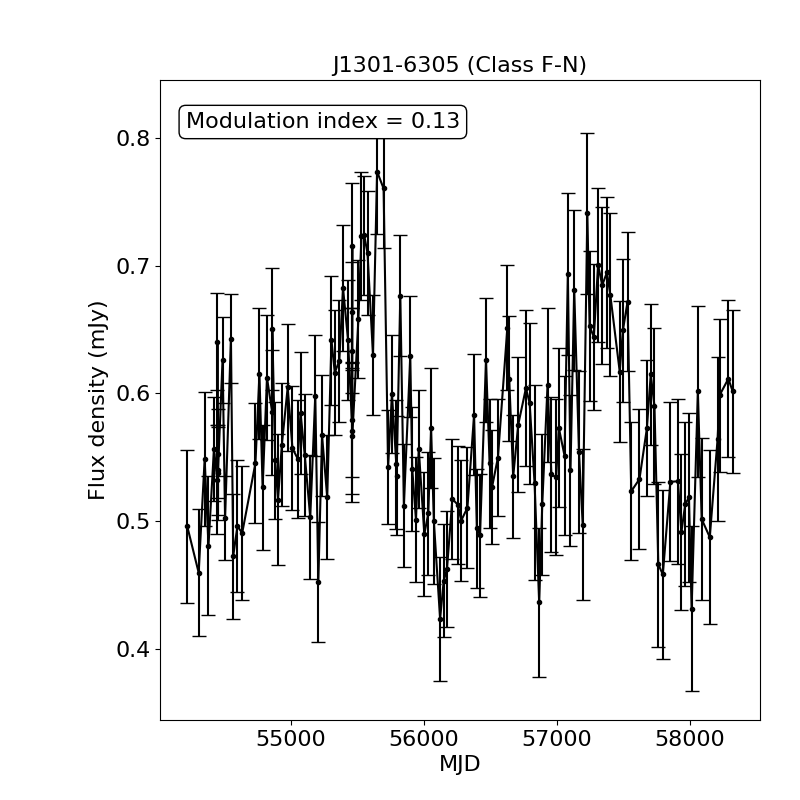}
\includegraphics[width=8cm]{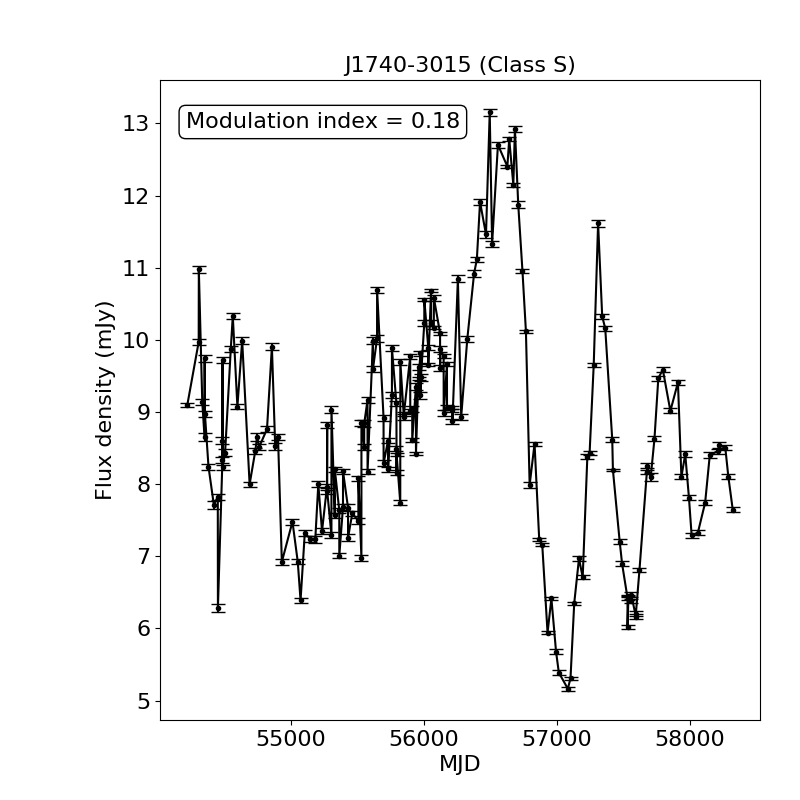}
\caption{Flux density as a function of MJD of PSRs J0627$+$0706, J0857$-$4424, J1301$-$6305 and J1740$-$3015. The modulation index is calculated using Eq.~\ref{eq;measured_mi}. See Section~\ref{section;res} for the pulsar classification.} 
\label{fig:flux}
\end{center}
\end{figure*}

Table~\ref{tab:all} lists the pulsar name, spin period ($P$), dispersion measure (DM) and distance (Dist), followed by the span of the observations and the number of data points ($N_{\rm obs}$). In Fig.~\ref{fig:flux} we show flux densities as a function of MJD of PSRs J0627$+$0706, J0857$-$4424, J1301$-$6305 and J1740$-$3015 as examples. 

\section{Structure function}
\label{section;str}

The structure function, $D(\tau)$, is defined by
\begin{equation}
    D(\tau) = \langle[S(t+\tau)-S(t)]^{2}\rangle,
\end{equation}
where $\tau$ is the time lag and the angle brackets denote an ensemble average.
For each pulsar, we have a discrete time series of flux density measurements, $S_{\rm i}$. The measured flux density, $S_{\rm i}$, includes a noise component, the standard deviation of which is $e_{\rm i}$. From this time series we can calculate the structure function, commonly used in studies of RISS. The structure function is not biased by the non-uniform sampling of the time series and allows us to measure the refractive time scales. The estimated structure function is biased by the noise in flux density measurements~\citep{You}, and it can be expanded as
\begin{eqnarray}
    \begin{aligned}
        D_{\rm est}(\tau) ~&=~ \frac{1}{\overline{S}^2N_{\rm p}} \biggl\{ \sum^{N_{\rm p}}_{\rm ij} (S_{\rm i}-S_{\rm j})^2\ +\ \sum^{N_{\rm p}}_{\rm ij}(e_{\rm i}^2 + e_{\rm j}^2) \\
        ~&+~  2 \sum^{N_{\rm p}}_{\rm ij} [e_{\rm i} - e_{\rm{j}}][S_{\rm i}-S_{\rm j}]\ -\ 2 \sum^{N_{\rm p}}_{\rm ij} e_{\rm i} e_{\rm j} \biggr\},
    \end{aligned}
\end{eqnarray}
where $D_{\rm{est}}(\tau)$ is the estimated structure function at a lag of $\tau$ days, $\overline{S}$ is the mean flux density and $N_{\rm p}$ is the number of all pairs ($S_{\rm i}$ and $S_{\rm j}$) with time lag of $\tau$.
The second term is the only error term that does not have zero mean and is subtracted to obtain the bias-corrected structure function $D(\tau)$ presented in this paper. Uncertainties in the estimated structure function caused by flux density measurement noise are calculated using Equation A3 in \citet{You}. The structure functions are also normalised by ${\overline{S}^2}$ and the bin size is chosen to have equal numbers of pairs within each bin. 

In the ideal case, we expect three different regimes within the structure function, created by the scintillation properties of the pulsar, and ultimately by the properties of the ISM along the line of sight. These regimes are: (1) a flat structure function at small lags, (2) a rising structure function at medium lags and (3) a flat structure function at large lags.
The amplitude of the structure function in the first regime results from a combination of the uncertainties on the measured flux densities, the variability of the pulsar itself, any unquenched DISS and the portion of the RISS fluctuation spectrum that is shorter than the average interval between observations for a particular pulsar. The rising part of the structure function can be characterised by a power-law function, $D(\tau)\propto\tau^{\gamma}$, with a logarithmic slope of $\gamma$. Previous results \citep{sc90} show that the value of the logarithmic slope $\gamma$ is $\sim1$ in many cases. The level of the structure function at large lags is set by the refractive scintillation properties of the ISM, and the time taken to reach saturation is related to the refractive timescale.

\section{Modulation index}
\label{section;flu}
In addition to the structure function, a useful parameter to characterise the variability is the modulation index, $m$, defined as 
\begin{eqnarray}
    \label{eq;measured_mi}
    m ~=~ \frac{\sigma_{S}}{\overline{S}}
\end{eqnarray}
where $\sigma_{S}$ is the standard deviation of the flux densities. 
The measured value of $m$ comes from a combination of several effects including the radiometer noise, the intrinsic intensity variability of the pulsar and the refractive and diffractive effects. We quantify these effects below.

The contribution from radiometer noise in the flux density measurements can be estimated as,
\begin{eqnarray}
    \label{eq;measured_mi_error}
    m_{\rm{n}} ~=~ \frac{\overline{e_{\rm{t}}}}{\overline{S}},
\end{eqnarray}
where $\overline{e_{\rm{t}}}$ is the mean uncertainty of flux density.
Although we expect that pulsars are intrinsically steady emitters in the long term (we return to this point later), short observations, such as our 180\,s observations, can be dominated by pulse-to-pulse flux variations, which typically have approximately a 100\% fluctuation level. We refer to noise caused by such pulse-to-pulse flux variations as `jitter noise' and estimate its contribution to the modulation index as 
\begin{eqnarray}
    m_{\rm j} ~=~ \sqrt{\frac{1}{N_{\rm pulse}}} ~=~ \sqrt{\frac{P}{t_{\rm obs}}},
\end{eqnarray}
where $N_{\rm pulse}$ is the number of pulse within the mean time span, $t_{\rm obs}$, of an individual observation. The measurement and jitter noise corrected modulation index can be estimated as
\begin{eqnarray}
    m_{\rm corr} ~=~ \sqrt{m^{2}-m^{2}_{\rm n}-m^{2}_{\rm j}}.
\end{eqnarray}

We can compute the expected modulation index from both diffractive and refractive scintillation, $m_{\rm d}$ and $m_{\rm r}$. This requires knowledge of the diffractive scintillation bandwidth, $\Delta\nu_{\rm d}$ and the scintillation timescale $t_{\rm d}$. Unfortunately our observational setup does not allow us to measure these values directly, because in the majority of our pulsars $\Delta\nu_{\rm d}$ is much smaller than our channel bandwidth. However, we can use models and empirical fits to data, along with models of the ISM turbulence to estimate the values of $\Delta\nu_{\rm d}$ and $t_{\rm d}$. Any such estimates should be treated with caution; observations show that our Galaxy does not conform to a simple model and a large spread in $\Delta\nu_{\rm d}$ and $t_{\rm d}$ can be obtained along similar lines of sight.

With these caveats in mind, we estimate the scattering time, $\tau_{\rm s}$, in $\mu$s scaled to 1.4~GHz using the empirical relation to DM obtained by~\citet{kmn15}:
\begin{eqnarray}
\label{eq;tau}
    \tau_{\rm s} ~=~ 1.2 \times 10^{-5} \,\, {\rm DM}^{2.2}(1.0 + 0.00194 {\rm DM}^2).
\end{eqnarray}
The value of $\Delta\nu_{\rm d}$ in MHz can be determined by
\begin{eqnarray}
    \label{eq;bandwidth}
    \Delta\nu_{\rm d} ~=~ \frac{1.16}{2\pi \tau_{\rm s}},
\end{eqnarray}
assuming a uniform medium with a Kolmogorov wave number spectrum~\citep{cr98} and noting the large amount of scatter around the relationship in equation~\ref{eq;tau} in the observational data.

Additionally, estimating $t_{\rm d}$ requires the pulsar velocity, $V_{\rm iss}$ and is given by \citet{jnk98} as
\begin{eqnarray}
\label{eq;td}
t_{\rm{d}} ~=~ \frac{3.85\times10^4 \,\, \sqrt{D \Delta\nu_{\rm{d}}}}{\nu \,\, V_{\rm{iss}}},
\end{eqnarray}
where $\nu$ is the observing central frequency in GHz. The pulsar distance, $D$ (in kpc), and transverse velocity, $V_{\rm{iss}}$ (in km\,s$^{-1}$), are obtained from the ATNF Pulsar Catalogue~\footnote{https://www.atnf.csiro.au/research/pulsar/psrcat/}~\citep{mhh05}. For pulsars without measured velocity, we used $V_{\rm iss} = 300$~km\,s$^{-1}$. Our observations sample the scintillation pattern over a finite bandwidth, $\Delta\nu$ (here 256~MHz) and the mean observing time $t_{\rm{obs}}$ (here typically 180~s). The modulation caused by diffractive scintillation, $m_{\rm{d}}$, is related to the number of scintles within this dynamic spectrum and can be roughly estimated as
\begin{eqnarray}
\label{eq:md}
m_{\rm{d}} ~=~ \sqrt{5 \,\, \frac{\Delta\nu_{\rm{d}}}{\Delta\nu} \,\, \frac{t_{\rm{d}}}{t_{\rm{obs}}}},
\end{eqnarray}
where the constant at the front is the filling factor accounting for the fact that the scintillation maximum fills only a fraction of the frequency-time plane~\citep{sc90}. The equation shows that if $\Delta\nu_{\rm{d}} \ll \Delta\nu$ then $m_{\rm{d}}$ will be small. Conversely if $\Delta\nu_{\rm{d}}$ is comparable or larger than $\Delta\nu$ then $m_{\rm{d}}$ can be larger than unity.

For a Kolmogorov spectrum with a small inner scale and a uniform scattering medium, the value of $m_r$ can been estimated as
\begin{eqnarray}
\label{eq;predicted_mr}
m_{\rm{r}} ~=~ 1.1 \biggl{(}\frac{\Delta\nu_{\rm{d}}}{\nu}\biggr{)}^{0.17},
\end{eqnarray}
where $\nu$ is observational frequency~\citep{grc93,Stinebring}. Finally the RISS time scale can be computed~\citep{sc90} via
\begin{eqnarray}
\label{eq;predicted_tr}
t_{\rm{r}} ~=~ \frac{4}{\pi} \,\, \frac{\nu}{\Delta\nu_{\rm{d}}} \,\,\, t_{\rm{d}}.
\end{eqnarray}

Columns~8 to 14 of Table~\ref{tab:all} list $m$, $m_{\rm{n}}$, $m_{\rm{j}}$, $m_{\rm{corr}}$, $m_{\rm{d}}$ and $m_{\rm{r}}$. We stress again that these are only estimates based on a simple model of the distribution of free electrons in the ISM. Considering uncertainties in pulsar distances, velocities and estimates of scattering time, these modulation index estimates can be an order of magnitude or more in error for any given pulsar. However, a comparison of the magnitudes of the various modulation indices allows an initial view to be made of likely dominant cause of the variability. 

\section{Results}
\label{section;res}
We calculated structure functions for our entire sample of pulsars and classified them according to the following scheme. Structure functions which show a rising slope with saturation or all three regimes identified above we denote as class ``S". A small fraction of pulsars show structure functions that continue to increase; this implies that the refractive timescale is longer than the maximum lag. We classify these as ``I". The bulk of the structure functions are flat, classified as ``F". These flat structure functions can arise for a number of reasons, where the modulation index is dominated by (as listed in Table~\ref{tab:all}):
\begin{itemize}
    \item measurement noise and/or weak pulsars, denoted as class ``F-N"
    \item jitter noise, denoted as class ``F-J"
    \item diffractive scintillation, class ``F-D"
    \item refractive scintillation, but with a refractive time that is short compared to our shortest lag. Class ``F-R"
    \item a combination of both refractive and diffractive scintillation, class ``F-DR"
\end{itemize}
The classification is shown in the final column in Table~\ref{tab:all}. We note that in all the ``F" cases, the value of the structure function gives an upper limit to the refractive modulation. Finally, we also created a class ``X" for pulsars with clearly anomalous values of the modulation index. These will be discussed further below.

\subsection{Refractive scintillation parameters}
\label{subsection;tim}
For structure functions showing saturation and/or a rising slope (classes ``S" and ``I"), we measured the logarithmic slope of the rising regime and levels of noise and saturation through fitting combinations of constant levels and a single power-law to structure functions. The Bayesian inference library {\sc BILBY}~\citep{Ashton} was used to fit the data and estimate the uncertainties in the parameters. We assumed a Gaussian likelihood function with a uniform prior for each parameter. Posterior distributions for the parameters were sampled using {\sc Bilby} as a wrapper for the {\sc Dynesty} sampling algorithm~\citep{spe20}. The uncertainty of each parameter was estimated as the average of deviations from the median at 16\% and 84\% of its posterior distribution. Since uncertainties of the estimated structure function only account for flux density measurement noise, large dips and scatters in the structure function can significantly affect our fitting and the estimate of uncertainties in parameters. In order to obtain reliable estimate of scintillation parameters, we gradually increase uncertainties of the estimate structure function and refit our models until the reduced $\chi$-squared of fitting is smaller than five. 

\begin{figure}
\begin{center}
\includegraphics[width=9cm]{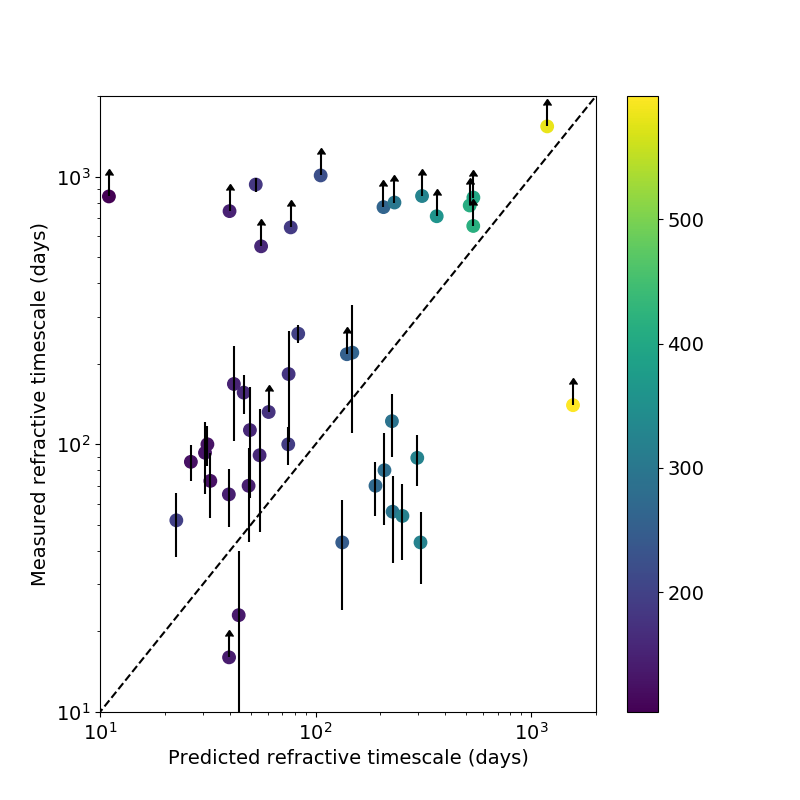}
\end{center}
\caption{Measured versus predicted refractive timescales for pulsars of classes ``S" and ``I". The predicted time-scale is estimated using Eq.~\ref{eq;bandwidth}, \ref{eq;td} and \ref{eq;predicted_tr}. We note that estimated refractive time-scales with these equations are only indicative and can be significantly different from their actual values. The colour bar represents pulsar DMs in the unit of cm$^{-3}$\,pc.}
\label{fig;TrPred}
\end{figure}

We list measured refractive scintillation parameters of 42 pulsars of classes ``S" and ``I" in Table~\ref{tab;ref}. The refractive time scale $T_{\rm r}$ (column 2) is estimated as the lag at which the structure function reaches half of its saturated value. For Class ``I" pulsars, only lower limits are presented. The slope of the rising regime is listed in column 4. 
Not all 42 pulsars show clear noise level at short lags and we only list those can be measured in column 3. The saturated values of structure function of Class ``S" pulsars are listed in column 5. In Fig.~\ref{fig;TrPred}, we compare the measured refractive timescale with predictions using Eq.~\ref{eq;predicted_tr}. 
%

\begin{table}
    \caption{Refractive parameters. Uncertainties are given in parenthesis. We quote two significant digits of the uncertainty of the refractive timescale, and one significant digit of the uncertainty of other parameters. }
	\label{tab;ref}
\begin{tabular}{lcccc}
\hline
Name & $T_{\rm r}$ &  Noise &  Slope  & Saturation  \\
     & [d]    &       &  [d$^{-1}$] & \\ 
\hline
\multicolumn{5}{c}{Class ``S"} \\
\hline
J0627$+$0706 & 23(17)  &            &  0.3(1)  &  0.184(3)  \\
J0631$+$1036 & 73(20)  &            &  1.5(5)  &  0.075(4)  \\
J0738$-$4042 & 151(26) &            &  0.69(9) &  0.0176(4)  \\
J0842$-$4851 & 52(14)  &            &  0.48(8) &  0.162(2)  \\
J0857$-$4424 & 934(56) &   0.064(3) &  0.89(5) &  0.284(5)  \\
             &               &             &             &   \\
J1001$-$5507 & 93(28)  &            &  0.46(8) &  0.0174(7)  \\
J1105$-$6107 & 70(16)  &            &  0.44(5) &  0.82(1)  \\
J1359$-$6038 & 56(20)  &            &  0.6(1)  &  0.0641(9)  \\
J1424$-$5822 & 43(13)  &            &  2(1)    &  0.0109(4)  \\
J1522$-$5829 & 259(20) &            &  0.54(2) &  0.00238(7)  \\
             &               &             &             &   \\
J1534$-$5405 & 100(16) &            &  0.71(9) &  0.0257(8)  \\
J1557$-$4258 & 70(27)  &   0.005(2) &  0.9(3)  &  0.0165(5)  \\
J1613$-$4714 & 91(44)  &   0.003(1) &  0.9(5)  &  0.0123(9)  \\
J1650$-$4502 & 54(17)  &            &  0.44(8) &  0.448(6)  \\
J1705$-$3423 & 65(16)  &            &  0.9(2)  &  0.0109(2)  \\
             &               &             &             &   \\
J1718$-$3825 & 43(19)  &            &  0.6(2)  &  0.117(2)  \\
J1722$-$3207 & 86(13)  &            &  1.4(3)  &  0.0157(6)  \\
J1730$-$3350 & 220(110)&   0.024(4) &  0.7(3)  &  0.056(2)  \\
J1740$-$3015 & 168(65) &   0.06(2)  &  0.6(2)  &  0.40(1)  \\
J1741$-$3927 & 113(50) &            &  0.7(2)  &  0.014(1)  \\
             &               &             &             &   \\
J1757$-$2421 & 183(81) &   0.020(5) &  0.7(3)  &  0.064(2)  \\
J1801$-$2451 & 122(32) &            &  0.8(2)  &  0.073(2)  \\
J1816$-$2650 & 100(17) &            &  0.52(5) &  0.032(1)  \\
J1841$-$0425 & 89(19)  &            &  0.47(6) &  0.0345(6)  \\
J1845$-$0743 & 80(30)  &            &  0.7(2)  &  0.141(4)  \\
\hline
\multicolumn{5}{c}{Class ``I"}\\
\hline
J0907$-$5157 & $>843(32)$    &  0.0070(4)  &  1.23(8)    &   \\
J1112$-$6103 & $>140(30)$    &             &	0.21(1)    &      \\
J1157$-$6224 & $>846(14)$    &  0.0055(3)  &  1.26(4)    &   \\
J1319$-$6056 & $>835(36)$   &  0.0005(1)  &  1.3(1) &   \\
J1357$-$62   & $>654(80)$    &  0.00100(5) &  1.0(2)     &   \\
             &               &             &             &   \\
J1420$-$6048 & $>711(96)$   &  0.025(5)   &  0.43(4) &   \\
J1539$-$5626 & $>132(34)$    &             & 	0.21(2)    &      \\
J1600$-$5044 & $>769(77)$    &  0.0084(2)  &  0.45(3)    &   \\
J1614$-$5048 & $>1540(100)$  &  0.0270(5)  &  0.79(6)    &   \\
J1708$-$3426 & $>646(23)$    &  0.0077(3)  &  0.91(4)    &   \\
             &               &             &             &   \\
J1723$-$3659 & $>217(48)$    &  0.028(3)   &  0.25(2)    &   \\
J1738$-$2955 & $>1010(120)$ &  0.066(5)   &  0.9(1) &   \\
J1741$-$2733 & $>743(30)$    &  0.017(1)   &  1.5(1)     &   \\
J1832$-$0827 & $>800(110)$   &  0.114(1)   &  0.46(4)    &   \\
J1844$-$0538 & $>780(79)$    &  0.0190(4)  &  0.45(3)    &   \\
             &               &             &             &   \\
J1847$-$0402 & $>16(10)$     &             &  0.13(1)    &   \\
J1848$-$0123 & $>549(15)$    &  0.0015(2)  &  0.73(2)    &   \\
\hline
\end{tabular}
\end{table}

\clearpage

\section{Discussion}
\label{section;dis}

\begin{figure*}
\begin{center}
\includegraphics[width=16cm]{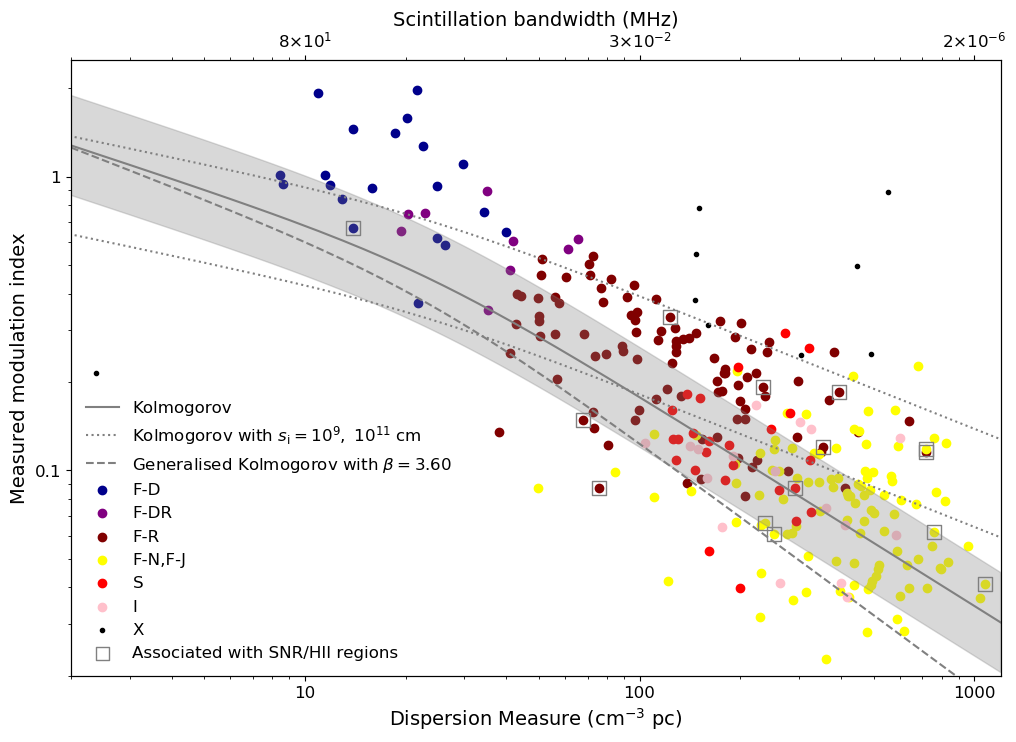}
\end{center}
\caption{Modulation index versus dispersion measure for all 286 pulsars. The colour coding represents the pulsar classification (see text for details). The top axis give an estimate of the scintillation bandwidth for a given DM using Eq.~\ref{eq;bandwidth}. The solid line shows the predicted refractive modulation index as a function of DM for a Kolmogorov spectrum (Eq.~\ref{eq;predicted_mr}). We show the region within a factor of 10 around the estimated scattering time scale (Eq.~\ref{eq;tau}) as the shaded region. The top dotted line shows predictions from a Kolmogorov model with an ``inner scale" of 10$^9$\,cm, while the bottom dotted line shows an ``inner scale" of 10$^{11}$\,cm. The dashed line shows a generalised Kolmogorov spectrum with an index of $\beta=3.6$. Pulsars potentially associated with SNRs and HII regions according to the ATNF Pulsar Catalogue are marked with empty squares.}
\label{fig;dmMd}
\end{figure*}

\subsection{Modulation index and scattering strength}

We presented well calibrated flux density measurements of 286 pulsars over a period of $\sim10$\,years. The wide DM range of these pulsars from $\sim10$ to $1000$\,cm$^{-3}$\,pc makes them extremely valuable for studies of RISS. Fig.~\ref{fig;dmMd} shows the modulation index corrected for the measurement noise, $\sqrt{m^{2}-m^{2}_{\rm{n}}}$, against DM for our sample of pulsars. The pulsars are colour coded according to the classification scheme described above. Pulsars in blue have $m_{\rm{c}}$ dominated by diffractive scintillation (class ``D"), purple for which both diffractive and refractive effects are important (class ``DR") and brown for refractive dominated (class ``R"). The pulsar which shows the classic three regimes of refractive scintillation (class ``S") are red, those for which the structure function has not yet saturated are shown pink (class ``I"). Finally, the weak pulsars are shown in yellow and these can be considered as upper limits on $m_{\rm{c}}$. The outlier pulsars (class ``X") are shown in black.

A clear colour trend can be seen across Fig.~\ref{fig;dmMd}.
At low DMs the modulation index is dominated by diffractive scintillation. As the DM increases, the scattering effect becomes stronger and we can no longer resolve the diffractive scintles in our data at which point refractive scintillation dominates the variability. At large DMs $>100$\,cm$^{-3}$\,pc, modulation indices are less than $\sim0.4$, consistent with previous results~\citep{Kaspi}, and showing that pulsar flux densities are intrinsically stable. We also see that the pulse jitter does not significantly affect our measurements of the modulation index. At extremely large DMs where the time span of data sets is shorter than $T_{\rm{r}}$, we are likely underestimating the refractive modulation index because the pulsars are weak and the modulation index is dominated by noise.
For a given DM, the scatter in the modulation index is large.
Compared with previous studies~\citep[e.g.,][]{Kaspi,Stinebring}, our sample of pulsars covers a much wider DM range, including 213 pulsars with DM larger than 100\,cm$^{-3}$\,pc. The DISS contribution to the variability is negligible for this sample of pulsars and they are particularly valuable for studying refractive scintillation at these observing frequencies.  

The modulation index of refractive scintillation allows us to probe the strength of interstellar scintillation and constrain the turbulence spectrum. Several previous studies of pulsar refractive scintillation compared the refractive modulation index with the diffractive scintillation bandwidth, and put constraints the power-law density inhomogeneity spectrum and the so-called ``inner scale"~\citep[e.g.,][]{cfr+87,Stinebring}. The observational setup used for this work is not ideal for measuring the diffractive scintillation bandwidth, especially for high DM pulsars, and we defer this to future studies with the new Parkes Ultra-Wideband Low receiver~\citep[UWL,][]{hmd+20}. In this work we compare the observed distribution of modulation indices over a wide DM range with theoretical predictions. 

In Figure~\ref{fig;dmMd}, the solid line shows the predicted refractive modulation index as a function of DM for a Kolmogorov spectrum (Eq.~\ref{eq;predicted_mr}). Large uncertainties are expected in estimating the scattering time scale using DM~\citep[e.g.,][]{gkk17}, and we show the region within a factor of 10 around the mean trend (Eq.~\ref{eq;tau}) as the shaded region. 
We also plot the prediction of a Kolmogorov spectrum with different inner scales~\citep[dotted lines,][]{grc93} and a generalised Kolmogorov spectrum with an index of $\beta=3.60$~\citep[dashed line,][]{Stinebring}.
We clearly see that a simple Kolmogorov model cannot explain the large scatter, even considering large uncertainties in estimates of scattering time based on the DM (shaded region). On the one hand, some pulsars show variability much larger than the prediction of a simple Kolmogorov model, and can only be explained by a Kolmogorov model with inner scales much larger than $10^9$\,cm. On the other hand, some pulsars show significantly lower variability than the Kolmogorov prediction, and might need $\beta<3.67$.
One possible explanation of these results is that the turbulence in the ISM is very different along individual lines of sight. For example, \citet{xz17} suggested that a short-wave-dominated density spectrum with the density structure formed at small scales due to shocks could enhance the scattering for some high DM pulsars.

Despite the large scatter, the inverse correlation between the modulation index and DM over such a wide DM range is remarkable. After rejecting outlier pulsars (class ``X"), we obtained a power-law fit with an index of $-0.67\pm0.03$ without correcting for jitter noise and $-0.72\pm0.03$ after correcting for jitter noise. These indices give us a power-law significantly steeper than the prediction of $-0.37$ for a Kolmogorov spectrum, and demonstrate the value of long term monitoring of high DM pulsars for studies of RISS. The very small modulation index of high DM pulsars, especially after correcting for measurement and jitter noise, also show that pulsars are nearly constant flux sources and our flux calibration is reliable.

Previous studies suggested that pulsars in supernova remnants (SNRs) or those that have lines of sight that pass through relatively nearby HII regions show stronger scattering. In our sample, 13 pulsars are potentially associated with SNRs according to the ATNF Pulsar Catalogue~\citep{mhh05} and only PSR J1825$-$1446 can be confirmed to be behind a known HII region~\citep{abb+14}. 
Although several of these pulsars show large modulation indices, some of them show variability weaker than the Kolmogorov model prediction. We also did not see any significant dependence of the modulation index on pulsar parameters, such as Galactic longitude and latitude. However, we note that our sample mainly consists of young pulsars and high spin-down rate pulsars. The observed scatter in modulation indices could reflect different properties of local interstellar environment of these pulsars. Comparing our results with similar studies of other pulsar populations, such as millisecond pulsars regularly observed by pulsar timing arrays~\citep[e.g.,][]{krh+20}, might allow us to reveal different local environments for different population of pulsars.  

As mentioned before, we did not measure the diffractive scintillation bandwidth but estimated it based on pulsar DMs. It is well known that there are large uncertainties in the relationship between DM and scattering time, which makes it challenging for us to constrain any models with only modulation indices. However, over a wide DM range (covering four orders of magnitudes in estimated scintillation bandwidth), the large scatter in modulation indices cannot be accounted for by just uncertainties in the estimate of scattering time. The Parkes project P574 has already been observing with the UWL receiver, which will allow us to measure the diffractive scintillation bandwidth of a significant fraction of our pulsars. Combining modulation indices measured here with future UWL scintillation bandwidth measurements, we will be able to investigate different turbulence models in more detail.

\subsection{Outlier pulsars}
We examined the table and the figure for obvious outliers; those pulsars for which the observed modulation index is significantly different to expectations. There are nine obvious standouts, classified as ``X" in Table~\ref{tab:all}.

\noindent
{\bf PSRs~J1302$-$6350 and  J1638$-$4725:} These are binary pulsars with long orbital period and long-duration eclipses \citep{jml+92,lfl+06}. This causes the large observed modulation indices which are much higher than expected from their DMs.

\noindent
{\bf PSRs~J1646$-$4346 and J1830$-$1059}: These pulsars show quasi-periodic modulation of their profiles \citep{khjs16,slk+19} which strongly affects the modulation index.

\noindent
{\bf PSRs~J1049$-$5833, J1701$-$3726 and J1727$-$2739:} Examination of the light curves and the individual observations shows evidence of nulling in these pulsars. For PSR~J1049$-$5833, we estimate the nulling time to be a few minutes with a nulling fraction of 30\%. For PSRs~J1701$-$3726 and J1727$-$2739 the nulling time is shorter, as both on and off states are observed within a 180~s observation. The nulling fractions are $\sim20$\% and 60\% respectively.

\noindent{\bf PSR~J1703$-$4851:} The profiles for this pulsar show two distinct states, a bright state dominated by the central component, and a weak state in which the components have almost equal intensity. There are roughly an equal number of observations in each of the two states, and the time scale for the state change must be shorter than 30 days.

\noindent
In addition, {\bf PSR~J0108$-$1431} is an extreme outlier. It has the lowest DM of the pulsar in our sample \citep{tnj+94} and yet also has a low modulation index. This pulsar is therefore the only one in our sample which is in the weak scintillation regime at 1.4~GHz.

\subsection{Intrinsic Variability}
\label{subsection;intrinsic}
Pulsars at greater distances from Earth are expected to have smaller values of $\Delta\nu_d$ (see equations 4 and 5), hence small values of $m_{\rm d}$ (Eq.~\ref{eq:md}) and $m_{\rm r}$ (Eq.~\ref{eq;predicted_mr}) in our data set. We should therefore see a correlation between the measured modulation index and the pulsar distance (or its proxy, the dispersion measure) provided that the pulsar itself shows no intrinsic modulation. Fig.~\ref{fig;dmMd} shows the expected trend, with the highest DM pulsars having the lowest $m$.

PSR~J1721$-$3532 has the smallest value of $m$, 0.019 and 18 pulsars have $m<0.05$. This shows that (at least some) pulsars are stable sources over a timescale of years. The fact that many pulsars exhibit low modulation indices also gives us confidence in the stability of our flux calibration over this long period of time.

\begin{figure}
\begin{center}
\includegraphics[width=8cm]{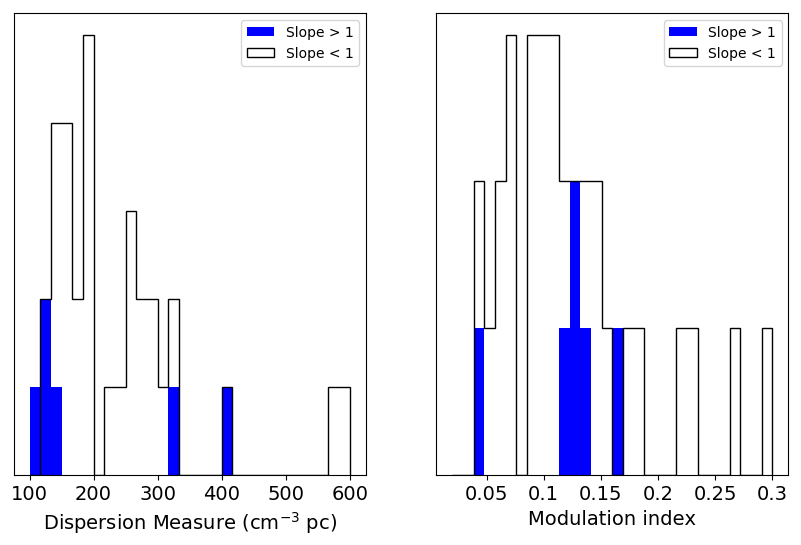}
\end{center}
\caption{Histogram of DM and modulation index for pulsars with measured slope larger (blue) and smaller (white) than one.}
\label{fig;slope}
\end{figure}

\subsection{Refractive time scales}

We measured refractive time scales for 25 pulsars and put lower limits on 17 pulsars (see Table~\ref{tab;ref}). Their structure functions together with the best fits are shown in the Appendix~\ref{app:s} and \ref{app:i}. In Fig.~\ref{fig;TrPred} we plot measured time scales against predictions with the colour bar showing pulsar DMs. Again, we see large scatter, but considering large uncertainties in pulsar distances and transverse velocities this is not surprising. We see some evidence that the simple model underestimates the time scale of low DM pulsars and overestimate that of high DM pulsars. However, we note that this could result from the time span of our data sets, which does not allow us to measure very short (< 30 days) or long (> 10 years) time scales.

Large dips have been observed in structure functions of many saturated pulsars at large lags. Similar features were reported by previous RISS studies~\citep[e.g.,][]{Kaspi}. These dips are caused by large modulations (or quasi-periodic structures) in the flux density time series. The flux density modulation is most significant for PSR J1740$-$3015, which is believed to be caused by ESEs~\citep{kcw+18}. While modulations in light curves of other pulsars are not as strong as that of PSR J1740$-$3015, those dips in structure functions show up at very similar lags, typically as $\sim1000$\,days. This indicates that ESE-like strong scattering events are likely to be causing observed large dips in other saturated pulsars. If this is the case then pulsar local ISM structures and non-Kolmogorov density fluctuations are playing an important role in the observed flux density variations of these pulsars. Detailed studies of ESE are beyond the scope of this paper, but our results show that the structure function can potentially be used to identify such events and well-calibrated pulsar monitoring is powerful in studying local ISM structures. Future wide-band observations with UWL will be more sensitive to such events as demonstrated by wide-band observations of extragalactic sources~\citep[e.g.,][]{bst+16}, and can provide additional information (e.g., DM and scintillation bandwidth measurements) during these extreme events.   

It has been suggested that the slope of the structure function reflects the distribution of ionised medium along the line of sight \citep[e.g.,][]{rnb86}. For a thin screen and a Kolmogorov spectrum, the expected slope is $\sim2.0$. If the ionised medium is evenly distributed along the line of sight, the slope approaches $\beta-3$ for small lags. \citet{Stinebring} found that none of the pulsars they analysed have slopes close to 2. Two-thirds of their sample show slopes in the range 0.4--1.0 and the rest clustered at 1.4--1.6. In Table~\ref{tab;ref} we list measured slopes of 39 pulsars. While a few pulsars show large slopes at $\sim1.7$--$1.8$, none of them is close to 2. In Fig.~\ref{fig;slope}, we show distributions of DMs and modulation indices for pulsars with a slope larger and smaller than one. We see evidence that pulsars with large slopes cluster at low DMs with narrow distributions of modulation index. This supports the picture that the slope approaches two for a thin scattering screen.  

\section{Conclusions}
\label{section;con}

We presented flux density measurements of 286 radio pulsars at 1.4\,GHz. This represents the largest sample of pulsars and the longest time span for studies of long-term flux density variation of pulsars. The sample covers a wide range of DM, and we find it to be correlated with both the measured modulation index and refractive scintillation time scale, although with a large scatter. This suggests that the ISM properties along different lines of sight in our Galaxy are significantly different. Monitoring of this sample of pulsars with the Parkes UWL receiver system is currently ongoing, which will enable us to measure the scintillation bandwidth of bright high DM pulsars at high frequencies. Combining our results with UWL data, we will be able to investigate different turbulence models and improve our understanding of RISS. The ultra-wide bandwidth will also allow us to detect and study ESEs and then probe local ISM structures and inhomogeneities along many lines of sight.

\section*{Acknowledgements}

The Parkes telescope is a part of the Australia Telescope National Facility which is funded by the Commonwealth of Australia for operation as a National Facility managed by CSIRO. The ATNF Pulsar Catalogue at \href{https://www.atnf.csiro.au/research/pulsar/psrcat/}{https://www.atnf.csiro.au/research/pulsar/psrcat/} was used for this work. Work at NRL is supported by NASA. R.M.S. is supported through ARC Future Fellowship FT190100155. KT is partially supported by JSPS KAKENHI Grant Number 16H05999 and 20H00180, and Bilateral Joint Research Projects of JSPS (KT).

\section*{Data availability}

The observations from the Parkes radio telescope are publicly available from \url{https://data.csiro.au/} after an 18 month embargo period. Note that all original data used in this paper are out of this embargo period and are available using the P574 observing project code. These data were processed and flux density measurements (which form the basis of this paper) are available from CSIRO Data Access Portal (DOI will be provided once accepted).

\bibliographystyle{mnras}
\bibliography{reference}

\begin{thebibliography}{}
\makeatletter
\relax
\def\mn@urlcharsother{\let\do\@makeother \do\$\do\&\do\#\do\^\do\_\do\%\do\~}
\def\mn@doi{\begingroup\mn@urlcharsother \@ifnextchar [ {\mn@doi@}
  {\mn@doi@[]}}
\def\mn@doi@[#1]#2{\def\@tempa{#1}\ifx\@tempa\@empty \href
  {http://dx.doi.org/#2} {doi:#2}\else \href {http://dx.doi.org/#2} {#1}\fi
  \endgroup}
\def\mn@eprint#1#2{\mn@eprint@#1:#2::\@nil}
\def\mn@eprint@arXiv#1{\href {http://arxiv.org/abs/#1} {{\tt arXiv:#1}}}
\def\mn@eprint@dblp#1{\href {http://dblp.uni-trier.de/rec/bibtex/#1.xml}
  {dblp:#1}}
\def\mn@eprint@#1:#2:#3:#4\@nil{\def\@tempa {#1}\def\@tempb {#2}\def\@tempc
  {#3}\ifx \@tempc \@empty \let \@tempc \@tempb \let \@tempb \@tempa \fi \ifx
  \@tempb \@empty \def\@tempb {arXiv}\fi \@ifundefined
  {mn@eprint@\@tempb}{\@tempb:\@tempc}{\expandafter \expandafter \csname
  mn@eprint@\@tempb\endcsname \expandafter{\@tempc}}}

\bibitem[\protect\citeauthoryear{{Anderson}, {Bania}, {Balser}, {Cunningham},
  {Wenger}, {Johnstone}  \& {Armentrout}}{{Anderson} et~al.}{2014}]{abb+14}
{Anderson} L.~D.,  {Bania} T.~M.,  {Balser} D.~S.,  {Cunningham} V.,  {Wenger}
  T.~V.,  {Johnstone} B.~M.,   {Armentrout} W.~P.,  2014, \mn@doi [\apjs]
  {10.1088/0067-0049/212/1/1}, \href
  {https://ui.adsabs.harvard.edu/abs/2014ApJS..212....1A} {212, 1}

\bibitem[\protect\citeauthoryear{{Armstrong}, {Rickett}  \&
  {Spangler}}{{Armstrong} et~al.}{1995}]{Armstrong}
{Armstrong} J.~W.,  {Rickett} B.~J.,   {Spangler} S.~R.,  1995, \mn@doi [\apj]
  {10.1086/175515}, \href
  {https://ui.adsabs.harvard.edu/abs/1995ApJ...443..209A} {443, 209}

\bibitem[\protect\citeauthoryear{{Ashton} et~al.,}{{Ashton}
  et~al.}{2019}]{Ashton}
{Ashton} G.,  et~al., 2019, \mn@doi [\apjs] {10.3847/1538-4365/ab06fc}, \href
  {https://ui.adsabs.harvard.edu/abs/2019ApJS..241...27A} {241, 27}

\bibitem[\protect\citeauthoryear{{Bannister}, {Stevens}, {Tuntsov}, {Walker},
  {Johnston}, {Reynolds}  \& {Bignall}}{{Bannister} et~al.}{2016}]{bst+16}
{Bannister} K.~W.,  {Stevens} J.,  {Tuntsov} A.~V.,  {Walker} M.~A.,
  {Johnston} S.,  {Reynolds} C.,   {Bignall} H.,  2016, \mn@doi [Science]
  {10.1126/science.aac7673}, \href
  {https://ui.adsabs.harvard.edu/abs/2016Sci...351..354B} {351, 354}

\bibitem[\protect\citeauthoryear{{Bhat}, {Gupta}  \& {Rao}}{{Bhat}
  et~al.}{1998}]{bgr98}
{Bhat} N.~D.~R.,  {Gupta} Y.,   {Rao} A.~P.,  1998, ApJ, 500, 262

\bibitem[\protect\citeauthoryear{Bhat, Rao  \& Gupta}{Bhat
  et~al.}{1999}]{brg99}
Bhat N. D.~R.,  Rao A.~P.,   Gupta Y.,  1999, ApJ, 514, 249

\bibitem[\protect\citeauthoryear{Blandford \& Narayan}{Blandford \&
  Narayan}{1985}]{bn85}
Blandford R.,  Narayan R.,  1985, MNRAS, 213, 591

\bibitem[\protect\citeauthoryear{{Burke-Spolaor} et~al.,}{{Burke-Spolaor}
  et~al.}{2012}]{bjb+12}
{Burke-Spolaor} S.,  et~al., 2012, \mn@doi [\mnras]
  {10.1111/j.1365-2966.2012.20998.x}, \href
  {https://ui.adsabs.harvard.edu/abs/2012MNRAS.423.1351B} {423, 1351}

\bibitem[\protect\citeauthoryear{Coles, Frehlich, Rickett  \& Codona}{Coles
  et~al.}{1987}]{cfr+87}
Coles W.~A.,  Frehlich R.~G.,  Rickett B.~J.,   Codona J.~L.,  1987, ApJ, 315,
  666

\bibitem[\protect\citeauthoryear{{Coles} et~al.,}{{Coles}
  et~al.}{2015}]{cks+15}
{Coles} W.~A.,  et~al., 2015, \mn@doi [\apj] {10.1088/0004-637X/808/2/113},
  \href {https://ui.adsabs.harvard.edu/abs/2015ApJ...808..113C} {808, 113}

\bibitem[\protect\citeauthoryear{{Cordes}}{{Cordes}}{1986}]{cor86}
{Cordes} J.~M.,  1986, ApJ, 311, 183

\bibitem[\protect\citeauthoryear{{Cordes} \& {Rickett}}{{Cordes} \&
  {Rickett}}{1998}]{cr98}
{Cordes} J.~M.,  {Rickett} B.~J.,  1998, ApJ, 507, 846

\bibitem[\protect\citeauthoryear{{Faucher-Gigu{\`e}re} \&
  {Kaspi}}{{Faucher-Gigu{\`e}re} \& {Kaspi}}{2006}]{fk06}
{Faucher-Gigu{\`e}re} C.-A.,  {Kaspi} V.~M.,  2006, \mn@doi [\apj]
  {10.1086/501516}, \href
  {https://ui.adsabs.harvard.edu/abs/2006ApJ...643..332F} {643, 332}

\bibitem[\protect\citeauthoryear{{Geyer} et~al.,}{{Geyer} et~al.}{2017}]{gkk17}
{Geyer} M.,  et~al., 2017, MNRAS, 470, 2659

\bibitem[\protect\citeauthoryear{Goodman \& Narayan}{Goodman \&
  Narayan}{1985}]{gn85}
Goodman J.,  Narayan R.,  1985, MNRAS, 214, 519

\bibitem[\protect\citeauthoryear{Gupta, Rickett  \& Coles}{Gupta
  et~al.}{1993}]{grc93}
Gupta Y.,  Rickett B.~J.,   Coles W.~A.,  1993, ApJ, 403, 183

\bibitem[\protect\citeauthoryear{{Hobbs} et~al.,}{{Hobbs}
  et~al.}{2020}]{hmd+20}
{Hobbs} G.,  et~al., 2020, \mn@doi [\pasa] {10.1017/pasa.2020.2}, \href
  {https://ui.adsabs.harvard.edu/abs/2020PASA...37...12H} {37, e012}

\bibitem[\protect\citeauthoryear{Johnston \& Kerr}{Johnston \&
  Kerr}{2018}]{Johnston2}
Johnston S.,  Kerr M.,  2018, MNRAS, 474, 4629

\bibitem[\protect\citeauthoryear{{Johnston}, {Manchester}, {Lyne}, {Bailes},
  {Kaspi}, {Qiao}  \& {D'Amico}}{{Johnston} et~al.}{1992}]{jml+92}
{Johnston} S.,  {Manchester} R.~N.,  {Lyne} A.~G.,  {Bailes} M.,  {Kaspi}
  V.~M.,  {Qiao} G.,   {D'Amico} N.,  1992, ApJ, 387, L37

\bibitem[\protect\citeauthoryear{{Johnston}, {Nicastro}  \&
  {Koribalski}}{{Johnston} et~al.}{1998}]{jnk98}
{Johnston} S.,  {Nicastro} L.,   {Koribalski} B.,  1998, MNRAS, 297, 108

\bibitem[\protect\citeauthoryear{Kaspi \& Stinebring}{Kaspi \&
  Stinebring}{1992}]{Kaspi}
Kaspi V.~M.,  Stinebring D.~R.,  1992, ApJ, 392, 530

\bibitem[\protect\citeauthoryear{{Keith} et~al.,}{{Keith}
  et~al.}{2013}]{kcs+13}
{Keith} M.~J.,  et~al., 2013, \mn@doi [\mnras] {10.1093/mnras/sts486}, \href
  {https://ui.adsabs.harvard.edu/abs/2013MNRAS.429.2161K} {429, 2161}

\bibitem[\protect\citeauthoryear{{Kerr}, {Hobbs}, {Johnston}  \&
  {Shannon}}{{Kerr} et~al.}{2016}]{khjs16}
{Kerr} M.,  {Hobbs} G.,  {Johnston} S.,   {Shannon} R.~M.,  2016, MNRAS, 455,
  1845

\bibitem[\protect\citeauthoryear{{Kerr}, {Coles}, {Ward}, {Johnston}, {Tuntsov}
   \& {Shannon}}{{Kerr} et~al.}{2018}]{kcw+18}
{Kerr} M.,  {Coles} W.~A.,  {Ward} C.~A.,  {Johnston} S.,  {Tuntsov} A.~V.,
  {Shannon} R.~M.,  2018, \mn@doi [\mnras] {10.1093/mnras/stx3101}, \href
  {https://ui.adsabs.harvard.edu/abs/2018MNRAS.474.4637K} {474, 4637}

\bibitem[\protect\citeauthoryear{{Kerr} et~al.,}{{Kerr} et~al.}{2020}]{krh+20}
{Kerr} M.,  et~al., 2020, \mn@doi [\pasa] {10.1017/pasa.2020.11}, \href
  {https://ui.adsabs.harvard.edu/abs/2020PASA...37...20K} {37, e020}

\bibitem[\protect\citeauthoryear{{Krishnakumar}, {Mitra}, {Naidu}, {Joshi}  \&
  {Manoharan}}{{Krishnakumar} et~al.}{2015}]{kmn15}
{Krishnakumar} M.~A.,  {Mitra} D.,  {Naidu} A.,  {Joshi} B.~C.,   {Manoharan}
  P.~K.,  2015, ApJ, 804, 23

\bibitem[\protect\citeauthoryear{{Labrecque}, {Rankin}  \&
  {Cordes}}{{Labrecque} et~al.}{1994}]{lrc94}
{Labrecque} D.~R.,  {Rankin} J.~M.,   {Cordes} J.~M.,  1994, \mn@doi [\aj]
  {10.1086/117198}, \href
  {https://ui.adsabs.harvard.edu/abs/1994AJ....108.1854L} {108, 1854}

\bibitem[\protect\citeauthoryear{{Lorimer} et~al.,}{{Lorimer}
  et~al.}{2006}]{lfl+06}
{Lorimer} D.~R.,  et~al., 2006, MNRAS, 372, 777

\bibitem[\protect\citeauthoryear{{Manchester}, {Hobbs}, {Teoh}  \&
  {Hobbs}}{{Manchester} et~al.}{2005}]{mhh05}
{Manchester} R.~N.,  {Hobbs} G.~B.,  {Teoh} A.,   {Hobbs} M.,  2005, \mn@doi
  [\aj] {10.1086/428488}, \href
  {https://ui.adsabs.harvard.edu/abs/2005AJ....129.1993M} {129, 1993}

\bibitem[\protect\citeauthoryear{{Pen} \& {King}}{{Pen} \& {King}}{2012}]{pk12}
{Pen} U.-L.,  {King} L.,  2012, \mn@doi [\mnras]
  {10.1111/j.1745-3933.2012.01223.x}, \href
  {https://ui.adsabs.harvard.edu/abs/2012MNRAS.421L.132P} {421, L132}

\bibitem[\protect\citeauthoryear{{Petroff}, {Keith}, {Johnston}, {van Straten}
  \& {Shannon}}{{Petroff} et~al.}{2013}]{pkj+13}
{Petroff} E.,  {Keith} M.~J.,  {Johnston} S.,  {van Straten} W.,   {Shannon}
  R.~M.,  2013, \mn@doi [\mnras] {10.1093/mnras/stt1401}, \href
  {https://ui.adsabs.harvard.edu/abs/2013MNRAS.435.1610P} {435, 1610}

\bibitem[\protect\citeauthoryear{{Reardon}, {Coles}, {Hobbs}, {Ord}, {Kerr},
  {Bailes}, {Bhat}  \& {Venkatraman Krishnan}}{{Reardon} et~al.}{2019}]{rch+19}
{Reardon} D.~J.,  {Coles} W.~A.,  {Hobbs} G.,  {Ord} S.,  {Kerr} M.,  {Bailes}
  M.,  {Bhat} N.~D.~R.,   {Venkatraman Krishnan} V.,  2019, \mn@doi [\mnras]
  {10.1093/mnras/stz643}, \href
  {https://ui.adsabs.harvard.edu/abs/2019MNRAS.485.4389R} {485, 4389}

\bibitem[\protect\citeauthoryear{{Rickett}}{{Rickett}}{1969}]{ric69}
{Rickett} B.~J.,  1969, \mn@doi [\nat] {10.1038/221158a0}, \href
  {https://ui.adsabs.harvard.edu/abs/1969Natur.221..158R} {221, 158}

\bibitem[\protect\citeauthoryear{{Rickett}, {Coles}  \& {Bourgois}}{{Rickett}
  et~al.}{1984}]{rcb84}
{Rickett} B.~J.,  {Coles} W.~A.,   {Bourgois} G.,  1984, A \& A, 134, 390

\bibitem[\protect\citeauthoryear{{Romani}, {Narayan}  \& {Blandford}}{{Romani}
  et~al.}{1986}]{rnb86}
{Romani} R.~W.,  {Narayan} R.,   {Blandford} R.,  1986, \mn@doi [\mnras]
  {10.1093/mnras/220.1.19}, \href
  {https://ui.adsabs.harvard.edu/abs/1986MNRAS.220...19R} {220, 19}

\bibitem[\protect\citeauthoryear{Sieber}{Sieber}{1982}]{Sieber}
Sieber W.,  1982, A \& A, 113, 311

\bibitem[\protect\citeauthoryear{{Speagle}}{{Speagle}}{2020}]{spe20}
{Speagle} J.~S.,  2020, \mn@doi [\mnras] {10.1093/mnras/staa278}, \href
  {https://ui.adsabs.harvard.edu/abs/2020MNRAS.493.3132S} {493, 3132}

\bibitem[\protect\citeauthoryear{{Stairs} et~al.,}{{Stairs}
  et~al.}{2019}]{slk+19}
{Stairs} I.~H.,  et~al., 2019, MNRAS, 485, 3230

\bibitem[\protect\citeauthoryear{Staveley-Smith et~al.,}{Staveley-Smith
  et~al.}{1996}]{Staveley-Smith}
Staveley-Smith L.,  et~al., 1996, Publications Astronomical Society of
  Australia, 13, 243

\bibitem[\protect\citeauthoryear{{Stinebring} \& {Condon}}{{Stinebring} \&
  {Condon}}{1990}]{sc90}
{Stinebring} D.~R.,  {Condon} J.~J.,  1990, ApJ, 352, 207

\bibitem[\protect\citeauthoryear{{Stinebring}, {Smirnova}, {Hankins}, {Hovis},
  {Kaspi}, {Kempner}, {Myers}  \& {Nice}}{{Stinebring}
  et~al.}{2000}]{Stinebring}
{Stinebring} D.~R.,  {Smirnova} T.~V.,  {Hankins} T.~H.,  {Hovis} J.~S.,
  {Kaspi} V.~M.,  {Kempner} J.~C.,  {Myers} E.,   {Nice} D.~J.,  2000, \mn@doi
  [\apj] {10.1086/309201}, \href
  {https://ui.adsabs.harvard.edu/abs/2000ApJ...539..300S} {539, 300}

\bibitem[\protect\citeauthoryear{{Tauris} et~al.,}{{Tauris}
  et~al.}{1994}]{tnj+94}
{Tauris} T.~M.,  et~al., 1994, ApJ, 428, L53

\bibitem[\protect\citeauthoryear{Thomas, Greene  \& James}{Thomas
  et~al.}{1990}]{Thomas}
Thomas B.~M.,  Greene K.~J.,   James G.~L.,  1990, IEEE Transactions on
  Antennas and Propagation, 38, 11

\bibitem[\protect\citeauthoryear{{Walker} \& {Wardle}}{{Walker} \&
  {Wardle}}{1998}]{ww98}
{Walker} M.,  {Wardle} M.,  1998, \mn@doi [\apjl] {10.1086/311332}, \href
  {https://ui.adsabs.harvard.edu/abs/1998ApJ...498L.125W} {498, L125}

\bibitem[\protect\citeauthoryear{{Walker}, {Tuntsov}, {Bignall}, {Reynolds},
  {Bannister}, {Johnston}, {Stevens}  \& {Ravi}}{{Walker}
  et~al.}{2017}]{wtb+17}
{Walker} M.~A.,  {Tuntsov} A.~V.,  {Bignall} H.,  {Reynolds} C.,  {Bannister}
  K.~W.,  {Johnston} S.,  {Stevens} J.,   {Ravi} V.,  2017, \mn@doi [\apj]
  {10.3847/1538-4357/aa705c}, \href
  {https://ui.adsabs.harvard.edu/abs/2017ApJ...843...15W} {843, 15}

\bibitem[\protect\citeauthoryear{{Wang}, {Manchester}, {Johnston}, {Rickett},
  {Zhang}, {Yusup}  \& {Chen}}{{Wang} et~al.}{2005}]{wmj+05}
{Wang} N.,  {Manchester} R.~N.,  {Johnston} S.,  {Rickett} B.,  {Zhang} J.,
  {Yusup} A.,   {Chen} M.,  2005, \mn@doi [\mnras]
  {10.1111/j.1365-2966.2005.08798.x}, \href
  {https://ui.adsabs.harvard.edu/abs/2005MNRAS.358..270W} {358, 270}

\bibitem[\protect\citeauthoryear{{Wang}, {Manchester}  \& {Johnston}}{{Wang}
  et~al.}{2007}]{wmj07}
{Wang} N.,  {Manchester} R.~N.,   {Johnston} S.,  2007, \mn@doi [\mnras]
  {10.1111/j.1365-2966.2007.11703.x}, \href
  {https://ui.adsabs.harvard.edu/abs/2007MNRAS.377.1383W} {377, 1383}

\bibitem[\protect\citeauthoryear{{Weltevrede}, {Edwards}  \&
  {Stappers}}{{Weltevrede} et~al.}{2006}]{wes06}
{Weltevrede} P.,  {Edwards} R.~T.,   {Stappers} B.~W.,  2006, \mn@doi [\aap]
  {10.1051/0004-6361:20053088}, \href
  {https://ui.adsabs.harvard.edu/abs/2006A&A...445..243W} {445, 243}

\bibitem[\protect\citeauthoryear{{Weltevrede} et~al.,}{{Weltevrede}
  et~al.}{2010}]{wjm+10}
{Weltevrede} P.,  et~al., 2010, \mn@doi [\pasa] {10.1071/AS09054}, \href
  {https://ui.adsabs.harvard.edu/abs/2010PASA...27...64W} {27, 64}

\bibitem[\protect\citeauthoryear{{Xie} et~al.,}{{Xie} et~al.}{2019}]{xwh+19}
{Xie} Y.-W.,  et~al., 2019, \mn@doi [Research in Astronomy and Astrophysics]
  {10.1088/1674-4527/19/7/103}, \href
  {https://ui.adsabs.harvard.edu/abs/2019RAA....19..103X} {19, 103}

\bibitem[\protect\citeauthoryear{{Xu} \& {Zhang}}{{Xu} \& {Zhang}}{2017}]{xz17}
{Xu} S.,  {Zhang} B.,  2017, \mn@doi [\apj] {10.3847/1538-4357/835/1/2}, \href
  {https://ui.adsabs.harvard.edu/abs/2017ApJ...835....2X} {835, 2}

\bibitem[\protect\citeauthoryear{You et~al.,}{You et~al.}{2007}]{You}
You X.~P.,  et~al., 2007, MNRAS, 378, 493

\makeatother
\end{thebibliography}

%
\clearpage
\appendix
\section{Table of observed pulsars}
\begin{table*}
\centering
\caption{Table of all 286 pulsars in our sample. $T$ is the time span of the data set. $N_{\rm{obs}}$ is the number of flux density measurements. $m$, $m_{\rm{n}}$, $m_{\rm{j}}$, $m_{\rm{d}}$ and $m_{\rm{r}}$ are measured modulation index and contributions from the measurement noise, pulse jitter, diffractive and refractive scintillation. $m_{\rm{corr}}$ is the noise and jitter corrected modulation index. $T_{\rm r}$ is the estimated refractive scintillation time-scale. Various entries in the ``Class" column are described in Section~\ref{section;res}.}
\label{tab:all}
\begin{tabular}{lcccccccccccccc}
\hline
$\rm{Name}$ & $P_{0}$ & $\rm{DM}$ & Dist & $T$ & $N_{\rm{obs}}$ & $\overline{S}$ & $m$ & $m_{\rm{n}}$ & $m_{\rm{j}}$ & $m_{\rm{corr}}$ & $m_{\rm{d}}$ & $m_{\rm{r}}$ & $T_{\rm{r}}$ & Class\\ 
& [s] & [$\rm{cm^{-3} pc}$] & [kpc] & [days] &  & [mJy] &  &  &  &  &  &  & [day] &  \\ 
\hline
\hline
J0034$-$0721 & 0.94 & 10.9 & 1.0 & 3754 & 77 & 11.21 & 1.93 & 0.01 & 0.04 & 1.93 & 2.41 & 0.66 & 1.0 & F-D\\ 
J0051$+$0423 & 0.35 & 13.9 & 1.2 & 3754 & 38 & 0.42 & 1.46 & 0.11 & 0.04 & 1.45 & 1.45 & 0.59 & 0.4 & F-D\\ 
J0108$-$1431 & 0.81 & 2.4 & 0.2 & 3845 & 144 & 1.49 & 0.22 & 0.03 & 0.04 & 0.21 & 16.58 & 1.20 & 0.0 & X\\ 
J0134$-$2937 & 0.14 & 21.8 & 25.0 & 3863 & 79 & 3.43 & 0.37 & 0.01 & 0.02 & 0.37 & 0.33 & 0.47 & 0.4 & F-D\\ 
J0152$-$1637 & 0.83 & 11.9 & 0.9 & 3760 & 62 & 2.03 & 0.93 & 0.01 & 0.05 & 0.93 & 2.25 & 0.63 & 0.6 & F-D\\ 
 & & & & & & & & & & & & & \\ 
J0206$-$4028 & 0.63 & 12.9 & 1.3 & 3760 & 72 & 0.80 & 0.84 & 0.03 & 0.05 & 0.84 & 1.84 & 0.61 & 0.6 & F-D\\ 
J0255$-$5304 & 0.45 & 15.9 & 1.5 & 2494 & 50 & 5.70 & 0.91 & 0.00 & 0.04 & 0.91 & 1.30 & 0.55 & 0.8 & F-D\\ 
J0343$-$3000 & 2.60 & 20.2 & 1.6 & 3803 & 32 & 1.07 & 1.59 & 0.03 & 0.10 & 1.59 & 0.57 & 0.49 & 0.7 & F-D\\ 
J0401$-$7608 & 0.55 & 21.7 & 1.0 & 4067 & 141 & 3.75 & 1.97 & 0.01 & 0.04 & 1.97 & 0.34 & 0.47 & 0.6 & F-D\\ 
J0448$-$2749 & 0.45 & 26.2 & 1.8 & 3807 & 36 & 1.72 & 0.59 & 0.02 & 0.04 & 0.59 & 0.34 & 0.42 & 1.1 & F-D\\ 
 & & & & & & & & & & & & & \\ 
J0452$-$1759 & 0.55 & 39.9 & 0.4 & 3784 & 85 & 15.17 & 0.65 & 0.00 & 0.04 & 0.65 & 0.20 & 0.33 & 13.7 & F-D\\ 
J0525$+$1115 & 0.35 & 79.4 & 1.8 & 1546 & 45 & 1.73 & 0.25 & 0.02 & 0.04 & 0.25 & 0.02 & 0.21 & 10.3 & F-R\\ 
J0536$-$7543 & 1.25 & 18.6 & 0.1 & 3955 & 125 & 12.01 & 1.41 & 0.00 & 0.06 & 1.41 & 0.92 & 0.51 & 1.3 & F-D\\ 
J0543$+$2329 & 0.25 & 77.7 & 1.6 & 3815 & 113 & 11.97 & 0.38 & 0.00 & 0.03 & 0.37 & 0.02 & 0.21 & 14.7 & F-R\\ 
J0601$-$0527 & 0.40 & 80.5 & 2.1 & 1546 & 46 & 2.50 & 0.12 & 0.01 & 0.05 & 0.11 & 0.02 & 0.21 & 18.2 & F-R\\ 
 & & & & & & & & & & & & & \\ 
J0614$+$2229 & 0.33 & 96.9 & 3.6 & 3815 & 114 & 3.25 & 0.15 & 0.01 & 0.05 & 0.14 & 0.05 & 0.18 & 269.2 & F-R\\ 
J0624$-$0424 & 1.04 & 70.8 & 1.9 & 1546 & 47 & 1.68 & 0.51 & 0.02 & 0.08 & 0.50 & 0.02 & 0.23 & 7.5 & F-R\\ 
J0627$+$0706 & 0.48 & 138.2 & 2.3 & 4002 & 120 & 1.32 & 0.18 & 0.02 & 0.04 & 0.18 & 0.00 & 0.14 & 32.2 & S\\ 
J0630$-$2834 & 1.24 & 34.4 & 0.3 & 3936 & 119 & 27.72 & 0.76 & 0.00 & 0.07 & 0.75 & 0.19 & 0.36 & 3.0 & F-D\\ 
J0631$+$1036 & 0.29 & 125.4 & 2.1 & 3782 & 31 & 0.92 & 0.16 & 0.03 & 0.03 & 0.16 & 0.00 & 0.15 & 25.2 & S\\ 
 & & & & & & & & & & & & & \\ 
J0659$+$1414 & 0.38 & 13.9 & 0.3 & 3815 & 113 & 2.82 & 0.67 & 0.02 & 0.04 & 0.67 & 2.26 & 0.59 & 0.9 & F-D\\ 
J0729$-$1448 & 0.25 & 91.7 & 2.7 & 4067 & 150 & 0.85 & 0.39 & 0.04 & 0.04 & 0.39 & 0.01 & 0.19 & 15.0 & F-R\\ 
J0729$-$1836 & 0.51 & 61.3 & 2.0 & 3703 & 85 & 1.97 & 0.57 & 0.02 & 0.04 & 0.57 & 0.04 & 0.25 & 9.8 & F-DR\\ 
J0738$-$4042 & 0.37 & 160.9 & 1.6 & 3727 & 126 & 102.19 & 0.05 & 0.00 & 0.02 & 0.05 & 0.00 & 0.13 & 76.4 & S\\ 
J0742$-$2822 & 0.17 & 73.7 & 2.0 & 4101 & 380 & 26.51 & 0.24 & 0.00 & 0.03 & 0.24 & 0.02 & 0.22 & 9.0 & F-R\\ 
 & & & & & & & & & & & & & \\ 
J0745$-$5353 & 0.21 & 121.4 & 0.6 & 4101 & 163 & 4.88 & 0.04 & 0.01 & 0.03 & 0.03 & 0.00 & 0.16 & 17.5 & F-J\\ 
J0809$-$4753 & 0.55 & 228.3 & 6.5 & 1647 & 51 & 2.65 & 0.03 & 0.01 & 0.06 & 100.00 & 0.00 & 0.10 & 62.1 & F-J\\ 
J0820$-$1350 & 1.24 & 40.9 & 1.9 & 1541 & 45 & 7.66 & 0.48 & 0.00 & 0.08 & 0.47 & 0.10 & 0.32 & 1.9 & F-DR\\ 
J0820$-$3826 & 0.12 & 195.6 & 4.1 & 3739 & 116 & 0.48 & 0.17 & 0.08 & 0.03 & 0.15 & 0.00 & 0.11 & 88.6 & F-R\\ 
J0820$-$4114 & 0.55 & 113.4 & 0.6 & 2346 & 53 & 7.34 & 0.28 & 0.01 & 0.05 & 0.27 & 0.00 & 0.16 & 10.7 & F-R\\ 
 & & & & & & & & & & & & & \\ 
J0834$-$4159 & 0.12 & 240.5 & 5.5 & 4067 & 132 & 0.27 & 0.27 & 0.09 & 0.02 & 0.25 & 0.00 & 0.10 & 158.5 & F-R\\ 
J0835$-$4510 & 0.09 & 68.0 & 0.3 & 4100 & 192 & 1081.78 & 0.15 & 0.00 & 0.02 & 0.15 & 0.03 & 0.23 & 10.3 & F-R\\ 
J0837$-$4135 & 0.75 & 147.3 & 1.5 & 1616 & 49 & 32.35 & 0.29 & 0.00 & 0.05 & 0.29 & 0.00 & 0.14 & 73.7 & F-R\\ 
J0842$-$4851 & 0.64 & 196.8 & 3.1 & 3838 & 90 & 1.07 & 0.23 & 0.02 & 0.05 & 0.22 & 0.00 & 0.11 & 78.2 & S\\ 
J0855$-$4644 & 0.06 & 236.4 & 5.6 & 4101 & 103 & 0.33 & 0.10 & 0.08 & 0.01 & 0.07 & 0.00 & 0.10 & 154.7 & F-N\\ 
 & & & & & & & & & & & & & \\ 
J0857$-$4424 & 0.33 & 184.4 & 2.8 & 4100 & 171 & 0.95 & 0.13 & 0.03 & 0.04 & 0.11 & 0.00 & 0.12 & 65.2 & S\\ 
J0901$-$4624 & 0.44 & 199.3 & 3.0 & 4100 & 132 & 0.45 & 0.18 & 0.06 & 0.05 & 0.17 & 0.00 & 0.11 & 79.4 & F-R\\ 
J0904$-$7459 & 0.55 & 51.1 & 1.1 & 2303 & 47 & 1.40 & 0.53 & 0.03 & 0.06 & 0.52 & 0.06 & 0.28 & 2.9 & F-R\\ 
J0905$-$5127 & 0.35 & 196.4 & 1.3 & 4100 & 137 & 1.08 & 0.20 & 0.03 & 0.05 & 0.19 & 0.00 & 0.11 & 51.0 & F-R\\ 
J0907$-$5157 & 0.25 & 103.7 & 0.3 & 1616 & 49 & 14.43 & 0.12 & 0.00 & 0.04 & 0.12 & 0.00 & 0.17 & 6.9 & I\\ 
 & & & & & & & & & & & & & \\ 
J0908$-$4913 & 0.11 & 180.4 & 1.0 & 4100 & 140 & 20.28 & 0.22 & 0.00 & 0.02 & 0.22 & 0.00 & 0.12 & 37.0 & F-R\\ 
J0924$-$5814 & 0.74 & 57.4 & 0.1 & 1616 & 51 & 5.34 & 0.37 & 0.01 & 0.06 & 0.37 & 0.02 & 0.26 & 1.2 & F-R\\ 
J0940$-$5428 & 0.09 & 134.6 & 0.4 & 4100 & 141 & 0.66 & 0.29 & 0.06 & 0.02 & 0.28 & 0.00 & 0.14 & 12.4 & F-R\\ 
J0942$-$5552 & 0.66 & 180.2 & 0.3 & 1616 & 50 & 7.36 & 0.21 & 0.01 & 0.06 & 0.21 & 0.00 & 0.12 & 20.2 & F-R\\ 
J0954$-$5430 & 0.47 & 201.6 & 0.4 & 4100 & 131 & 0.48 & 0.32 & 0.06 & 0.06 & 0.31 & 0.00 & 0.11 & 30.6 & F-R\\ 
 & & & & & & & & & & & & & \\ 
J1001$-$5507 & 1.44 & 130.3 & 0.3 & 1616 & 50 & 9.02 & 0.13 & 0.00 & 0.09 & 0.09 & 0.00 & 0.15 & 10.3 & S\\ 
J1003$-$4747 & 0.31 & 98.5 & 0.4 & 4100 & 175 & 1.40 & 0.35 & 0.02 & 0.04 & 0.34 & 0.01 & 0.18 & 6.4 & F-R\\ 
J1015$-$5719 & 0.14 & 278.1 & 2.7 & 4100 & 138 & 3.47 & 0.11 & 0.04 & 0.03 & 0.09 & 0.00 & 0.09 & 151.2 & F-R\\ 
J1016$-$5819 & 0.09 & 252.2 & 2.6 & 4100 & 131 & 0.36 & 0.14 & 0.08 & 0.02 & 0.12 & 0.00 & 0.09 & 120.0 & F-N\\ 
J1016$-$5857 & 0.11 & 394.5 & 3.2 & 4100 & 142 & 0.87 & 0.19 & 0.06 & 0.03 & 0.18 & 0.00 & 0.07 & 338.3 & F-R\\ 
\hline
\end{tabular}

\end{table*}

\begin{table*}
\centering
\contcaption{}
\begin{tabular}{lcccccccccccccc}
\hline
$\rm{Name}$ & $P_{0}$ & $\rm{DM}$ & Dist & $T$ & $N_{\rm{obs}}$ & $\overline{S}$ & $m$ & $m_{\rm{n}}$ & $m_{\rm{j}}$ & $m_{\rm{corr}}$ & $m_{\rm{d}}$ & $m_{\rm{r}}$ & $T_{\rm{r}}$ & Class\\ 
& [s] & [$\rm{cm^{-3} pc}$] & [kpc] & [days] &  & [mJy] &  &  &  &  &  &  & [day] &  \\ 
\hline
\hline
J1019$-$5749 & 0.16 & 1040.0 & 10.9 & 4018 & 138 & 4.64 & 0.06 & 0.05 & 0.03 & 0.02 & 0.00 & 0.03 & 4807.0 & F-N\\ 
J1020$-$6026 & 0.14 & 441.5 & 3.3 & 2202 & 76 & 0.29 & 0.13 & 0.10 & 0.01 & 0.08 & 0.00 & 0.06 & 435.1 & F-N\\ 
J1028$-$5819 & 0.09 & 96.5 & 1.4 & 3756 & 117 & 0.20 & 0.43 & 0.04 & 0.02 & 0.43 & 0.01 & 0.18 & 12.1 & F-R\\ 
J1034$-$3224 & 1.15 & 50.8 & 1.6 & 1616 & 56 & 5.29 & 0.46 & 0.02 & 0.05 & 0.46 & 0.04 & 0.28 & 3.6 & F-R\\ 
J1038$-$5831 & 0.66 & 72.7 & 0.9 & 1546 & 48 & 1.30 & 0.54 & 0.03 & 0.06 & 0.53 & 0.02 & 0.22 & 5.5 & F-R\\ 
 & & & & & & & & & & & & & \\ 
J1043$-$6116 & 0.29 & 448.9 & 3.1 & 4100 & 138 & 1.26 & 0.14 & 0.02 & 0.05 & 0.13 & 0.00 & 0.06 & 436.6 & F-R\\ 
J1047$-$6709 & 0.20 & 116.2 & 1.8 & 3910 & 52 & 2.66 & 0.30 & 0.01 & 0.03 & 0.30 & 0.00 & 0.16 & 19.7 & F-R\\ 
J1048$-$5832 & 0.12 & 128.7 & 2.9 & 4100 & 150 & 10.00 & 0.25 & 0.01 & 0.03 & 0.25 & 0.00 & 0.15 & 31.3 & F-R\\ 
J1049$-$5833 & 2.20 & 446.8 & 3.0 & 1531 & 49 & 0.68 & 0.50 & 0.03 & 0.07 & 0.49 & 0.00 & 0.06 & 428.0 & X\\ 
J1052$-$5954 & 0.18 & 491.9 & 3.1 & 2488 & 92 & 0.15 & 0.14 & 0.13 & 0.02 & 0.04 & 0.00 & 0.06 & 535.7 & F-N\\ 
 & & & & & & & & & & & & & \\ 
J1055$-$6022 & 0.95 & 590.0 & 3.6 & 2234 & 70 & 0.20 & 0.16 & 0.10 & 0.05 & 0.11 & 0.00 & 0.05 & 840.1 & F-N\\ 
J1055$-$6028 & 0.10 & 636.5 & 3.8 & 3815 & 160 & 1.01 & 0.15 & 0.05 & 0.03 & 0.15 & 0.00 & 0.05 & 1016.1 & F-R\\ 
J1057$-$5226 & 0.20 & 29.7 & 0.1 & 4100 & 150 & 8.11 & 1.10 & 0.01 & 0.02 & 1.10 & 0.28 & 0.39 & 5.2 & F-D\\ 
J1105$-$6107 & 0.06 & 271.2 & 2.4 & 4100 & 153 & 1.18 & 0.30 & 0.04 & 0.02 & 0.29 & 0.00 & 0.09 & 133.4 & S\\ 
J1112$-$6103 & 0.06 & 599.1 & 4.5 & 4100 & 162 & 2.91 & 0.14 & 0.05 & 0.02 & 0.13 & 0.00 & 0.05 & 970.0 & I\\ 
 & & & & & & & & & & & & & \\ 
J1115$-$6052 & 0.26 & 226.9 & 2.2 & 4100 & 131 & 0.43 & 0.11 & 0.07 & 0.04 & 0.07 & 0.00 & 0.10 & 88.3 & F-N\\ 
J1123$-$6259 & 0.27 & 223.3 & 2.2 & 4100 & 131 & 0.50 & 0.22 & 0.05 & 0.04 & 0.21 & 0.00 & 0.10 & 85.5 & F-R\\ 
J1138$-$6207 & 0.12 & 520.4 & 7.2 & 2488 & 83 & 0.68 & 0.09 & 0.08 & 0.02 & 0.04 & 0.00 & 0.06 & 913.0 & F-N\\ 
J1146$-$6030 & 0.27 & 111.7 & 1.6 & 1616 & 48 & 4.87 & 0.38 & 0.01 & 0.04 & 0.38 & 0.01 & 0.16 & 17.5 & F-R\\ 
J1156$-$5707 & 0.29 & 243.2 & 2.9 & 4100 & 131 & 0.26 & 0.29 & 0.08 & 0.04 & 0.27 & 0.00 & 0.09 & 116.7 & F-R\\ 
 & & & & & & & & & & & & & \\ 
J1157$-$6224 & 0.40 & 325.2 & 4.0 & 1616 & 49 & 12.99 & 0.14 & 0.00 & 0.05 & 0.13 & 0.00 & 0.08 & 253.9 & I\\ 
J1216$-$6223 & 0.37 & 790.0 & 18.3 & 2488 & 81 & 0.21 & 0.11 & 0.10 & 0.03 & 0.04 & 0.00 & 0.04 & 3494.6 & F-N\\ 
J1224$-$6407 & 0.22 & 97.7 & 4.0 & 4100 & 247 & 9.69 & 0.30 & 0.00 & 0.03 & 0.29 & 0.01 & 0.18 & 16.4 & F-R\\ 
J1243$-$6423 & 0.39 & 297.2 & 2.0 & 2346 & 58 & 34.65 & 0.20 & 0.00 & 0.05 & 0.20 & 0.00 & 0.08 & 148.7 & F-R\\ 
J1248$-$6344 & 0.20 & 433.0 & 10.7 & 2488 & 80 & 0.20 & 0.24 & 0.12 & 0.02 & 0.21 & 0.00 & 0.06 & 756.8 & F-N\\ 
 & & & & & & & & & & & & & \\ 
J1301$-$6305 & 0.18 & 374.0 & 10.7 & 4100 & 130 & 0.57 & 0.13 & 0.09 & 0.02 & 0.09 & 0.00 & 0.07 & 557.2 & F-N\\ 
J1302$-$6350 & 0.05 & 146.7 & 2.6 & 4100 & 142 & 4.12 & 0.38 & 0.03 & 0.02 & 0.38 & 0.01 & 0.14 & 133.6 & X\\ 
J1305$-$6203 & 0.43 & 471.0 & 11.6 & 4100 & 124 & 0.65 & 0.12 & 0.06 & 0.05 & 0.08 & 0.00 & 0.06 & 940.1 & F-N\\ 
J1306$-$6617 & 0.47 & 436.9 & 15.9 & 1616 & 53 & 4.78 & 0.04 & 0.01 & 0.05 & 100.00 & 0.00 & 0.06 & 939.2 & F-J\\ 
J1317$-$6302 & 0.26 & 678.1 & 12.8 & 1606 & 35 & 1.55 & 0.06 & 0.03 & 0.04 & 0.03 & 0.00 & 0.05 & 2119.1 & F-N\\ 
 & & & & & & & & & & & & & \\ 
J1319$-$6056 & 0.28 & 400.9 & 11.8 & 1606 & 48 & 1.29 & 0.05 & 0.02 & 0.04 & 0.01 & 0.00 & 0.07 & 677.8 & I\\ 
J1320$-$5359 & 0.28 & 97.1 & 2.2 & 4100 & 127 & 2.04 & 0.33 & 0.02 & 0.04 & 0.32 & 0.01 & 0.18 & 8.8 & F-R\\ 
J1326$-$5859 & 0.48 & 287.3 & 3.0 & 1606 & 48 & 17.67 & 0.04 & 0.00 & 0.05 & 100.00 & 0.00 & 0.08 & 169.6 & F-J\\ 
J1326$-$6408 & 0.79 & 502.7 & 12.0 & 1606 & 48 & 1.99 & 0.07 & 0.02 & 0.07 & 0.03 & 0.00 & 0.06 & 1095.7 & F-J\\ 
J1326$-$6700 & 0.54 & 209.6 & 6.7 & 2346 & 51 & 13.02 & 0.21 & 0.01 & 0.04 & 0.21 & 0.00 & 0.11 & 131.4 & F-R\\ 
 & & & & & & & & & & & & & \\ 
J1327$-$6222 & 0.53 & 318.8 & 4.0 & 1606 & 48 & 20.18 & 0.05 & 0.00 & 0.05 & 100.00 & 0.00 & 0.08 & 243.6 & F-J\\ 
J1327$-$6301 & 0.20 & 294.9 & 6.2 & 1606 & 48 & 4.54 & 0.13 & 0.01 & 0.03 & 0.13 & 0.00 & 0.08 & 258.0 & F-R\\ 
J1327$-$6400 & 0.28 & 679.0 & 13.0 & 2488 & 78 & 0.21 & 0.26 & 0.12 & 0.02 & 0.23 & 0.00 & 0.05 & 2143.2 & F-N\\ 
J1328$-$4357 & 0.53 & 42.0 & 1.4 & 3235 & 51 & 4.19 & 0.61 & 0.01 & 0.05 & 0.60 & 0.09 & 0.32 & 1.9 & F-DR\\ 
J1338$-$6204 & 1.24 & 640.3 & 12.4 & 1606 & 34 & 6.05 & 0.04 & 0.01 & 0.08 & 100.00 & 0.00 & 0.05 & 1848.3 & F-J\\ 
 & & & & & & & & & & & & & \\ 
J1340$-$6456 & 0.38 & 77.0 & 1.4 & 1606 & 35 & 0.99 & 0.42 & 0.03 & 0.05 & 0.41 & 0.02 & 0.21 & 7.6 & F-R\\ 
J1341$-$6220 & 0.19 & 719.6 & 12.6 & 4100 & 177 & 2.71 & 0.12 & 0.03 & 0.04 & 0.11 & 0.00 & 0.04 & 2384.8 & F-R\\ 
J1349$-$6130 & 0.26 & 284.5 & 5.5 & 4100 & 164 & 0.86 & 0.08 & 0.05 & 0.04 & 0.04 & 0.00 & 0.08 & 224.8 & F-N\\ 
J1352$-$6803 & 0.63 & 214.6 & 9.9 & 1606 & 36 & 1.06 & 0.26 & 0.03 & 0.06 & 0.25 & 0.00 & 0.10 & 167.4 & F-R\\ 
J1356$-$5521 & 0.51 & 174.2 & 7.3 & 1606 & 34 & 1.02 & 0.33 & 0.04 & 0.05 & 0.32 & 0.00 & 0.12 & 93.1 & F-R\\ 
 & & & & & & & & & & & & & \\ 
J1357$-$62 & 0.46 & 416.8 & 6.5 & 1606 & 53 & 12.94 & 0.04 & 0.00 & 0.05 & 100.00 & 0.00 & 0.06 & 543.7 & I\\ 
J1357$-$6429 & 0.17 & 128.5 & 3.1 & 4100 & 178 & 0.58 & 0.29 & 0.09 & 0.02 & 0.27 & 0.00 & 0.15 & 32.2 & F-R\\ 
J1359$-$6038 & 0.13 & 293.7 & 5.0 & 4100 & 186 & 12.73 & 0.07 & 0.00 & 0.02 & 0.06 & 0.00 & 0.08 & 229.4 & S\\ 
J1401$-$6357 & 0.84 & 98.0 & 1.8 & 1606 & 49 & 7.15 & 0.24 & 0.00 & 0.07 & 0.23 & 0.01 & 0.18 & 14.0 & F-R\\ 
J1406$-$6121 & 0.21 & 537.8 & 7.3 & 2488 & 89 & 0.49 & 0.14 & 0.12 & 0.02 & 0.06 & 0.00 & 0.05 & 985.0 & F-N\\ 
\hline
\end{tabular}

\end{table*}
\begin{table*}
\centering
\contcaption{}
\begin{tabular}{lcccccccccccccc}
\hline
$\rm{Name}$ & $P_{0}$ & $\rm{DM}$ & Dist & $T$ & $N_{\rm{obs}}$ & $\overline{S}$ & $m$ & $m_{\rm{n}}$ & $m_{\rm{j}}$ & $m_{\rm{corr}}$ & $m_{\rm{d}}$ & $m_{\rm{r}}$ & $T_{\rm{r}}$ & Class\\ 
& [s] & [$\rm{cm^{-3} pc}$] & [kpc] & [days] &  & [mJy] &  &  &  &  &  &  & [day] &  \\ 
\hline
\hline
J1410$-$6132 & 0.05 & 961.0 & 13.5 & 3967 & 149 & 3.20 & 0.08 & 0.06 & 0.01 & 0.05 & 0.00 & 0.04 & 4531.6 & F-N\\ 
J1412$-$6145 & 0.32 & 514.4 & 7.1 & 4100 & 159 & 0.82 & 0.10 & 0.09 & 0.04 & 0.02 & 0.00 & 0.06 & 885.5 & F-N\\ 
J1413$-$6141 & 0.29 & 670.6 & 8.5 & 4100 & 147 & 0.92 & 0.10 & 0.07 & 0.03 & 0.07 & 0.00 & 0.05 & 1688.0 & F-N\\ 
J1418$-$3921 & 1.10 & 60.5 & 4.1 & 1606 & 49 & 1.04 & 0.46 & 0.03 & 0.08 & 0.45 & 0.05 & 0.25 & 8.0 & F-R\\ 
J1420$-$6048 & 0.07 & 360.1 & 5.6 & 4100 & 170 & 1.21 & 0.09 & 0.05 & 0.02 & 0.07 & 0.00 & 0.07 & 373.1 & I\\ 
 & & & & & & & & & & & & & \\ 
J1424$-$5822 & 0.37 & 323.9 & 6.7 & 1606 & 50 & 1.37 & 0.11 & 0.03 & 0.05 & 0.10 & 0.00 & 0.08 & 325.9 & S\\ 
J1428$-$5530 & 0.57 & 82.4 & 1.9 & 1606 & 50 & 6.45 & 0.45 & 0.01 & 0.06 & 0.45 & 0.02 & 0.20 & 10.1 & F-R\\ 
J1430$-$6623 & 0.79 & 65.3 & 1.3 & 1606 & 49 & 14.94 & 0.62 & 0.00 & 0.07 & 0.61 & 0.03 & 0.24 & 6.7 & F-DR\\ 
J1435$-$5954 & 0.47 & 44.3 & 1.1 & 1606 & 50 & 1.55 & 0.39 & 0.03 & 0.05 & 0.39 & 0.08 & 0.31 & 2.2 & F-R\\ 
J1452$-$5851 & 0.39 & 260.5 & 4.5 & 2488 & 78 & 0.33 & 0.13 & 0.08 & 0.03 & 0.09 & 0.00 & 0.09 & 169.3 & F-N\\ 
 & & & & & & & & & & & & & \\ 
J1452$-$6036 & 0.15 & 349.5 & 6.1 & 4100 & 135 & 1.85 & 0.12 & 0.02 & 0.03 & 0.11 & 0.00 & 0.07 & 365.7 & F-R\\ 
J1453$-$6413 & 0.18 & 71.2 & 2.8 & 4100 & 173 & 19.71 & 0.46 & 0.00 & 0.03 & 0.46 & 0.02 & 0.22 & 7.8 & F-R\\ 
J1456$-$6843 & 0.26 & 8.6 & 0.4 & 3782 & 131 & 63.03 & 0.95 & 0.00 & 0.03 & 0.94 & 4.01 & 0.72 & 0.4 & F-D\\ 
J1509$-$5850 & 0.09 & 142.1 & 3.4 & 4100 & 129 & 0.22 & 0.18 & 0.15 & 0.01 & 0.08 & 0.00 & 0.14 & 41.4 & F-N\\ 
J1512$-$5759 & 0.13 & 627.5 & 6.8 & 4100 & 131 & 7.76 & 0.05 & 0.01 & 0.03 & 0.04 & 0.00 & 0.05 & 1317.8 & F-J\\ 
 & & & & & & & & & & & & & \\ 
J1513$-$5908 & 0.15 & 252.5 & 4.4 & 4100 & 136 & 1.58 & 0.09 & 0.06 & 0.03 & 0.05 & 0.00 & 0.09 & 156.8 & F-N\\ 
J1514$-$5925 & 0.15 & 194.0 & 3.9 & 2488 & 79 & 0.25 & 0.15 & 0.12 & 0.02 & 0.09 & 0.00 & 0.11 & 85.3 & F-N\\ 
J1515$-$5720 & 0.29 & 480.6 & 6.1 & 4100 & 130 & 0.23 & 0.19 & 0.10 & 0.03 & 0.16 & 0.00 & 0.06 & 709.4 & F-N\\ 
J1522$-$5829 & 0.40 & 199.9 & 3.9 & 1606 & 50 & 6.08 & 0.04 & 0.01 & 0.05 & 100.00 & 0.00 & 0.11 & 90.4 & S\\ 
J1524$-$5625 & 0.08 & 152.2 & 3.4 & 4100 & 147 & 1.34 & 0.11 & 0.05 & 0.02 & 0.09 & 0.00 & 0.13 & 47.8 & F-R\\ 
 & & & & & & & & & & & & & \\ 
J1524$-$5706 & 1.12 & 832.0 & 7.4 & 4100 & 131 & 0.42 & 0.08 & 0.07 & 0.09 & 100.00 & 0.00 & 0.04 & 2483.2 & F-N\\ 
J1530$-$5327 & 0.28 & 49.6 & 1.1 & 4100 & 158 & 0.78 & 0.39 & 0.05 & 0.04 & 0.38 & 0.07 & 0.29 & 2.8 & F-R\\ 
J1531$-$5610 & 0.08 & 110.4 & 2.8 & 4100 & 139 & 0.87 & 0.10 & 0.06 & 0.02 & 0.08 & 0.01 & 0.17 & 22.5 & F-N\\ 
J1534$-$5334 & 1.37 & 24.8 & 0.8 & 1606 & 50 & 7.39 & 0.62 & 0.00 & 0.09 & 0.61 & 0.60 & 0.44 & 1.8 & F-D\\ 
J1534$-$5405 & 0.29 & 190.8 & 3.8 & 1606 & 36 & 1.44 & 0.11 & 0.03 & 0.04 & 0.10 & 0.00 & 0.11 & 80.6 & S\\ 
 & & & & & & & & & & & & & \\ 
J1536$-$5433 & 0.88 & 147.5 & 3.3 & 1606 & 36 & 1.81 & 0.13 & 0.02 & 0.07 & 0.11 & 0.00 & 0.14 & 44.3 & F-J\\ 
J1538$-$5551 & 0.10 & 604.6 & 6.0 & 2488 & 79 & 0.45 & 0.10 & 0.10 & 0.01 & 0.01 & 0.00 & 0.05 & 1140.8 & F-N\\ 
J1539$-$5626 & 0.24 & 175.8 & 3.5 & 4100 & 129 & 4.86 & 0.07 & 0.01 & 0.04 & 0.05 & 0.00 & 0.12 & 19.4 & I\\ 
J1541$-$5535 & 0.30 & 426.1 & 5.2 & 4100 & 132 & 0.27 & 0.13 & 0.10 & 0.04 & 0.07 & 0.00 & 0.06 & 508.2 & F-N\\ 
J1543$-$5459 & 0.38 & 345.9 & 4.7 & 4100 & 129 & 0.89 & 0.10 & 0.07 & 0.05 & 0.05 & 0.00 & 0.07 & 311.9 & F-N\\ 
 & & & & & & & & & & & & & \\ 
J1544$-$5308 & 0.18 & 35.2 & 0.9 & 1547 & 49 & 4.75 & 0.35 & 0.01 & 0.03 & 0.35 & 0.15 & 0.36 & 1.4 & F-DR\\ 
J1548$-$5607 & 0.17 & 314.7 & 5.7 & 4100 & 129 & 1.45 & 0.05 & 0.04 & 0.03 & 0.02 & 0.00 & 0.08 & 282.9 & F-N\\ 
J1549$-$4848 & 0.29 & 56.0 & 1.3 & 4100 & 130 & 1.37 & 0.39 & 0.03 & 0.04 & 0.39 & 0.05 & 0.26 & 3.9 & F-R\\ 
J1551$-$5310 & 0.45 & 491.6 & 5.9 & 2488 & 80 & 0.93 & 0.11 & 0.08 & 0.04 & 0.06 & 0.00 & 0.06 & 732.2 & F-N\\ 
J1555$-$3134 & 0.52 & 73.0 & 5.3 & 3235 & 53 & 3.72 & 0.14 & 0.01 & 0.05 & 0.13 & 0.03 & 0.22 & 13.3 & F-R\\ 
 & & & & & & & & & & & & & \\ 
J1557$-$4258 & 0.33 & 144.5 & 8.6 & 1606 & 50 & 2.93 & 0.13 & 0.01 & 0.04 & 0.13 & 0.00 & 0.14 & 48.2 & S\\ 
J1559$-$4438 & 0.26 & 56.1 & 2.3 & 1606 & 50 & 37.32 & 0.29 & 0.00 & 0.04 & 0.29 & 0.07 & 0.26 & 10.8 & F-R\\ 
J1600$-$5044 & 0.19 & 262.8 & 6.9 & 4100 & 132 & 21.45 & 0.04 & 0.00 & 0.03 & 0.03 & 0.00 & 0.09 & 140.5 & I\\ 
J1600$-$5751 & 0.19 & 176.6 & 4.2 & 4100 & 129 & 2.38 & 0.19 & 0.02 & 0.04 & 0.18 & 0.00 & 0.12 & 72.4 & F-R\\ 
J1601$-$5335 & 0.29 & 195.2 & 3.6 & 2488 & 81 & 0.22 & 0.25 & 0.13 & 0.03 & 0.22 & 0.00 & 0.11 & 82.6 & F-N\\ 
 & & & & & & & & & & & & & \\ 
J1602$-$5100 & 0.86 & 170.8 & 8.0 & 4067 & 129 & 7.38 & 0.13 & 0.01 & 0.06 & 0.12 & 0.00 & 0.12 & 93.5 & F-R\\ 
J1604$-$4909 & 0.33 & 140.8 & 3.2 & 1606 & 49 & 5.99 & 0.28 & 0.01 & 0.04 & 0.28 & 0.00 & 0.14 & 52.0 & F-R\\ 
J1605$-$5257 & 0.66 & 35.1 & 0.9 & 1606 & 49 & 20.10 & 0.90 & 0.00 & 0.06 & 0.89 & 0.15 & 0.36 & 1.3 & F-DR\\ 
J1611$-$5209 & 0.18 & 127.3 & 3.0 & 4100 & 130 & 1.56 & 0.31 & 0.03 & 0.04 & 0.30 & 0.01 & 0.15 & 30.9 & F-R\\ 
J1613$-$4714 & 0.38 & 161.2 & 3.5 & 1606 & 49 & 1.42 & 0.13 & 0.02 & 0.05 & 0.12 & 0.00 & 0.13 & 55.0 & S\\ 
 & & & & & & & & & & & & & \\ 
J1614$-$5048 & 0.23 & 582.4 & 5.2 & 4100 & 153 & 3.60 & 0.06 & 0.02 & 0.04 & 0.04 & 0.00 & 0.05 & 977.9 & I\\ 
J1617$-$5055 & 0.07 & 467.0 & 4.7 & 2488 & 129 & 0.45 & 0.15 & 0.14 & 0.01 & 0.07 & 0.00 & 0.06 & 590.3 & F-N\\ 
J1623$-$4256 & 0.36 & 295.0 & 21.6 & 1606 & 49 & 2.77 & 0.10 & 0.02 & 0.05 & 0.08 & 0.00 & 0.08 & 480.6 & F-J\\ 
J1626$-$4537 & 0.37 & 237.0 & 5.0 & 1606 & 48 & 1.17 & 0.18 & 0.03 & 0.05 & 0.17 & 0.00 & 0.10 & 146.1 & F-R\\ 
J1626$-$4807 & 0.29 & 817.0 & 7.5 & 2909 & 80 & 0.54 & 0.12 & 0.09 & 0.02 & 0.08 & 0.00 & 0.04 & 2393.3 & F-N\\ 
\hline
\end{tabular}

\end{table*}
\begin{table*}
\centering
\contcaption{}
\begin{tabular}{lcccccccccccccc}
\hline
$\rm{Name}$ & $P_{0}$ & $\rm{DM}$ & Dist & $T$ & $N_{\rm{obs}}$ & $\overline{S}$ & $m$ & $m_{\rm{n}}$ & $m_{\rm{j}}$ & $m_{\rm{corr}}$ & $m_{\rm{d}}$ & $m_{\rm{r}}$ & $T_{\rm{r}}$ & Class\\ 
& [s] & [$\rm{cm^{-3} pc}$] & [kpc] & [days] &  & [mJy] &  &  &  &  &  &  & [day] &  \\ 
\hline
\hline
J1627$-$4706 & 0.14 & 456.1 & 6.4 & 2160 & 70 & 0.21 & 0.14 & 0.13 & 0.01 & 0.06 & 0.00 & 0.06 & 653.3 & F-N\\ 
J1630$-$4733 & 0.58 & 498.0 & 5.0 & 1606 & 35 & 9.68 & 0.10 & 0.02 & 0.06 & 0.08 & 0.00 & 0.06 & 693.8 & F-J\\ 
J1632$-$4621 & 1.71 & 562.9 & 10.1 & 1606 & 50 & 0.79 & 0.10 & 0.03 & 0.10 & 100.00 & 0.00 & 0.05 & 1275.0 & F-J\\ 
J1632$-$4757 & 0.23 & 574.2 & 4.8 & 2488 & 78 & 0.53 & 0.12 & 0.10 & 0.02 & 0.07 & 0.00 & 0.05 & 920.2 & F-N\\ 
J1632$-$4818 & 0.81 & 758.0 & 5.3 & 4100 & 115 & 0.47 & 0.10 & 0.08 & 0.04 & 0.05 & 0.00 & 0.04 & 1726.4 & F-N\\ 
 & & & & & & & & & & & & & \\ 
J1633$-$4453 & 0.44 & 474.1 & 14.9 & 1606 & 49 & 2.55 & 0.12 & 0.01 & 0.05 & 0.11 & 0.00 & 0.06 & 1081.7 & F-J\\ 
J1633$-$5015 & 0.35 & 398.4 & 6.0 & 1606 & 50 & 7.60 & 0.04 & 0.00 & 0.04 & 100.00 & 0.00 & 0.07 & 476.0 & F-J\\ 
J1636$-$4440 & 0.21 & 449.0 & 12.4 & 2347 & 72 & 0.30 & 0.14 & 0.11 & 0.02 & 0.09 & 0.00 & 0.06 & 880.9 & F-N\\ 
J1637$-$4553 & 0.12 & 193.2 & 3.4 & 4100 & 161 & 1.60 & 0.29 & 0.03 & 0.03 & 0.28 & 0.00 & 0.11 & 79.3 & F-R\\ 
J1637$-$4642 & 0.15 & 419.1 & 4.4 & 4100 & 129 & 0.94 & 0.11 & 0.08 & 0.03 & 0.08 & 0.00 & 0.06 & 453.7 & F-N\\ 
 & & & & & & & & & & & & & \\ 
J1638$-$4417 & 0.12 & 436.4 & 12.0 & 4100 & 129 & 0.28 & 0.12 & 0.11 & 0.02 & 0.04 & 0.00 & 0.06 & 815.8 & F-N\\ 
J1638$-$4608 & 0.28 & 423.1 & 4.6 & 4100 & 130 & 0.42 & 0.08 & 0.07 & 0.03 & 0.02 & 0.00 & 0.06 & 471.2 & F-N\\ 
J1638$-$4725 & 0.76 & 552.1 & 4.7 & 3885 & 132 & 0.55 & 0.89 & 0.12 & 0.04 & 0.89 & 0.00 & 0.05 & 838.7 & X\\ 
J1640$-$4715 & 0.52 & 586.3 & 4.9 & 4100 & 129 & 1.57 & 0.07 & 0.05 & 0.06 & 100.00 & 0.00 & 0.05 & 968.4 & F-N\\ 
J1643$-$4505 & 0.24 & 478.6 & 4.7 & 3815 & 122 & 0.43 & 0.12 & 0.08 & 0.03 & 0.09 & 0.00 & 0.06 & 621.5 & F-N\\ 
 & & & & & & & & & & & & & \\ 
J1644$-$4559 & 0.46 & 478.8 & 4.5 & 3989 & 67 & 324.82 & 0.03 & 0.00 & 0.03 & 100.00 & 0.00 & 0.06 & 606.0 & F-J\\ 
J1646$-$4346 & 0.23 & 490.4 & 6.2 & 4100 & 131 & 1.45 & 0.26 & 0.05 & 0.04 & 0.25 & 0.00 & 0.06 & 749.8 & X\\ 
J1646$-$6831 & 1.79 & 43.0 & 1.2 & 3235 & 51 & 4.43 & 0.40 & 0.01 & 0.09 & 0.39 & 0.09 & 0.31 & 2.3 & F-R\\ 
J1648$-$3256 & 0.72 & 128.3 & 6.7 & 1606 & 31 & 0.78 & 0.27 & 0.03 & 0.06 & 0.26 & 0.01 & 0.15 & 47.3 & F-R\\ 
J1648$-$4611 & 0.16 & 392.3 & 4.5 & 4100 & 125 & 0.62 & 0.13 & 0.09 & 0.02 & 0.09 & 0.00 & 0.07 & 397.7 & F-N\\ 
 & & & & & & & & & & & & & \\ 
J1649$-$4653 & 0.56 & 331.0 & 5.0 & 4100 & 126 & 0.39 & 0.16 & 0.11 & 0.05 & 0.10 & 0.00 & 0.08 & 293.4 & F-N\\ 
J1650$-$4502 & 0.38 & 320.2 & 3.9 & 4100 & 127 & 0.58 & 0.27 & 0.07 & 0.05 & 0.26 & 0.00 & 0.08 & 243.3 & S\\ 
J1650$-$4921 & 0.16 & 229.3 & 5.3 & 3772 & 117 & 0.26 & 0.15 & 0.10 & 0.03 & 0.11 & 0.00 & 0.10 & 140.8 & F-N\\ 
J1651$-$4246 & 0.84 & 482.0 & 5.2 & 2304 & 52 & 21.19 & 0.04 & 0.00 & 0.07 & 100.00 & 0.00 & 0.06 & 660.6 & F-J\\ 
J1651$-$5222 & 0.64 & 179.1 & 6.3 & 3235 & 50 & 3.93 & 0.21 & 0.01 & 0.06 & 0.21 & 0.00 & 0.12 & 91.4 & F-R\\ 
 & & & & & & & & & & & & & \\ 
J1652$-$2404 & 1.70 & 68.4 & 3.4 & 1606 & 46 & 1.55 & 0.29 & 0.02 & 0.10 & 0.27 & 0.03 & 0.23 & 9.3 & F-R\\ 
J1653$-$3838 & 0.31 & 207.2 & 5.4 & 1606 & 48 & 1.79 & 0.08 & 0.02 & 0.04 & 0.07 & 0.00 & 0.11 & 115.3 & F-R\\ 
J1653$-$4249 & 0.61 & 416.1 & 4.5 & 1606 & 46 & 1.60 & 0.07 & 0.02 & 0.06 & 0.03 & 0.00 & 0.06 & 450.5 & F-J\\ 
J1700$-$3312 & 1.36 & 167.0 & 7.4 & 1606 & 47 & 1.50 & 0.24 & 0.02 & 0.09 & 0.23 & 0.00 & 0.12 & 85.7 & F-R\\ 
J1701$-$3726 & 2.45 & 303.4 & 12.9 & 2346 & 53 & 4.16 & 0.25 & 0.01 & 0.06 & 0.24 & 0.00 & 0.08 & 395.1 & X\\ 
 & & & & & & & & & & & & & \\ 
J1702$-$4128 & 0.18 & 367.1 & 4.0 & 4100 & 129 & 1.30 & 0.18 & 0.06 & 0.03 & 0.17 & 0.00 & 0.07 & 326.1 & F-R\\ 
J1702$-$4306 & 0.22 & 538.4 & 5.0 & 3815 & 107 & 0.40 & 0.11 & 0.08 & 0.03 & 0.08 & 0.00 & 0.05 & 820.4 & F-N\\ 
J1702$-$4310 & 0.24 & 377.6 & 4.3 & 4100 & 127 & 0.89 & 0.10 & 0.06 & 0.04 & 0.08 & 0.00 & 0.07 & 361.3 & F-N\\ 
J1703$-$3241 & 1.21 & 110.3 & 3.2 & 1606 & 48 & 8.44 & 0.13 & 0.01 & 0.08 & 0.10 & 0.01 & 0.17 & 23.8 & F-J\\ 
J1703$-$4851 & 1.40 & 150.3 & 4.1 & 1606 & 47 & 1.17 & 0.78 & 0.02 & 0.09 & 0.78 & 0.00 & 0.13 & 51.3 & X\\ 
 & & & & & & & & & & & & & \\ 
J1705$-$1906 & 0.30 & 22.9 & 0.8 & 4067 & 125 & 8.17 & 0.75 & 0.01 & 0.04 & 0.75 & 0.38 & 0.46 & 0.5 & F-DR\\ 
J1705$-$3423 & 0.26 & 146.4 & 3.8 & 1606 & 48 & 5.35 & 0.10 & 0.01 & 0.04 & 0.09 & 0.00 & 0.14 & 47.0 & S\\ 
J1705$-$3950 & 0.32 & 207.2 & 3.4 & 4100 & 126 & 1.60 & 0.17 & 0.04 & 0.05 & 0.16 & 0.00 & 0.11 & 91.6 & F-R\\ 
J1707$-$4053 & 0.58 & 360.0 & 4.0 & 3777 & 34 & 11.35 & 0.02 & 0.01 & 0.06 & 100.00 & 0.00 & 0.07 & 314.2 & F-J\\ 
J1707$-$4729 & 0.27 & 268.3 & 12.8 & 1606 & 48 & 2.54 & 0.08 & 0.02 & 0.04 & 0.07 & 0.00 & 0.09 & 304.0 & F-J\\ 
 & & & & & & & & & & & & & \\ 
J1708$-$3426 & 0.69 & 190.7 & 4.7 & 1606 & 46 & 2.22 & 0.11 & 0.02 & 0.06 & 0.09 & 0.00 & 0.11 & 90.3 & I\\ 
J1709$-$1640 & 0.65 & 24.9 & 0.6 & 1547 & 46 & 11.25 & 0.93 & 0.00 & 0.06 & 0.93 & 0.51 & 0.44 & 1.4 & F-D\\ 
J1709$-$4429 & 0.10 & 75.7 & 2.6 & 4100 & 149 & 11.42 & 0.09 & 0.01 & 0.03 & 0.08 & 0.02 & 0.22 & 10.0 & F-R\\ 
J1715$-$3903 & 0.28 & 314.0 & 3.7 & 4100 & 127 & 0.70 & 0.18 & 0.09 & 0.04 & 0.15 & 0.00 & 0.08 & 228.1 & F-N\\ 
J1715$-$4034 & 2.07 & 254.0 & 3.9 & 1606 & 47 & 1.78 & 0.13 & 0.02 & 0.11 & 0.07 & 0.00 & 0.09 & 149.8 & F-J\\ 
 & & & & & & & & & & & & & \\ 
J1717$-$3425 & 0.66 & 587.7 & 25.0 & 1606 & 48 & 3.78 & 0.03 & 0.01 & 0.06 & 100.00 & 0.00 & 0.05 & 2195.9 & F-J\\ 
J1717$-$5800 & 0.32 & 125.2 & 11.2 & 2314 & 68 & 0.44 & 0.24 & 0.07 & 0.03 & 0.23 & 0.01 & 0.15 & 58.1 & F-R\\ 
J1718$-$3825 & 0.07 & 247.5 & 3.5 & 4100 & 125 & 1.93 & 0.15 & 0.04 & 0.02 & 0.14 & 0.00 & 0.09 & 133.9 & S\\ 
J1719$-$4006 & 0.19 & 386.6 & 9.8 & 1606 & 33 & 1.38 & 0.06 & 0.03 & 0.03 & 0.04 & 0.00 & 0.07 & 572.0 & F-N\\ 
J1720$-$2933 & 0.62 & 42.6 & 1.1 & 1606 & 48 & 1.96 & 0.32 & 0.02 & 0.06 & 0.31 & 0.09 & 0.32 & 2.1 & F-R\\ 
\hline
\end{tabular}

\end{table*}
\begin{table*}
\centering
\contcaption{}
\begin{tabular}{lcccccccccccccc}
\hline
$\rm{Name}$ & $P_{0}$ & $\rm{DM}$ & Dist & $T$ & $N_{\rm{obs}}$ & $\overline{S}$ & $m$ & $m_{\rm{n}}$ & $m_{\rm{j}}$ & $m_{\rm{corr}}$ & $m_{\rm{d}}$ & $m_{\rm{r}}$ & $T_{\rm{r}}$ & Class\\ 
& [s] & [$\rm{cm^{-3} pc}$] & [kpc] & [days] &  & [mJy] &  &  &  &  &  &  & [day] &  \\ 
\hline
\hline
J1721$-$3532 & 0.28 & 496.0 & 4.6 & 4100 & 166 & 16.53 & 0.04 & 0.01 & 0.04 & 100.00 & 0.00 & 0.06 & 659.8 & F-J\\ 
J1722$-$3207 & 0.48 & 126.1 & 2.9 & 1606 & 49 & 5.32 & 0.13 & 0.01 & 0.05 & 0.12 & 0.00 & 0.15 & 16.3 & S\\ 
J1722$-$3632 & 0.40 & 416.2 & 4.0 & 1606 & 48 & 2.83 & 0.09 & 0.02 & 0.05 & 0.07 & 0.00 & 0.06 & 425.4 & F-J\\ 
J1722$-$3712 & 0.24 & 99.5 & 2.5 & 4100 & 130 & 3.61 & 0.16 & 0.01 & 0.04 & 0.16 & 0.01 & 0.18 & 17.0 & F-R\\ 
J1723$-$3659 & 0.20 & 254.4 & 3.5 & 4100 & 128 & 2.09 & 0.10 & 0.03 & 0.04 & 0.09 & 0.00 & 0.09 & 142.0 & I\\ 
 & & & & & & & & & & & & & \\ 
J1726$-$3530 & 1.11 & 718.0 & 4.7 & 3941 & 102 & 0.39 & 0.16 & 0.11 & 0.05 & 0.11 & 0.00 & 0.04 & 1452.6 & F-N\\ 
J1727$-$2739 & 1.29 & 147.0 & 4.0 & 1606 & 48 & 1.93 & 0.55 & 0.02 & 0.05 & 0.54 & 0.00 & 0.14 & 48.4 & X\\ 
J1730$-$3350 & 0.14 & 261.3 & 3.5 & 4100 & 135 & 4.20 & 0.09 & 0.02 & 0.03 & 0.08 & 0.00 & 0.09 & 150.0 & S\\ 
J1731$-$4744 & 0.83 & 123.1 & 0.7 & 4100 & 130 & 28.01 & 0.33 & 0.00 & 0.06 & 0.33 & 0.00 & 0.15 & 8.4 & F-R\\ 
J1733$-$2228 & 0.87 & 41.1 & 1.1 & 1606 & 48 & 3.69 & 0.25 & 0.02 & 0.07 & 0.24 & 0.10 & 0.32 & 2.0 & F-R\\ 
 & & & & & & & & & & & & & \\ 
J1733$-$3716 & 0.34 & 153.2 & 3.1 & 4100 & 126 & 3.38 & 0.13 & 0.02 & 0.05 & 0.12 & 0.00 & 0.13 & 14.9 & F-R\\ 
J1734$-$3333 & 1.17 & 578.0 & 4.5 & 2451 & 80 & 0.86 & 0.19 & 0.09 & 0.04 & 0.15 & 0.00 & 0.05 & 895.7 & F-N\\ 
J1735$-$3258 & 0.35 & 758.0 & 5.0 & 2451 & 81 & 0.65 & 0.17 & 0.11 & 0.02 & 0.13 & 0.00 & 0.04 & 1670.2 & F-N\\ 
J1737$-$3137 & 0.45 & 488.1 & 4.2 & 4100 & 138 & 0.93 & 0.11 & 0.07 & 0.05 & 0.07 & 0.00 & 0.06 & 606.7 & F-N\\ 
J1738$-$2955 & 0.44 & 223.4 & 3.5 & 2488 & 84 & 0.19 & 0.22 & 0.14 & 0.03 & 0.16 & 0.00 & 0.10 & 107.6 & I\\ 
 & & & & & & & & & & & & & \\ 
J1738$-$3211 & 0.77 & 49.6 & 1.3 & 1606 & 47 & 2.26 & 0.09 & 0.02 & 0.07 & 0.06 & 0.06 & 0.29 & 3.0 & F-J\\ 
J1739$-$2903 & 0.32 & 138.6 & 2.9 & 4100 & 128 & 4.26 & 0.09 & 0.01 & 0.03 & 0.08 & 0.00 & 0.14 & 36.5 & F-R\\ 
J1739$-$3023 & 0.11 & 170.5 & 3.1 & 4100 & 130 & 0.81 & 0.21 & 0.07 & 0.03 & 0.20 & 0.00 & 0.12 & 57.7 & F-R\\ 
J1739$-$3131 & 0.53 & 600.1 & 4.4 & 1606 & 31 & 6.41 & 0.04 & 0.01 & 0.05 & 100.00 & 0.00 & 0.05 & 963.6 & F-J\\ 
J1740$-$3015 & 0.61 & 152.0 & 0.4 & 4100 & 174 & 8.73 & 0.18 & 0.00 & 0.05 & 0.17 & 0.00 & 0.13 & 16.4 & S\\ 
 & & & & & & & & & & & & & \\ 
J1741$-$2733 & 0.89 & 149.2 & 3.1 & 1606 & 32 & 1.67 & 0.12 & 0.03 & 0.07 & 0.09 & 0.00 & 0.13 & 43.7 & I\\ 
J1741$-$3016 & 1.89 & 382.0 & 3.9 & 1606 & 49 & 1.93 & 0.25 & 0.03 & 0.10 & 0.23 & 0.00 & 0.07 & 350.0 & F-R\\ 
J1741$-$3927 & 0.51 & 158.5 & 4.6 & 1606 & 44 & 5.11 & 0.12 & 0.01 & 0.05 & 0.10 & 0.00 & 0.13 & 39.9 & S\\ 
J1743$-$3150 & 2.41 & 193.1 & 3.3 & 1606 & 45 & 2.16 & 0.11 & 0.02 & 0.12 & 100.00 & 0.00 & 0.11 & 77.8 & F-J\\ 
J1745$-$3040 & 0.37 & 88.4 & 0.2 & 4100 & 151 & 19.26 & 0.26 & 0.00 & 0.04 & 0.26 & 0.02 & 0.19 & 24.3 & F-R\\ 
 & & & & & & & & & & & & & \\ 
J1749$-$3002 & 0.61 & 509.4 & 12.7 & 1548 & 44 & 4.14 & 0.05 & 0.02 & 0.06 & 100.00 & 0.00 & 0.06 & 1160.8 & F-J\\ 
J1750$-$3157 & 0.91 & 206.3 & 4.3 & 1606 & 32 & 1.65 & 0.15 & 0.03 & 0.07 & 0.13 & 0.00 & 0.11 & 102.1 & F-R\\ 
J1751$-$4657 & 0.74 & 20.4 & 0.7 & 2140 & 49 & 6.62 & 0.75 & 0.00 & 0.06 & 0.74 & 0.63 & 0.49 & 0.7 & F-DR\\ 
J1752$-$2806 & 0.56 & 50.4 & 0.2 & 1548 & 42 & 41.10 & 0.29 & 0.00 & 0.06 & 0.28 & 0.27 & 0.28 & 61.1 & F-R\\ 
J1757$-$2421 & 0.23 & 179.5 & 3.1 & 4100 & 135 & 7.09 & 0.09 & 0.01 & 0.04 & 0.08 & 0.00 & 0.12 & 64.7 & S\\ 
 & & & & & & & & & & & & & \\ 
J1801$-$2154 & 0.38 & 386.0 & 4.1 & 2488 & 77 & 0.22 & 0.17 & 0.12 & 0.03 & 0.12 & 0.00 & 0.07 & 369.5 & F-N\\ 
J1801$-$2304 & 0.42 & 1073.9 & 4.0 & 4100 & 139 & 8.23 & 0.06 & 0.04 & 0.05 & 100.00 & 0.00 & 0.03 & 3113.4 & F-N\\ 
J1801$-$2451 & 0.12 & 291.6 & 3.8 & 4100 & 146 & 1.46 & 0.10 & 0.04 & 0.03 & 0.08 & 0.00 & 0.08 & 296.8 & S\\ 
J1803$-$2137 & 0.13 & 234.0 & 4.4 & 4100 & 139 & 15.21 & 0.19 & 0.01 & 0.03 & 0.19 & 0.00 & 0.10 & 102.3 & F-R\\ 
J1806$-$2125 & 0.48 & 747.0 & 4.9 & 4100 & 121 & 0.82 & 0.12 & 0.10 & 0.05 & 0.02 & 0.00 & 0.04 & 1603.3 & F-N\\ 
 & & & & & & & & & & & & & \\ 
J1807$-$0847 & 0.16 & 112.4 & 1.5 & 1547 & 49 & 18.16 & 0.18 & 0.00 & 0.03 & 0.17 & 0.02 & 0.16 & 140.6 & F-R\\ 
J1809$-$1917 & 0.08 & 197.1 & 3.3 & 4100 & 127 & 2.59 & 0.12 & 0.05 & 0.02 & 0.11 & 0.00 & 0.11 & 29.1 & F-R\\ 
J1815$-$1738 & 0.20 & 724.6 & 4.9 & 2487 & 79 & 0.44 & 0.09 & 0.08 & 0.02 & 0.03 & 0.00 & 0.04 & 1507.2 & F-N\\ 
J1816$-$2650 & 0.59 & 128.1 & 3.6 & 1606 & 35 & 2.38 & 0.11 & 0.02 & 0.06 & 0.09 & 0.00 & 0.15 & 34.5 & S\\ 
J1817$-$3618 & 0.39 & 94.3 & 4.4 & 1647 & 47 & 3.19 & 0.34 & 0.01 & 0.05 & 0.33 & 0.01 & 0.18 & 11.8 & F-R\\ 
 & & & & & & & & & & & & & \\ 
J1820$-$0427 & 0.60 & 84.4 & 2.9 & 3907 & 72 & 9.03 & 0.10 & 0.00 & 0.06 & 0.08 & 0.02 & 0.20 & 16.5 & F-J\\ 
J1820$-$1529 & 0.33 & 768.5 & 5.2 & 3865 & 71 & 0.98 & 0.12 & 0.08 & 0.04 & 0.08 & 0.00 & 0.04 & 1756.8 & F-N\\ 
J1822$-$2256 & 1.87 & 121.2 & 3.3 & 3937 & 52 & 3.89 & 0.19 & 0.02 & 0.07 & 0.18 & 0.00 & 0.16 & 29.3 & F-R\\ 
J1822$-$4209 & 0.46 & 72.5 & 3.7 & 1647 & 33 & 0.87 & 0.16 & 0.04 & 0.05 & 0.15 & 0.03 & 0.22 & 10.9 & F-R\\ 
J1823$-$3106 & 0.28 & 50.2 & 1.6 & 1647 & 46 & 5.52 & 0.34 & 0.00 & 0.04 & 0.33 & 0.07 & 0.28 & 4.3 & F-R\\ 
 & & & & & & & & & & & & & \\ 
J1824$-$1945 & 0.19 & 224.4 & 3.7 & 4100 & 133 & 7.25 & 0.11 & 0.00 & 0.03 & 0.11 & 0.00 & 0.10 & 121.8 & F-R\\ 
J1825$-$0935 & 0.77 & 19.4 & 0.3 & 3526 & 103 & 11.69 & 0.65 & 0.01 & 0.04 & 0.65 & 1.17 & 0.50 & 3.8 & F-DR\\ 
J1825$-$1446 & 0.28 & 352.2 & 4.4 & 4100 & 126 & 3.35 & 0.12 & 0.02 & 0.04 & 0.11 & 0.00 & 0.07 & 151.5 & F-R\\ 
J1826$-$1334 & 0.10 & 231.0 & 3.6 & 4100 & 130 & 4.69 & 0.05 & 0.02 & 0.03 & 0.04 & 0.00 & 0.10 & 88.6 & F-N\\ 
J1828$-$1057 & 0.25 & 249.0 & 3.6 & 2487 & 80 & 0.33 & 0.16 & 0.12 & 0.02 & 0.10 & 0.00 & 0.09 & 138.7 & F-N\\ 
\hline
\end{tabular}

\end{table*}
\begin{table*}
\centering
\contcaption{}
\begin{tabular}{lcccccccccccccc}
\hline
$\rm{Name}$ & $P_{0}$ & $\rm{DM}$ & Dist & $T$ & $N_{\rm{obs}}$ & $\overline{S}$ & $m$ & $m_{\rm{n}}$ & $m_{\rm{j}}$ & $m_{\rm{corr}}$ & $m_{\rm{d}}$ & $m_{\rm{r}}$ & $T_{\rm{r}}$ & Class\\ 
& [s] & [$\rm{cm^{-3} pc}$] & [kpc] & [days] &  & [mJy] &  &  &  &  &  &  & [day] &  \\ 
\hline
\hline
J1828$-$1101 & 0.07 & 605.0 & 4.8 & 4100 & 127 & 3.46 & 0.09 & 0.05 & 0.02 & 0.08 & 0.00 & 0.05 & 1019.4 & F-N\\ 
J1829$-$1751 & 0.31 & 217.1 & 5.9 & 3907 & 73 & 10.81 & 0.10 & 0.00 & 0.04 & 0.09 & 0.00 & 0.10 & 9.3 & F-R\\ 
J1830$-$1059 & 0.41 & 159.7 & 3.1 & 4100 & 151 & 1.29 & 0.31 & 0.02 & 0.04 & 0.31 & 0.00 & 0.13 & 51.0 & X\\ 
J1832$-$0827 & 0.65 & 300.9 & 5.2 & 4015 & 124 & 3.82 & 0.15 & 0.01 & 0.07 & 0.13 & 0.00 & 0.08 & 146.8 & I\\ 
J1833$-$0827 & 0.09 & 410.9 & 4.5 & 4015 & 124 & 6.53 & 0.09 & 0.01 & 0.02 & 0.08 & 0.00 & 0.07 & 184.4 & F-R\\ 
 & & & & & & & & & & & & & \\ 
J1834$-$0731 & 0.51 & 294.0 & 4.1 & 2402 & 76 & 1.49 & 0.10 & 0.07 & 0.06 & 0.02 & 0.00 & 0.08 & 208.1 & F-N\\ 
J1835$-$0643 & 0.31 & 467.9 & 5.0 & 4015 & 97 & 2.85 & 0.10 & 0.05 & 0.05 & 0.07 & 0.00 & 0.06 & 611.1 & F-N\\ 
J1835$-$0944 & 0.15 & 276.2 & 4.2 & 1314 & 38 & 0.60 & 0.09 & 0.07 & 0.02 & 0.06 & 0.00 & 0.09 & 185.3 & F-N\\ 
J1835$-$1106 & 0.17 & 132.7 & 3.2 & 4014 & 126 & 2.46 & 0.18 & 0.02 & 0.03 & 0.18 & 0.00 & 0.15 & 11.2 & F-R\\ 
J1837$-$0559 & 0.20 & 319.5 & 4.3 & 4015 & 118 & 0.56 & 0.14 & 0.11 & 0.04 & 0.08 & 0.00 & 0.08 & 254.3 & F-N\\ 
 & & & & & & & & & & & & & \\ 
J1837$-$0604 & 0.10 & 459.3 & 4.8 & 2402 & 92 & 1.11 & 0.13 & 0.09 & 0.01 & 0.09 & 0.00 & 0.06 & 571.8 & F-N\\ 
J1838$-$0453 & 0.38 & 617.2 & 6.6 & 2402 & 86 & 0.42 & 0.10 & 0.10 & 0.04 & 100.00 & 0.00 & 0.05 & 1254.2 & F-N\\ 
J1838$-$0549 & 0.24 & 276.6 & 4.1 & 2402 & 78 & 0.34 & 0.20 & 0.13 & 0.03 & 0.15 & 0.00 & 0.09 & 182.3 & F-N\\ 
J1839$-$0321 & 0.24 & 452.6 & 7.9 & 2402 & 68 & 0.22 & 0.18 & 0.12 & 0.03 & 0.13 & 0.00 & 0.06 & 712.7 & F-N\\ 
J1839$-$0905 & 0.42 & 344.5 & 6.8 & 1515 & 50 & 0.20 & 0.14 & 0.09 & 0.02 & 0.11 & 0.00 & 0.07 & 374.1 & F-N\\ 
 & & & & & & & & & & & & & \\ 
J1841$-$0345 & 0.20 & 194.3 & 3.8 & 3841 & 37 & 1.22 & 0.08 & 0.04 & 0.03 & 0.06 & 0.00 & 0.11 & 84.1 & F-N\\ 
J1841$-$0425 & 0.19 & 325.5 & 4.4 & 4015 & 126 & 3.02 & 0.07 & 0.01 & 0.04 & 0.06 & 0.00 & 0.08 & 329.0 & S\\ 
J1841$-$0524 & 0.45 & 284.5 & 4.1 & 2402 & 85 & 0.18 & 0.18 & 0.13 & 0.03 & 0.12 & 0.00 & 0.08 & 194.8 & F-N\\ 
J1842$-$0905 & 0.34 & 343.4 & 9.7 & 4014 & 126 & 0.91 & 0.10 & 0.04 & 0.05 & 0.08 & 0.00 & 0.07 & 443.4 & F-N\\ 
J1843$-$0355 & 0.13 & 797.7 & 5.8 & 2402 & 77 & 0.86 & 0.11 & 0.10 & 0.02 & 0.04 & 0.00 & 0.04 & 2008.4 & F-N\\ 
 & & & & & & & & & & & & & \\ 
J1844$-$0256 & 0.27 & 826.0 & 6.0 & 1515 & 53 & 0.79 & 0.17 & 0.11 & 0.02 & 0.12 & 0.00 & 0.04 & 2203.3 & F-N\\ 
J1844$-$0538 & 0.26 & 411.7 & 5.4 & 4014 & 124 & 2.97 & 0.07 & 0.02 & 0.04 & 0.05 & 0.00 & 0.07 & 483.7 & I\\ 
J1845$-$0434 & 0.49 & 230.8 & 4.1 & 3982 & 125 & 2.59 & 0.07 & 0.02 & 0.06 & 0.03 & 0.00 & 0.10 & 125.3 & F-J\\ 
J1845$-$0743 & 0.10 & 280.9 & 7.1 & 4014 & 125 & 3.26 & 0.16 & 0.01 & 0.03 & 0.15 & 0.00 & 0.09 & 249.2 & S\\ 
J1847$-$0402 & 0.60 & 142.0 & 3.4 & 4014 & 125 & 4.55 & 0.12 & 0.01 & 0.05 & 0.11 & 0.00 & 0.14 & 95.5 & I\\ 
 & & & & & & & & & & & & & \\ 
J1848$-$0123 & 0.66 & 159.5 & 4.4 & 1547 & 48 & 14.21 & 0.09 & 0.00 & 0.06 & 0.07 & 0.00 & 0.13 & 32.7 & I\\ 
J1852$-$0635 & 0.52 & 171.0 & 4.5 & 1547 & 53 & 9.18 & 0.19 & 0.01 & 0.05 & 0.18 & 0.00 & 0.12 & 70.3 & F-R\\ 
J1852$-$2610 & 0.34 & 56.8 & 2.5 & 1647 & 32 & 1.38 & 0.21 & 0.03 & 0.04 & 0.20 & 0.05 & 0.26 & 5.5 & F-R\\ 
J1853$+$0011 & 0.40 & 568.8 & 5.8 & 1224 & 34 & 0.19 & 0.14 & 0.10 & 0.02 & 0.09 & 0.00 & 0.05 & 988.4 & F-N\\ 
J1853$-$0004 & 0.10 & 437.5 & 5.3 & 4014 & 124 & 0.88 & 0.15 & 0.07 & 0.03 & 0.14 & 0.00 & 0.06 & 546.4 & F-N\\ 
 & & & & & & & & & & & & & \\ 
J1900$-$2600 & 0.61 & 38.0 & 0.7 & 1647 & 44 & 15.39 & 0.14 & 0.00 & 0.06 & 0.12 & 0.15 & 0.34 & 2.4 & F-R\\ 
J1913$-$0440 & 0.83 & 89.4 & 4.0 & 1547 & 47 & 7.14 & 0.26 & 0.00 & 0.07 & 0.25 & 0.02 & 0.19 & 31.8 & F-R\\ 
J1941$-$2602 & 0.40 & 50.0 & 3.6 & 1647 & 45 & 2.11 & 0.32 & 0.01 & 0.05 & 0.32 & 0.09 & 0.28 & 5.9 & F-R\\ 
J2048$-$1616 & 1.96 & 11.5 & 0.9 & 1537 & 42 & 26.01 & 1.01 & 0.00 & 0.06 & 1.01 & 0.95 & 0.64 & 0.1 & F-D\\ 
J2330$-$2005 & 1.64 & 8.5 & 0.9 & 1618 & 47 & 3.40 & 1.01 & 0.01 & 0.09 & 1.01 & 3.54 & 0.73 & 0.2 & F-D\\ 
 & & & & & & & & & & & & & \\ 
J2346$-$0609 & 1.18 & 22.5 & 3.7 & 3914 & 45 & 2.66 & 1.28 & 0.01 & 0.05 & 1.27 & 0.28 & 0.46 & 0.5 & F-D\\ 
\hline
\end{tabular}

\end{table*}

\section{Structure function of Class ``S"}

Plots of structure functions of pulsars classified as ``S". Black solid lines represent the fitting results of three regimes as described in Section~\ref{section;res}. The red shaded regions represent the measure refractive time scale and its uncertainty.

\label{app:s}
\begin{figure*}
\begin{center}
\includegraphics[width=8cm]{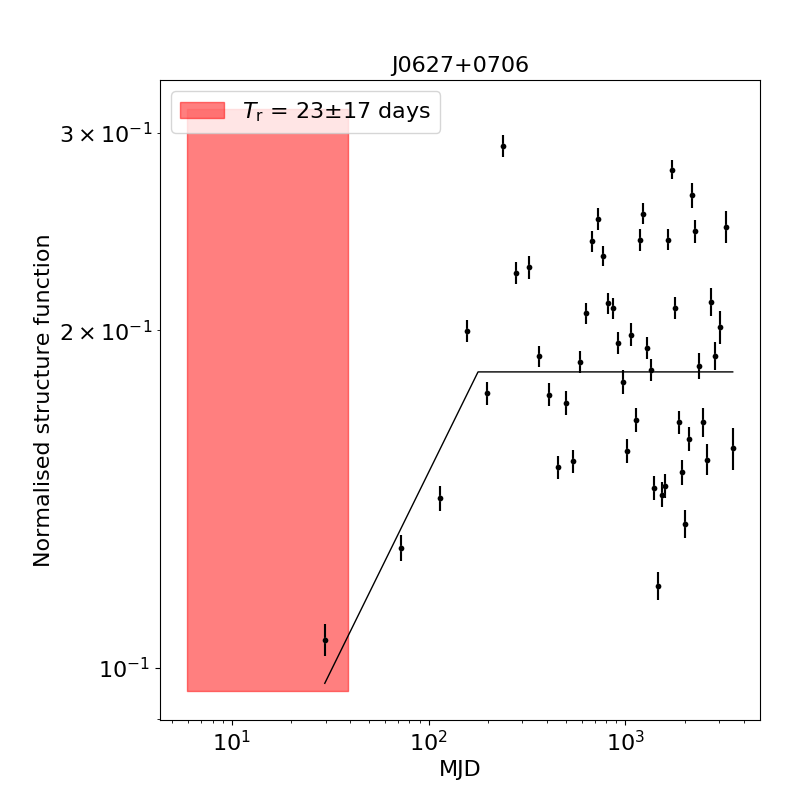}
\includegraphics[width=8cm]{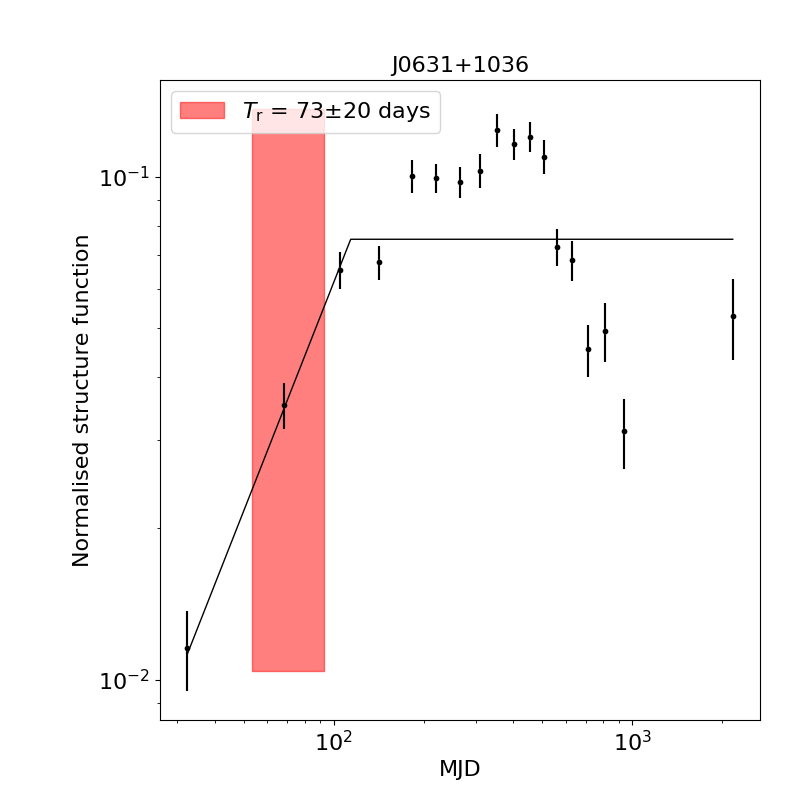}
\end{center}
\caption{Structure functions of PSRs J0627+0706 and J0631+1036.}
\end{figure*}

\begin{figure*}
\begin{center}
\includegraphics[width=8cm]{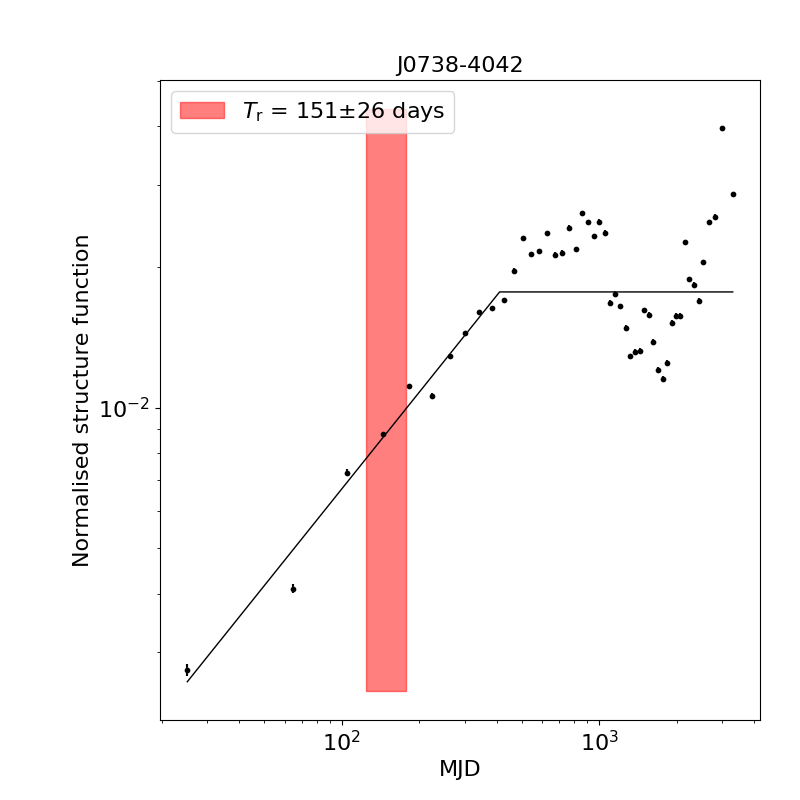}
\includegraphics[width=8cm]{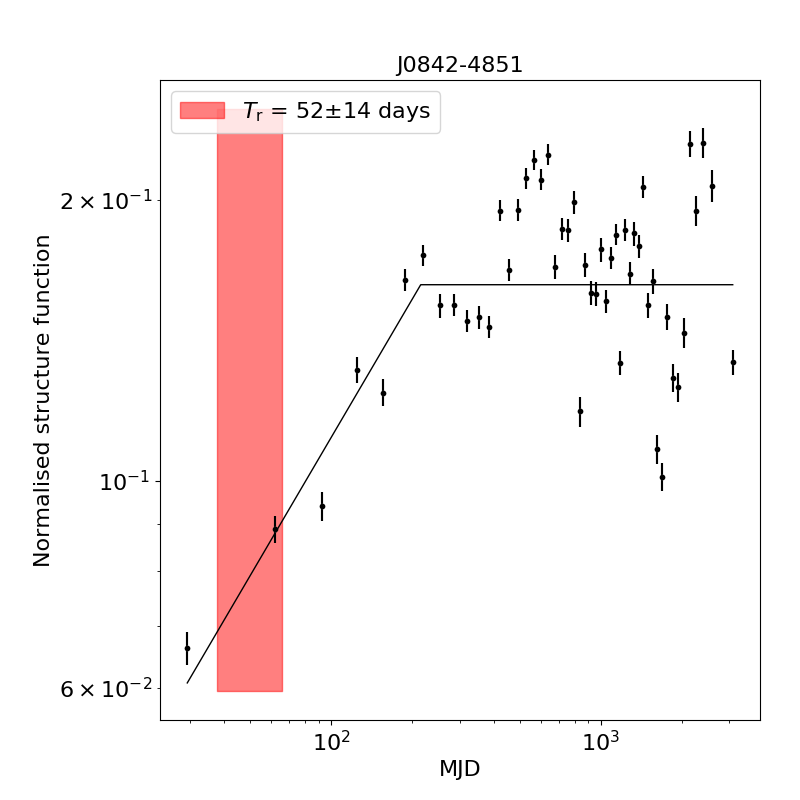}
\end{center}
\caption{Structure functions of PSRs J0738$-$4042 and J0842$-$4851.}
\end{figure*}

\begin{figure*}
\begin{center}
\includegraphics[width=8cm]{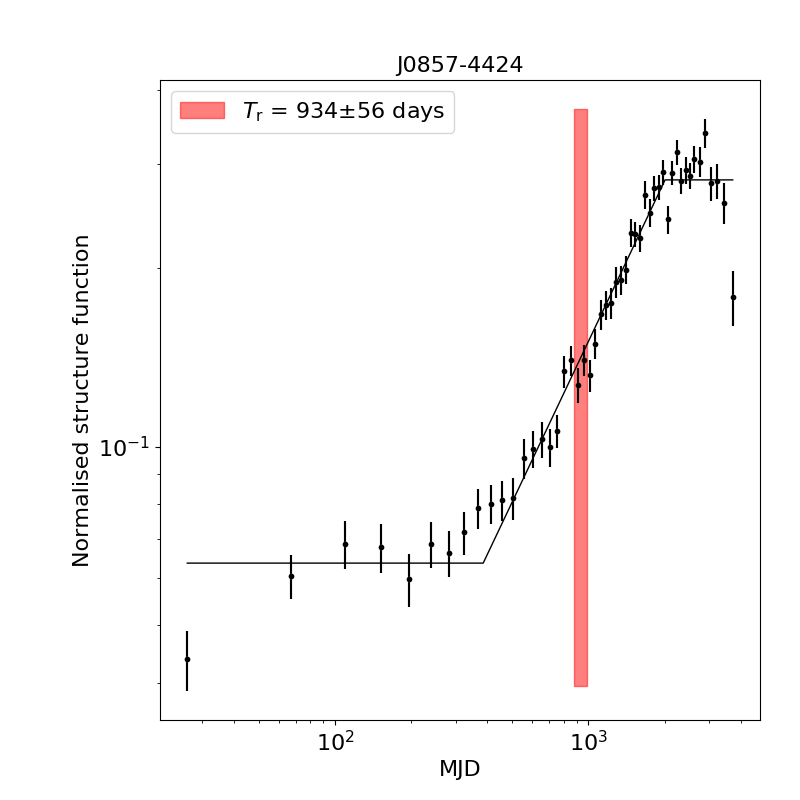}
\includegraphics[width=8cm]{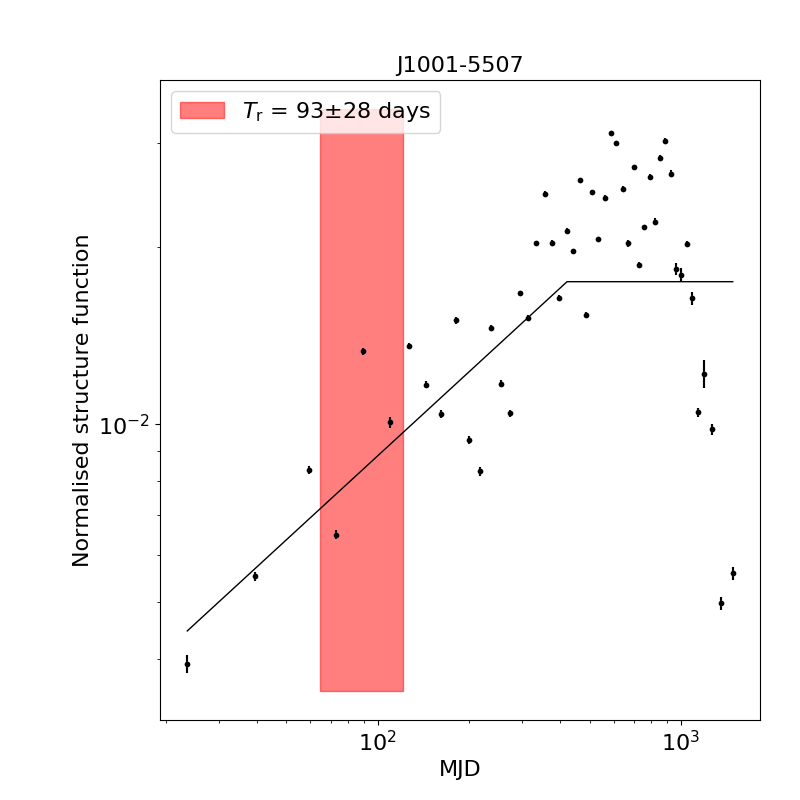}
\end{center}
\caption{Structure functions of PSRs J0857$-$4424 and J1001$-$5507.}
\end{figure*}

\begin{figure*}
\begin{center}
\includegraphics[width=8cm]{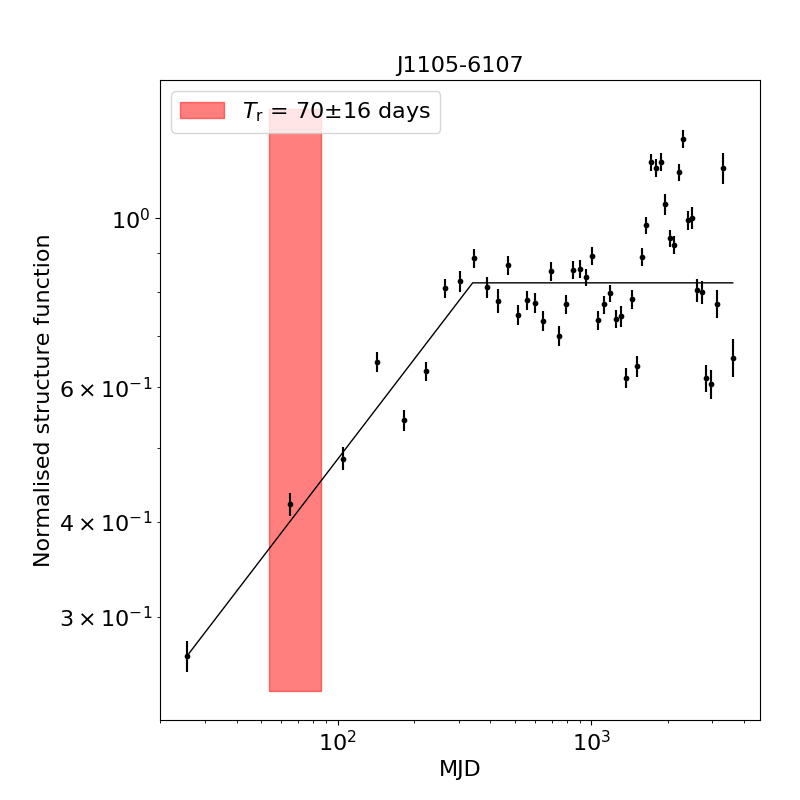}
\includegraphics[width=8cm]{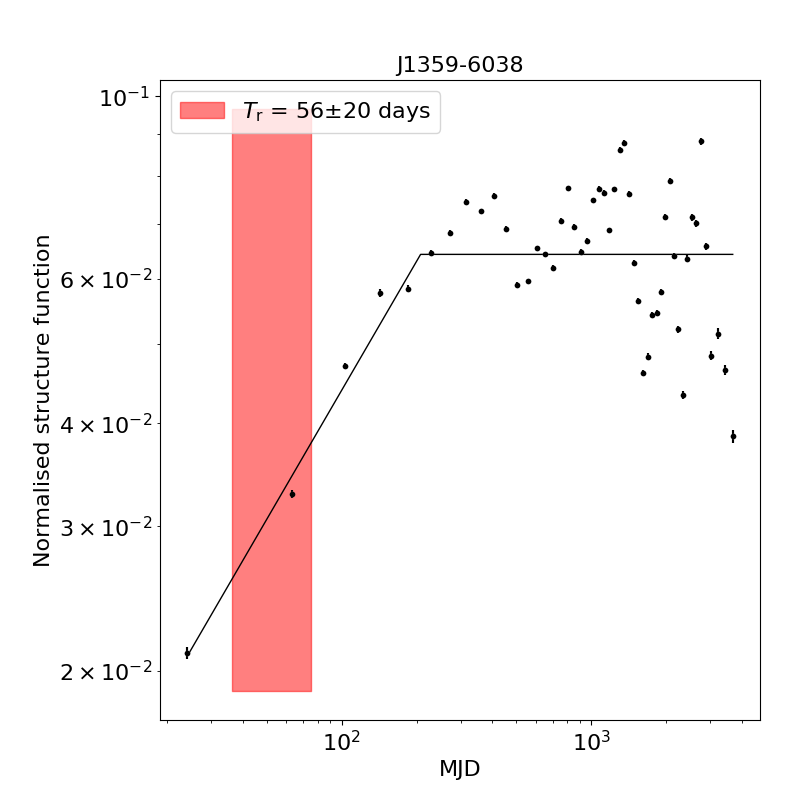}
\end{center}
\caption{Structure functions of PSRs J1105$-$6107 and J1359$-$6038.}
\end{figure*}

\begin{figure*}
\begin{center}
\includegraphics[width=8cm]{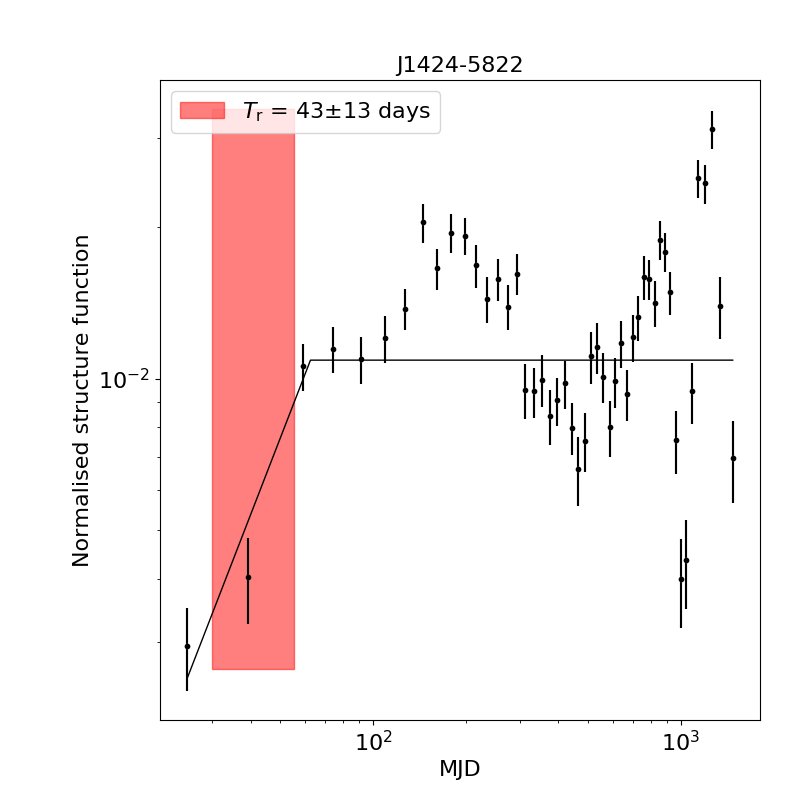}
\includegraphics[width=8cm]{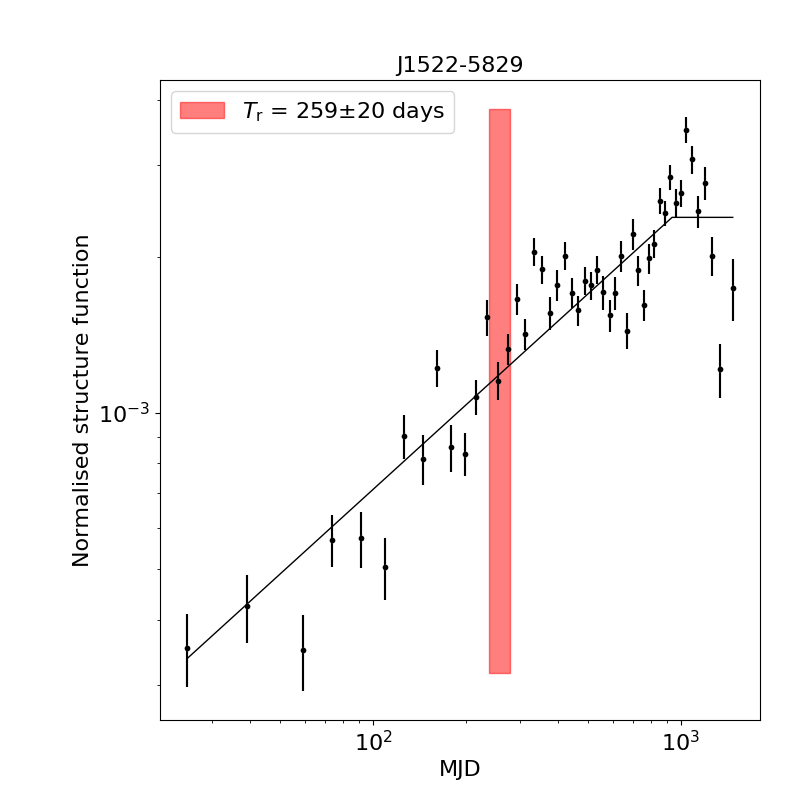}
\end{center}
\caption{Structure functions of PSRs J1424$-$5822 and J1522$-$5829.}
\end{figure*}

\begin{figure*}
\begin{center}
\includegraphics[width=8cm]{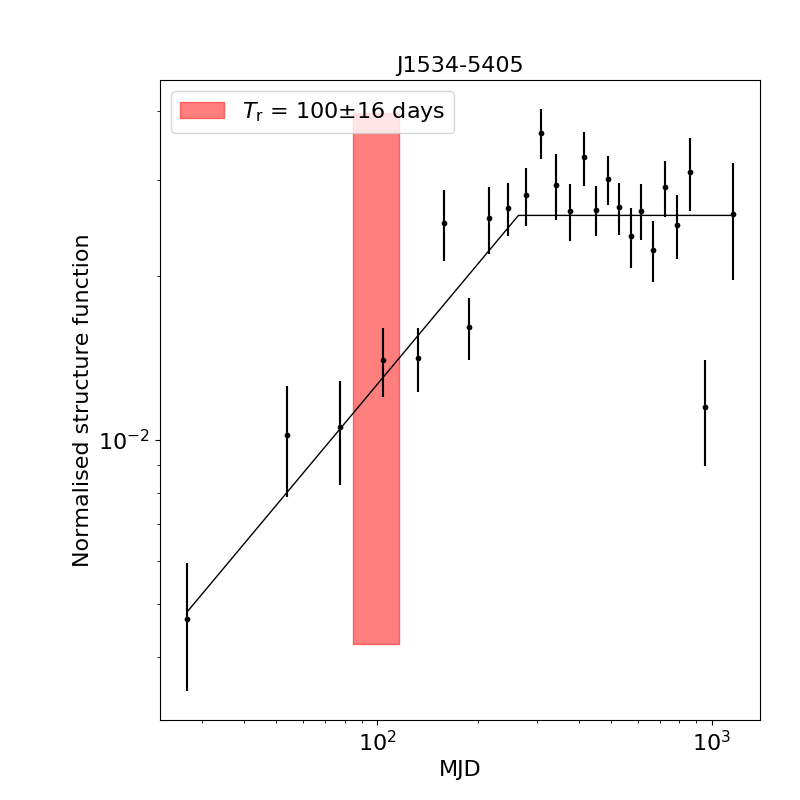}
\includegraphics[width=8cm]{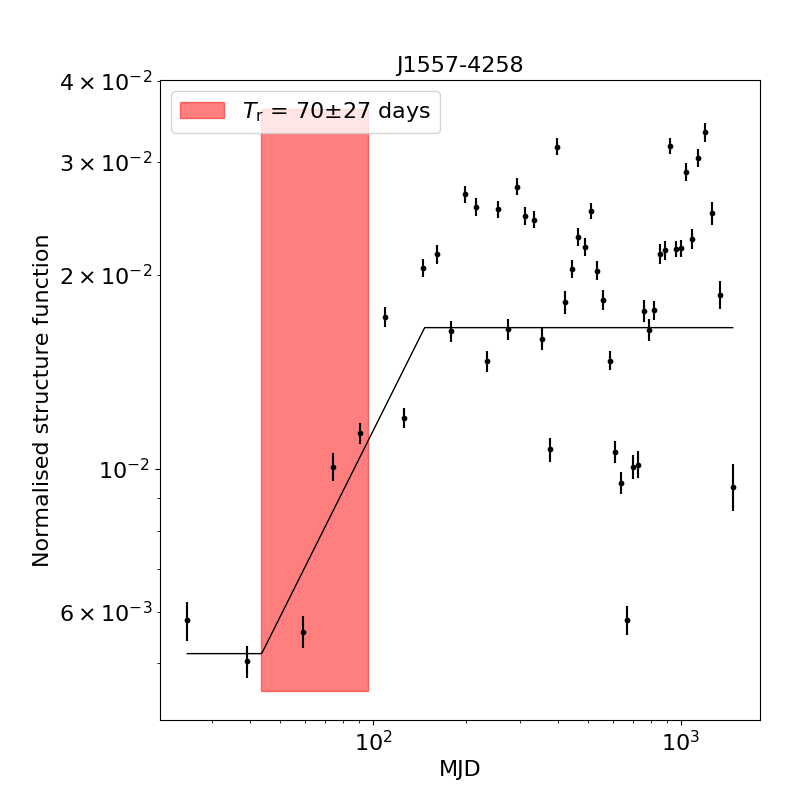}
\end{center}
\caption{Structure functions of PSRs J1534$-$5405 and J1557$-$4258.}
\end{figure*}

\begin{figure*}
\begin{center}
\includegraphics[width=8cm]{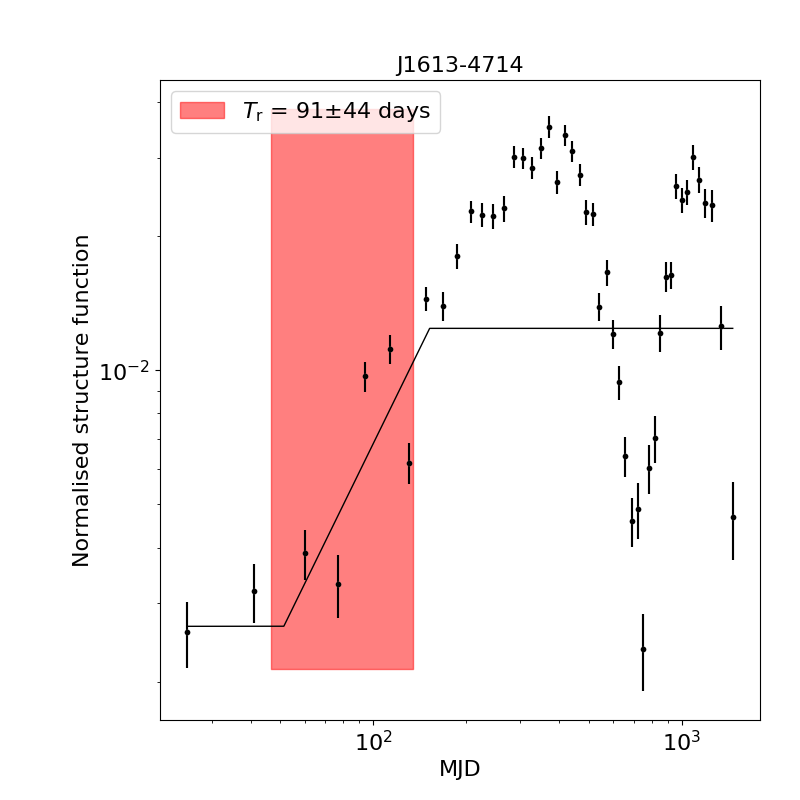}
\includegraphics[width=8cm]{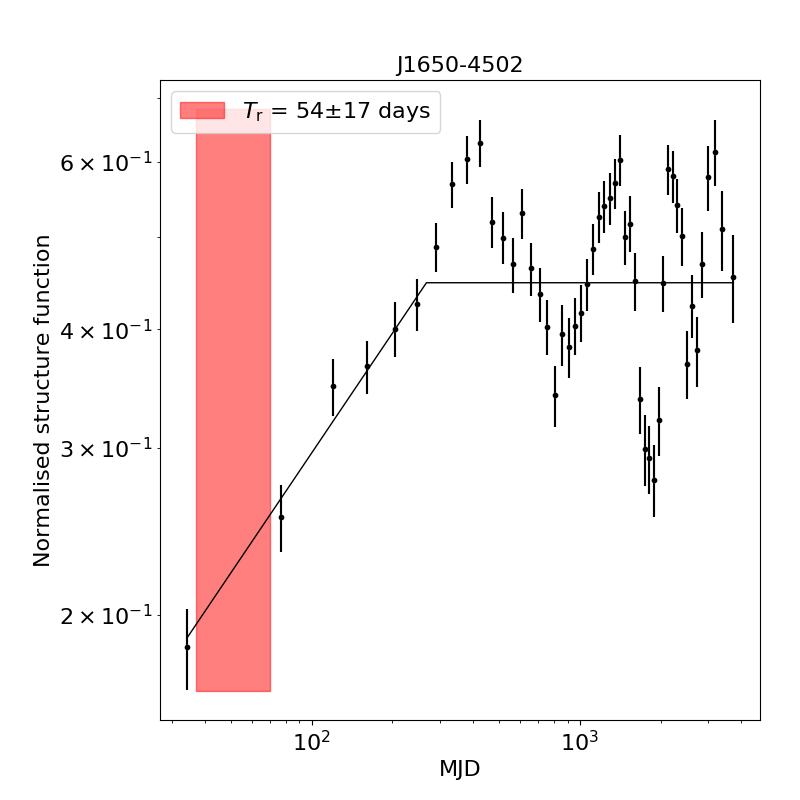}
\end{center}
\caption{Structure functions of PSRs J1613$-$4714 and J1650$-$4502.}
\end{figure*}

\begin{figure*}
\begin{center}
\includegraphics[width=8cm]{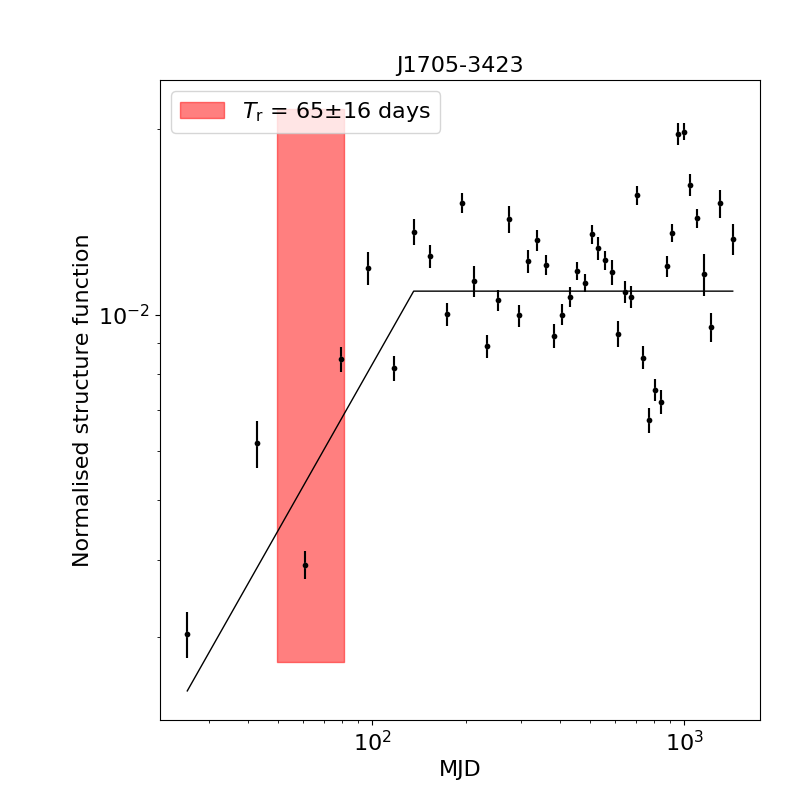}
\includegraphics[width=8cm]{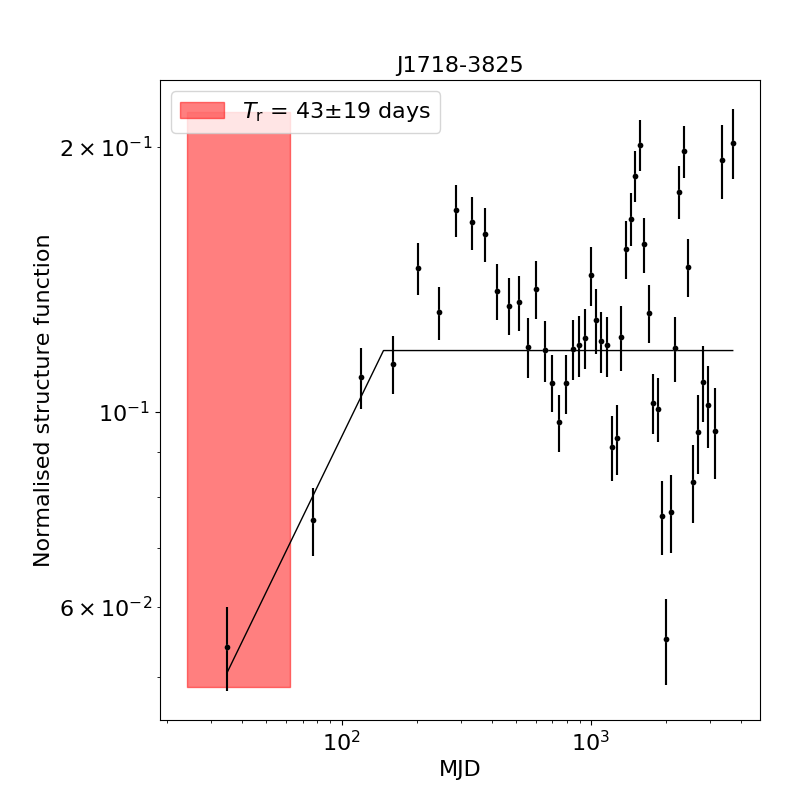}
\end{center}
\caption{Structure functions of PSRs J1705$-$3423 and J1718$-$3825.}
\end{figure*}

\begin{figure*}
\begin{center}
\includegraphics[width=8cm]{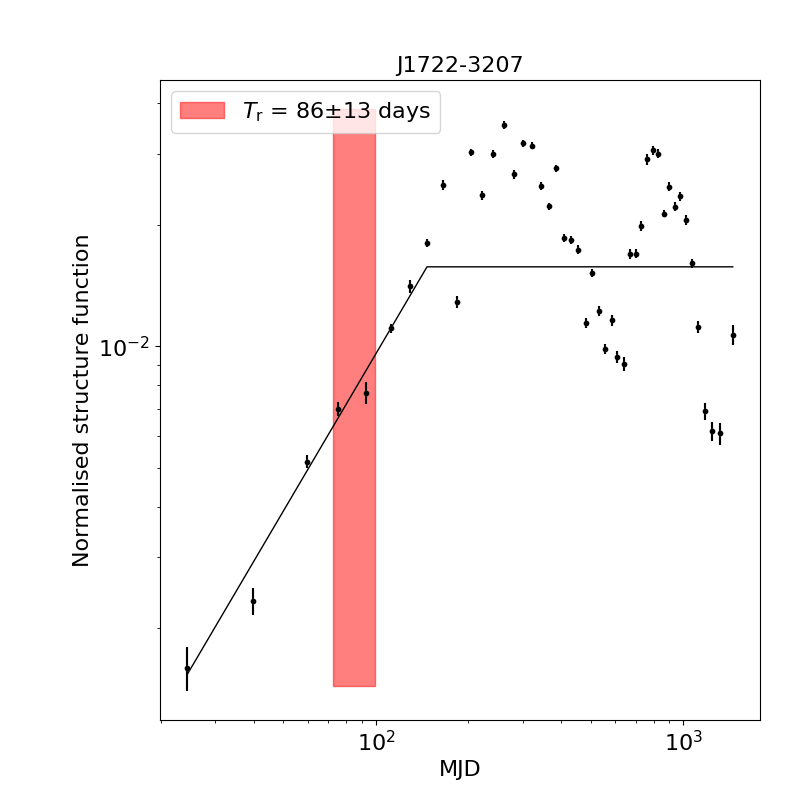}
\includegraphics[width=8cm]{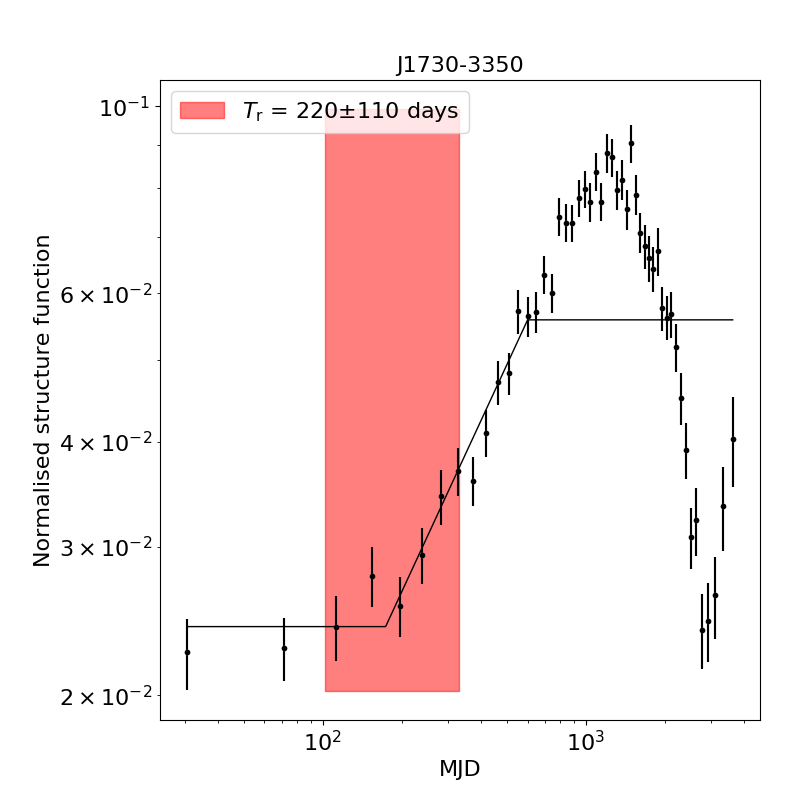}
\end{center}
\caption{Structure functions of PSRs J1722$-$3207 and J1730$-$3350.}
\end{figure*}

\begin{figure*}
\begin{center}
\includegraphics[width=8cm]{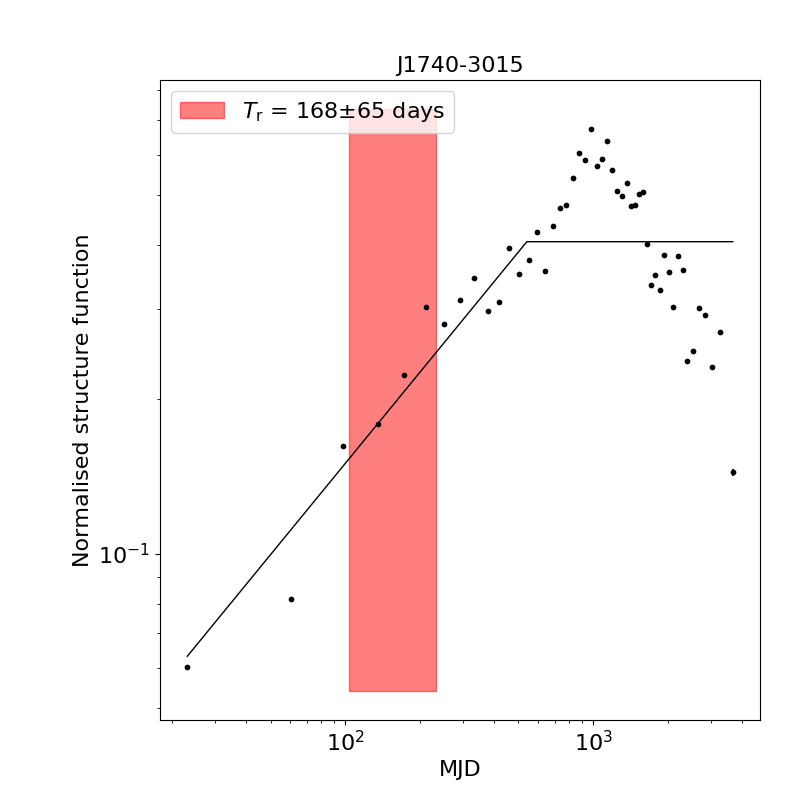}
\includegraphics[width=8cm]{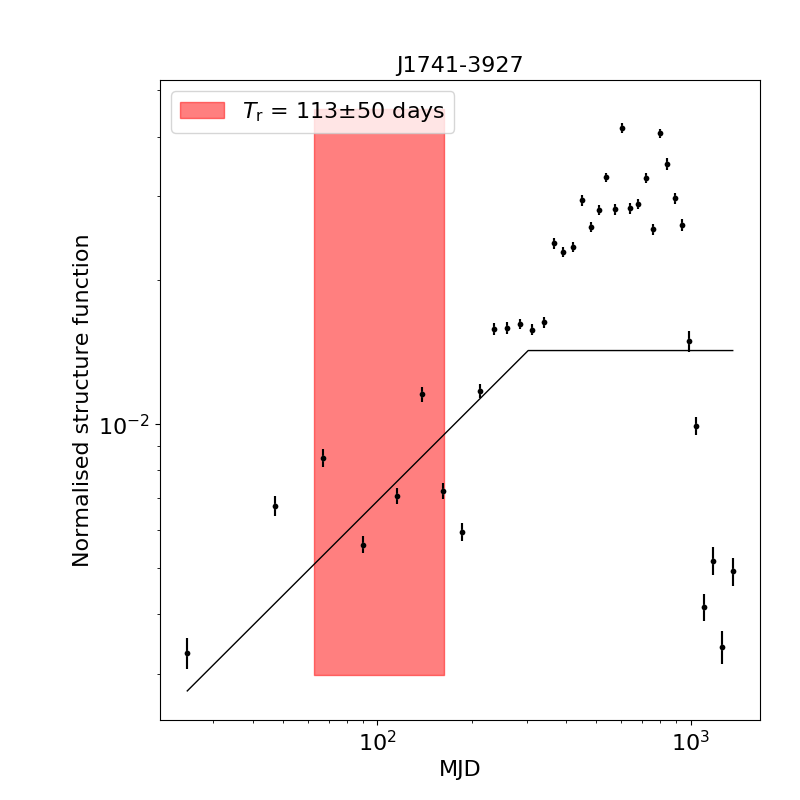}
\end{center}
\caption{Structure functions of PSRs J1740$-$3015 and J1741$-$3927.}
\end{figure*}

\begin{figure*}
\begin{center}
\includegraphics[width=8cm]{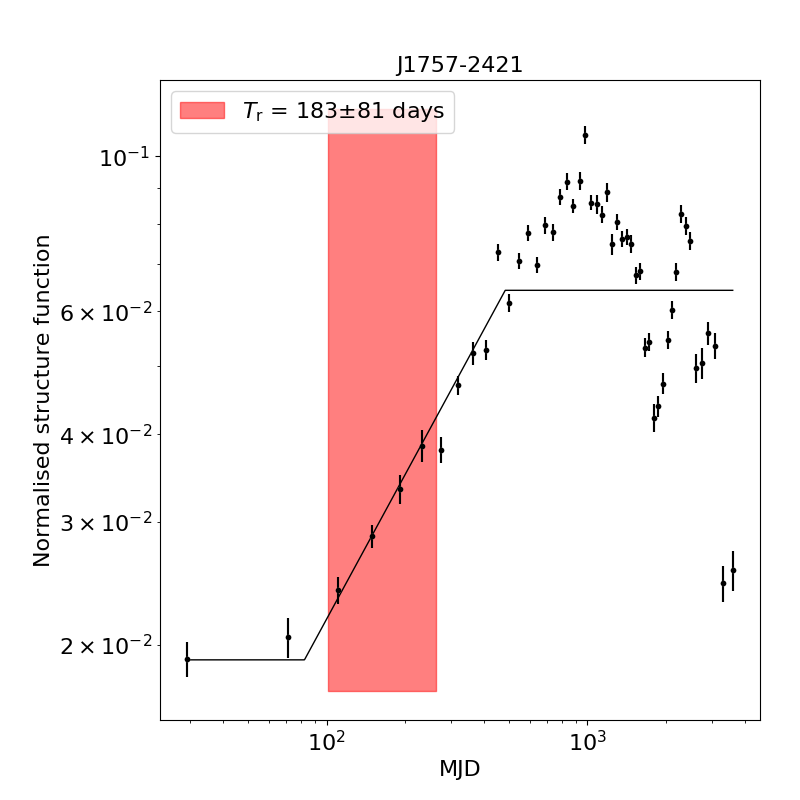}
\includegraphics[width=8cm]{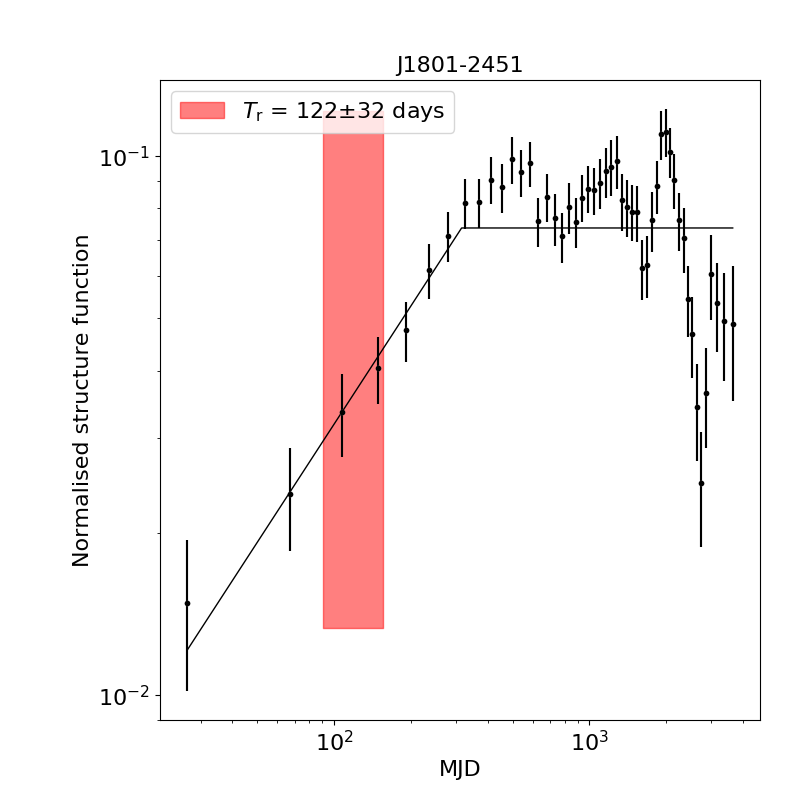}
\end{center}
\caption{Structure functions of PSRs J1757$-$2421 and J1801$-$2451.}
\end{figure*}

\begin{figure*}
\begin{center}
\includegraphics[width=8cm]{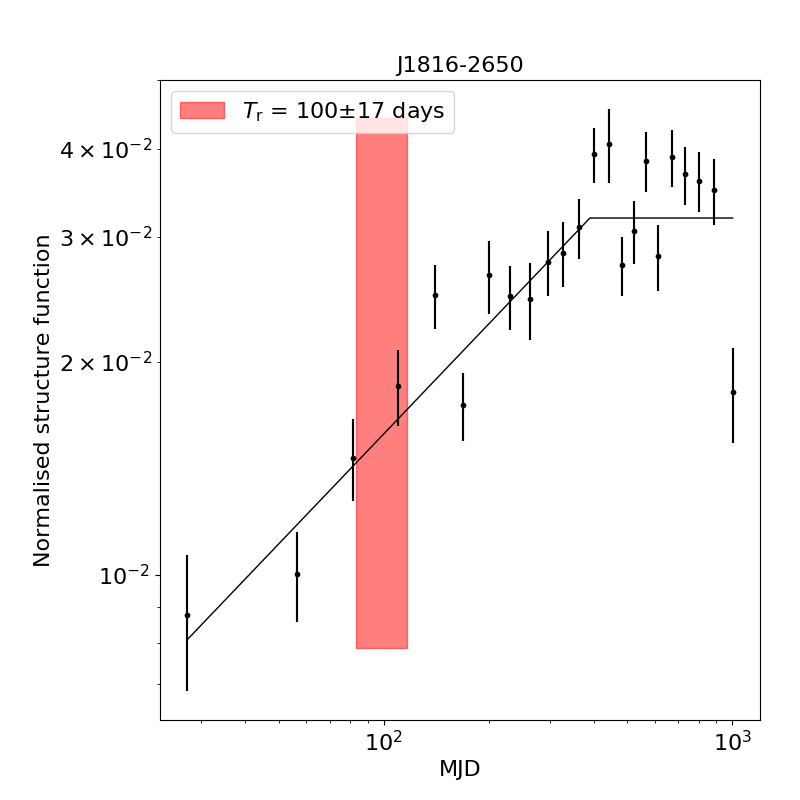}
\includegraphics[width=8cm]{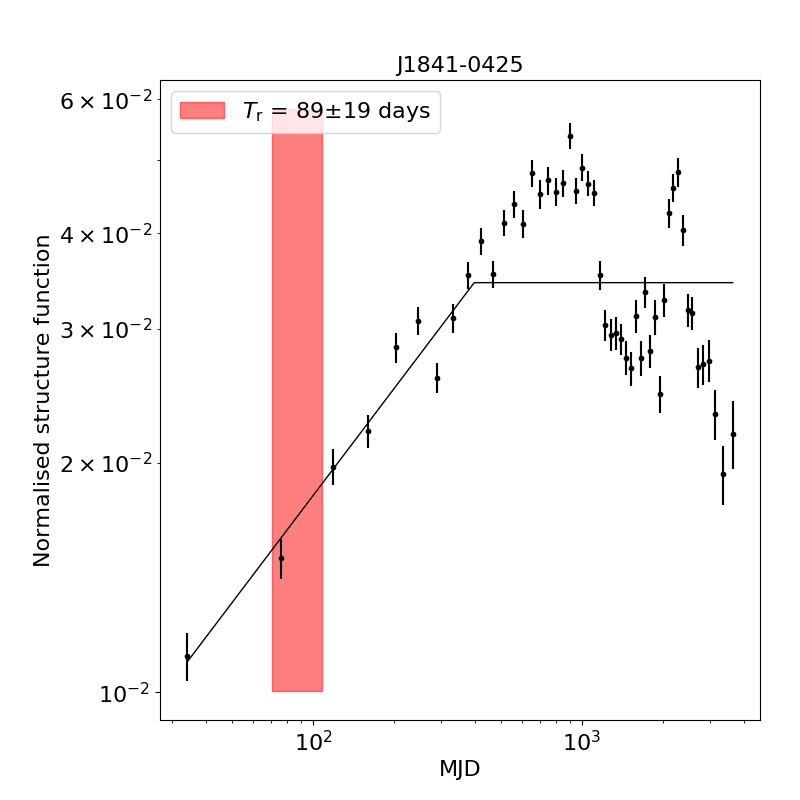}
\end{center}
\caption{Structure functions of PSRs J1816$-$2650 and J1841$-$0425.}
\end{figure*}

\begin{figure*}
\begin{center}
\includegraphics[width=8cm]{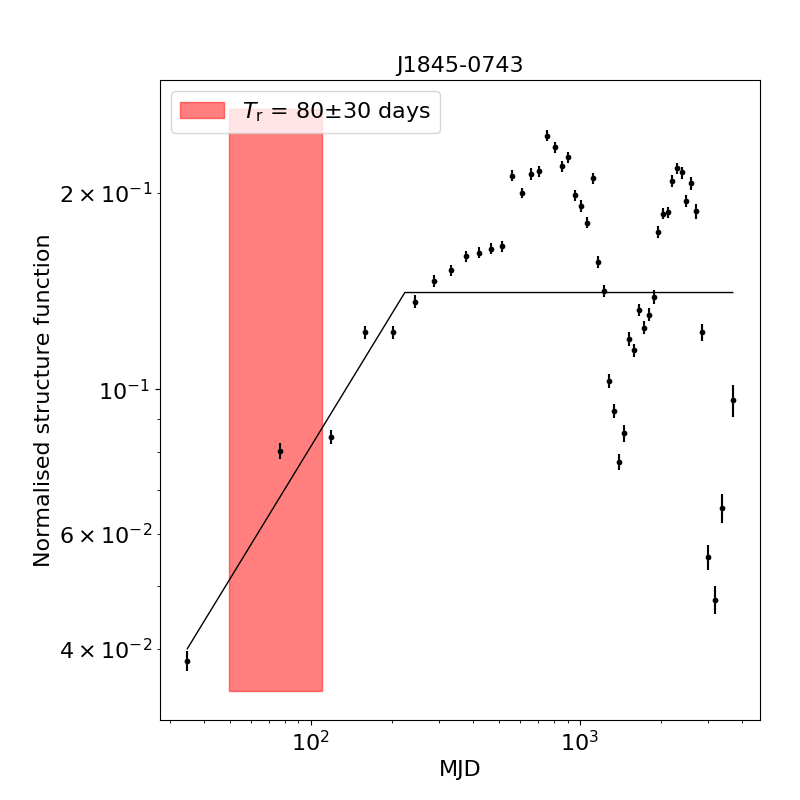}
\end{center}
\caption{Structure function of PSR J1845$-$0743.}
\end{figure*}

\section{Structure function of Class ``I"}
\label{app:i}
Plots of structure functions of pulsars classified as ``I". Black solid lines represent the fitting results of rising and saturation of the structure function. The grey shaded regions represent the lower limit of the refractive time scale.

\begin{figure*}
\begin{center}
\includegraphics[width=8cm]{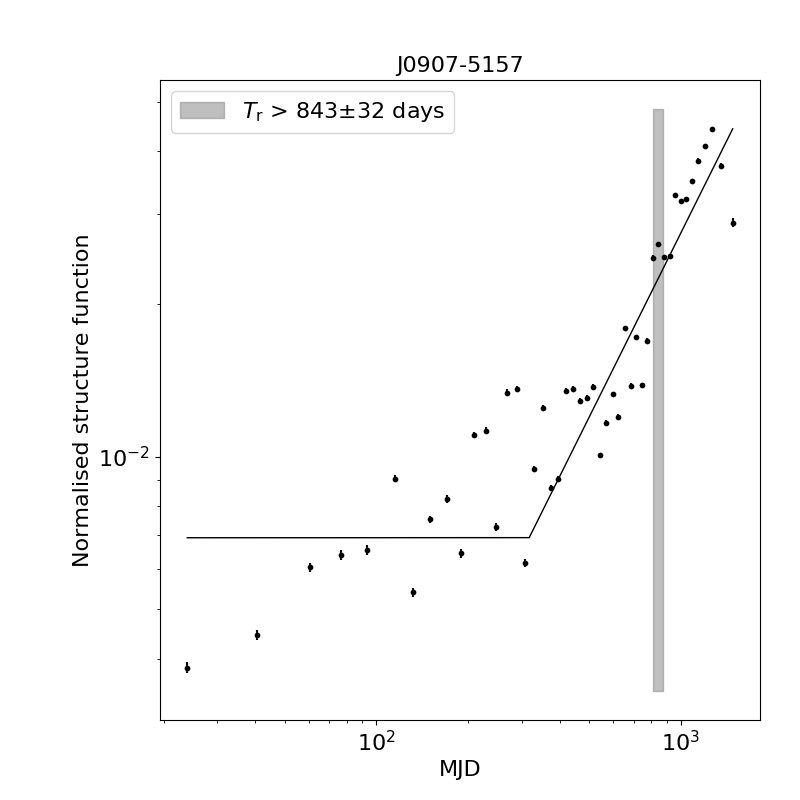}
\includegraphics[width=8cm]{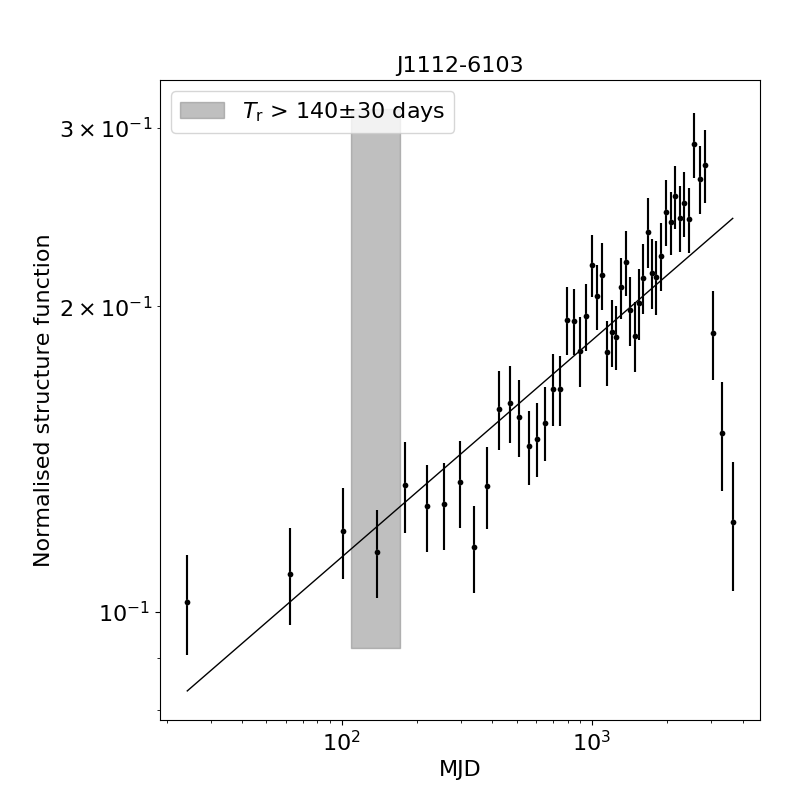}
\end{center}
\caption{Structure functions of PSRs J0907$-$5157 and J1112$-$6103.}
\end{figure*}

\begin{figure*}
\begin{center}
\includegraphics[width=8cm]{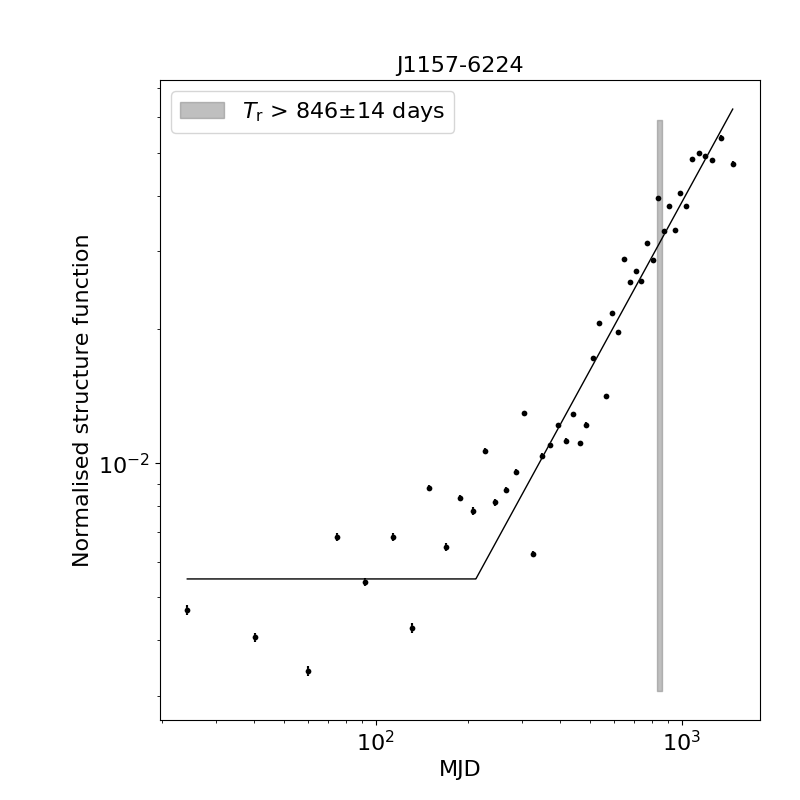}
\includegraphics[width=8cm]{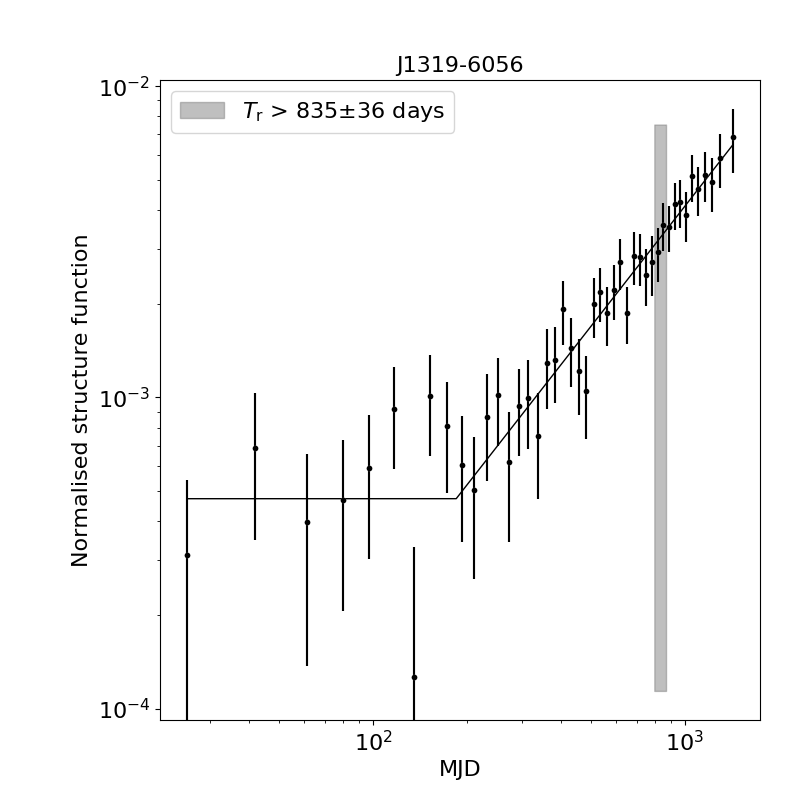}
\end{center}
\caption{Structure functions of PSRs J1157$-$6224 and J1319$-$6056.}
\end{figure*}

\begin{figure*}
\begin{center}
\includegraphics[width=8cm]{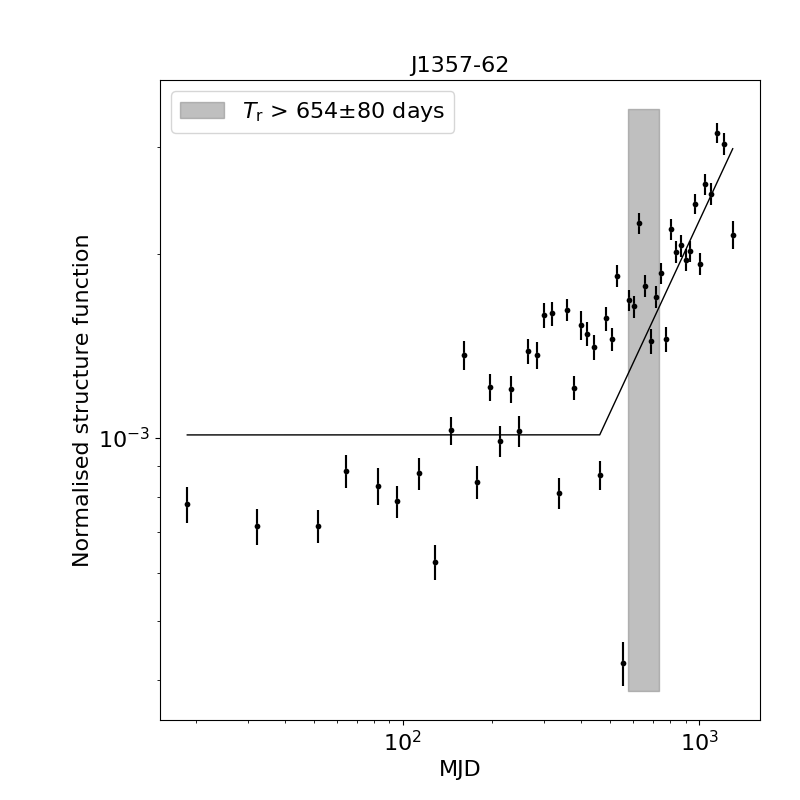}
\includegraphics[width=8cm]{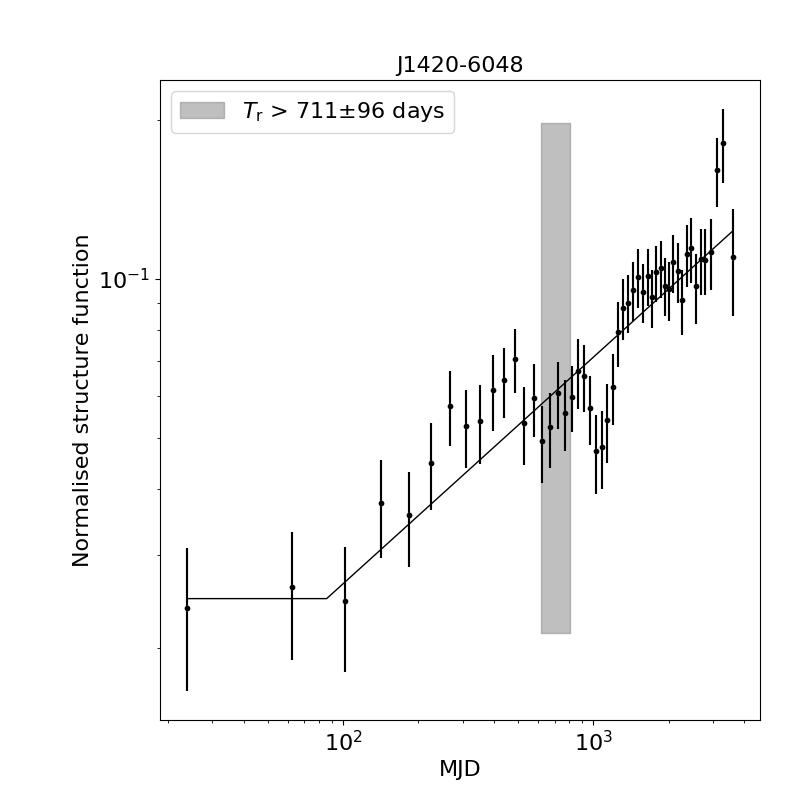}
\end{center}
\caption{Structure functions of PSRs J1357$-$62 and J1420$-$6048.}
\end{figure*}

\begin{figure*}
\begin{center}
\includegraphics[width=8cm]{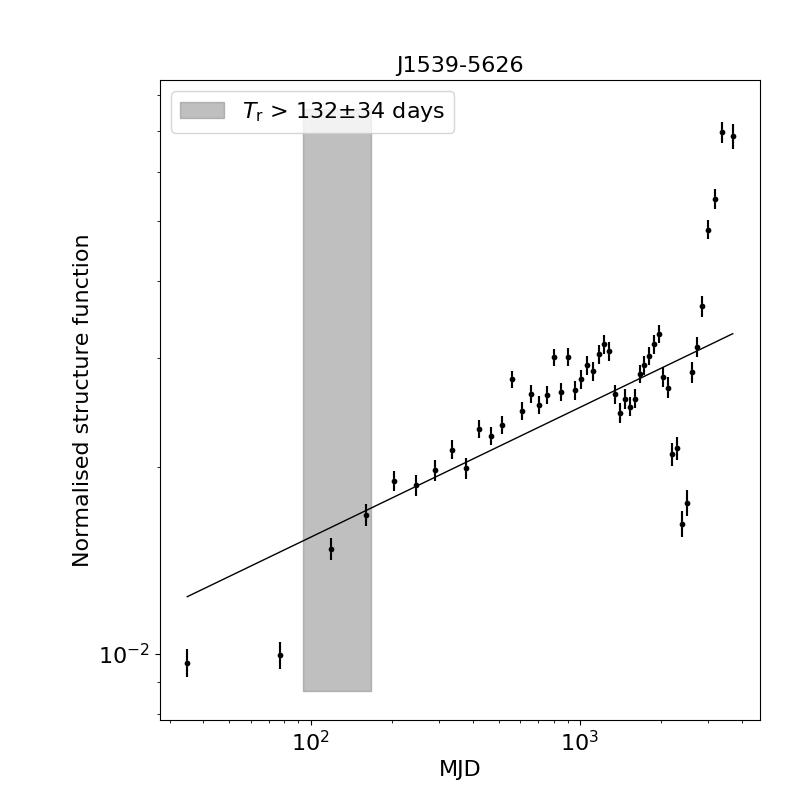}
\includegraphics[width=8cm]{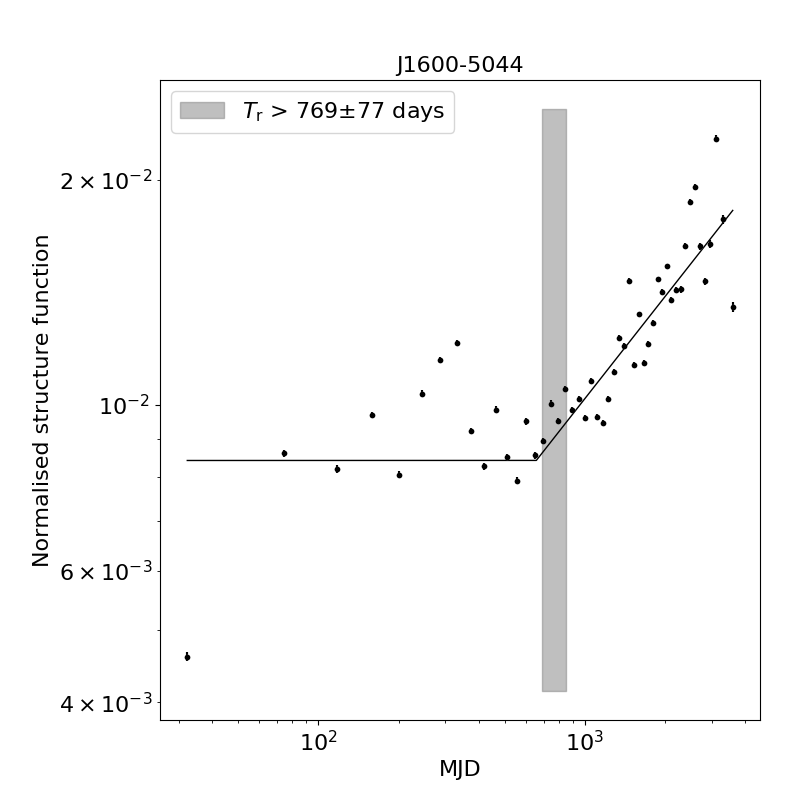}
\end{center}
\caption{Structure functions of PSRs J1539$-$5626 and J1600$-$5044.}
\end{figure*}

\begin{figure*}
\begin{center}
\includegraphics[width=8cm]{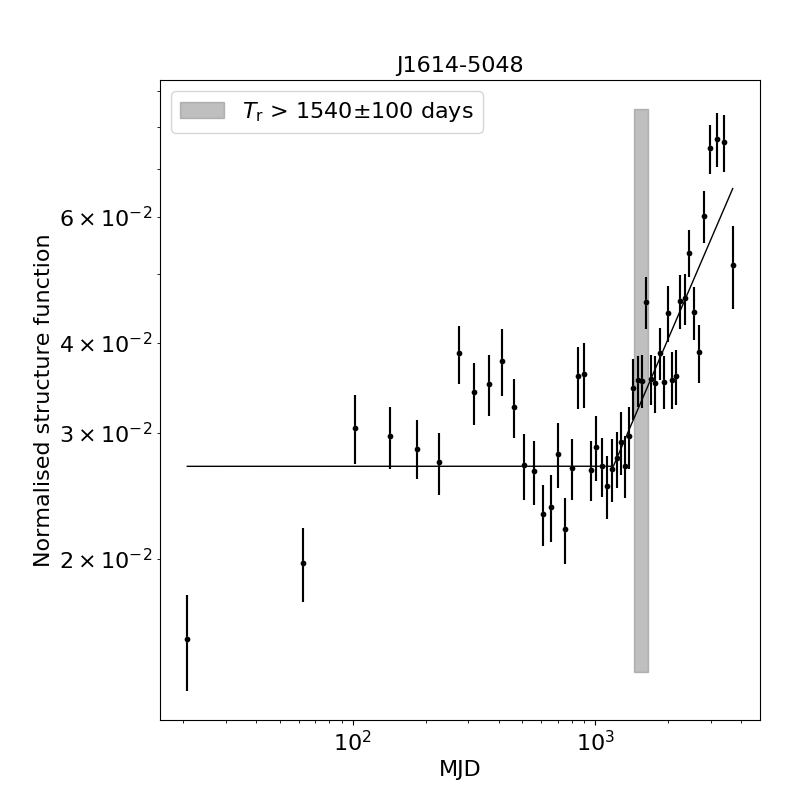}
\includegraphics[width=8cm]{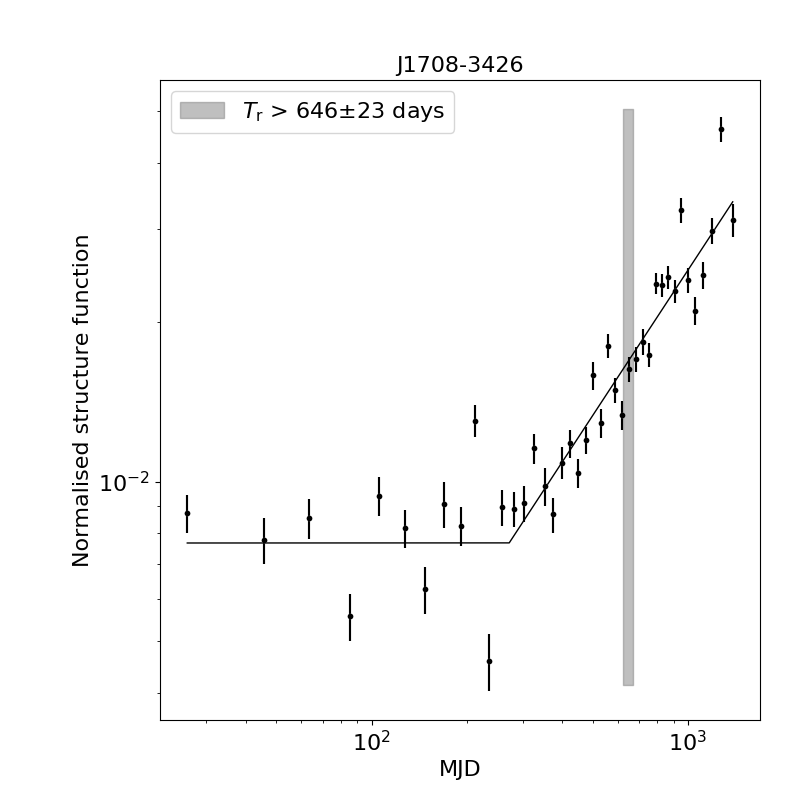}
\end{center}
\caption{Structure functions of PSRs J1614$-$5048 and J1708$-$3426.}
\end{figure*}

\begin{figure*}
\begin{center}
\includegraphics[width=8cm]{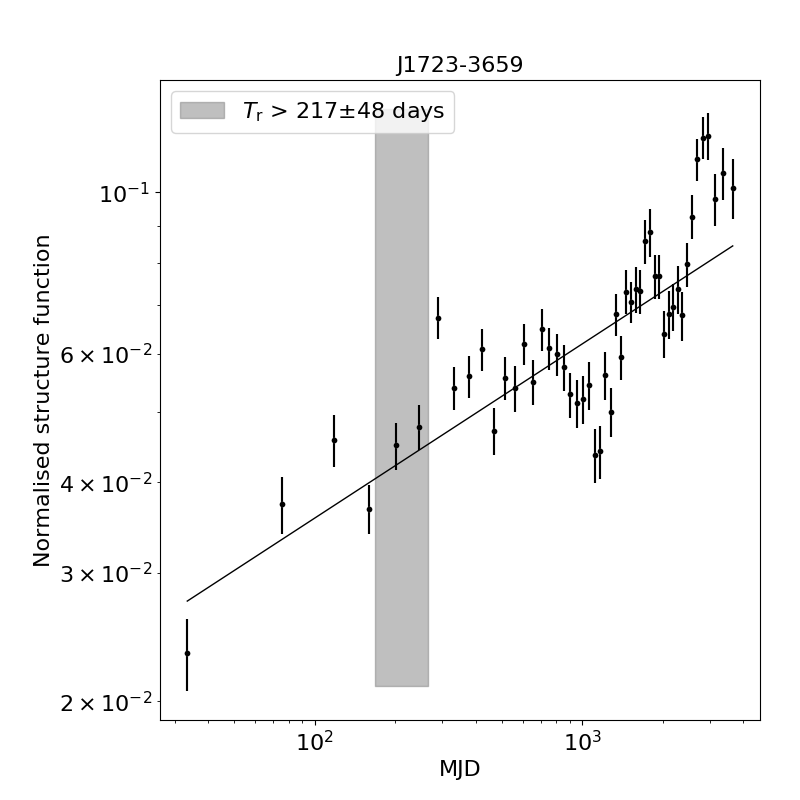}
\includegraphics[width=8cm]{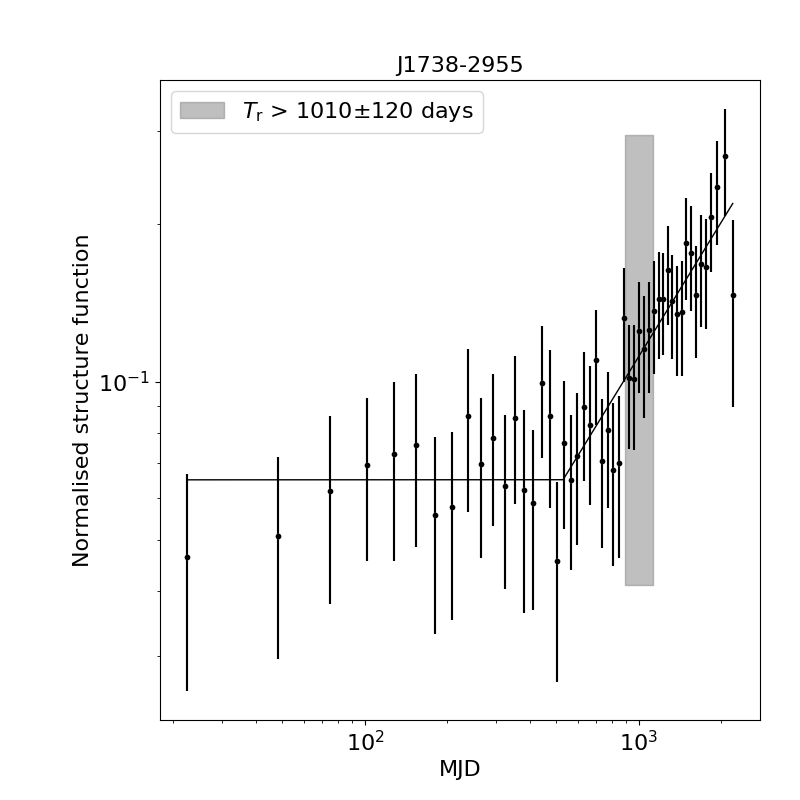}
\end{center}
\caption{Structure functions of PSRs J1723$-$3659 and J1738$-$2955.}
\end{figure*}

\begin{figure*}
\begin{center}
\includegraphics[width=8cm]{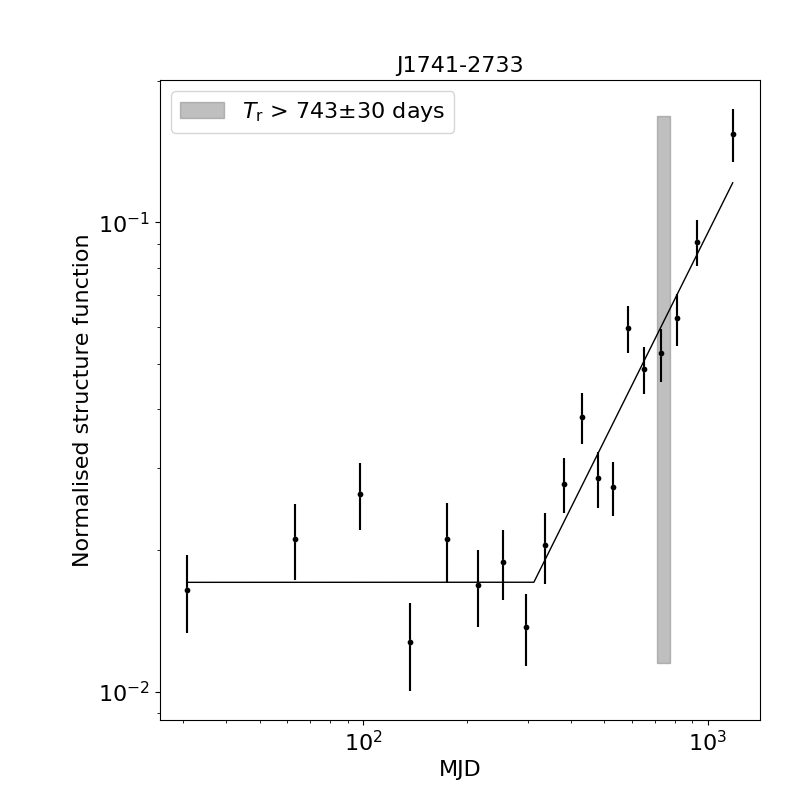}
\includegraphics[width=8cm]{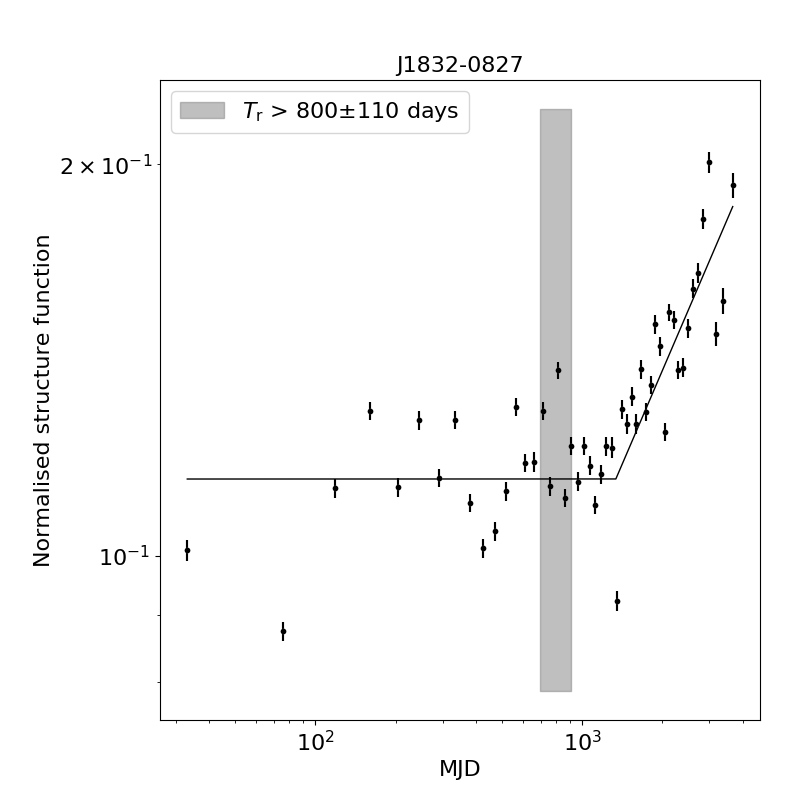}
\end{center}
\caption{Structure functions of PSRs J1741$-$2733 and J1832$-$0827.}
\end{figure*}

\begin{figure*}
\begin{center}
\includegraphics[width=8cm]{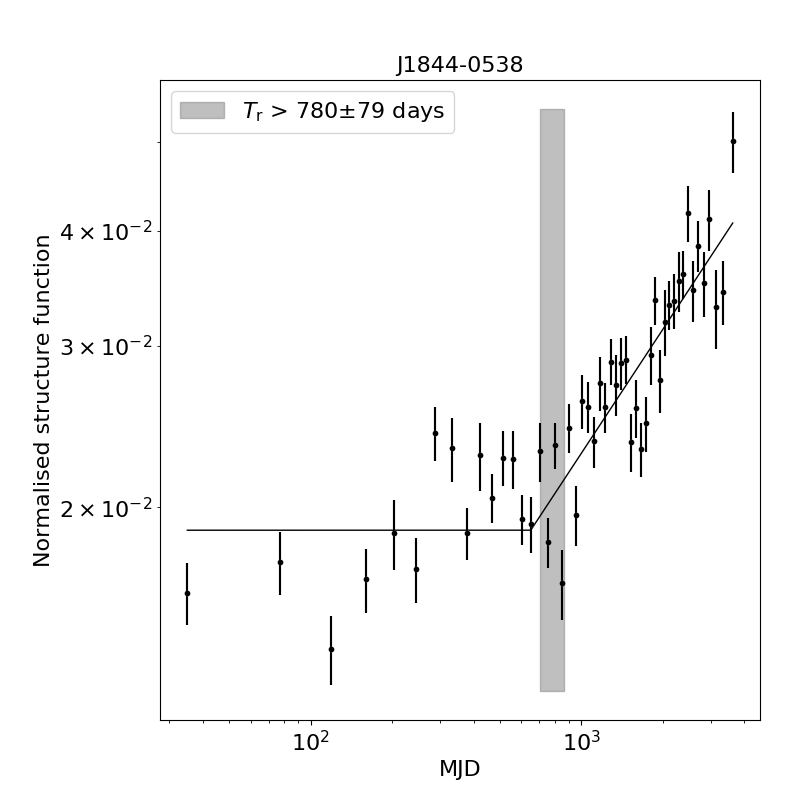}
\includegraphics[width=8cm]{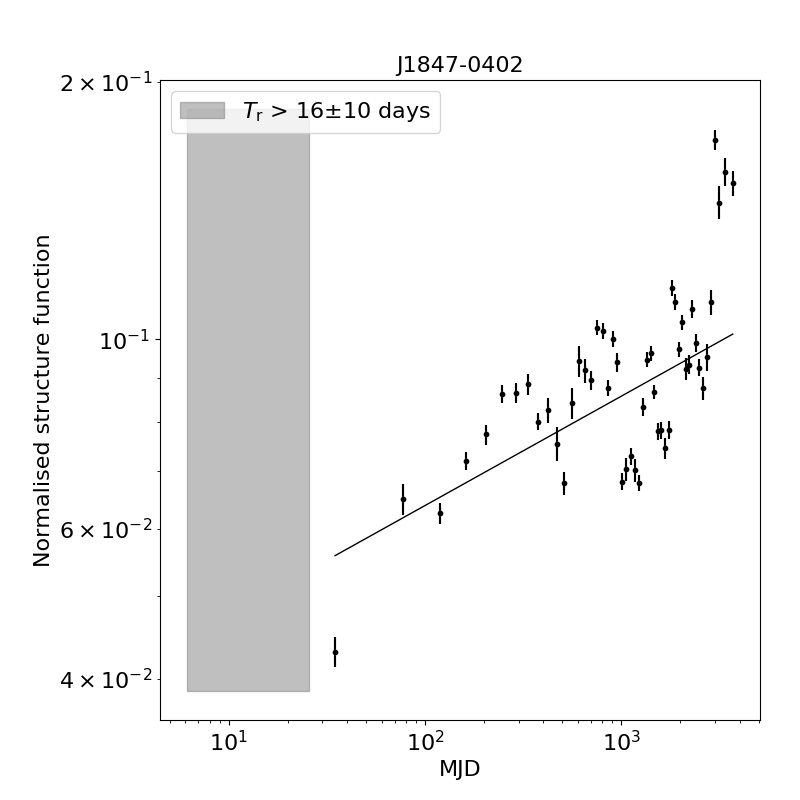}
\end{center}
\caption{Structure functions of PSRs J1844$-$0538 and J1847$-$0402.}
\end{figure*}

\begin{figure*}
\begin{center}
\includegraphics[width=8cm]{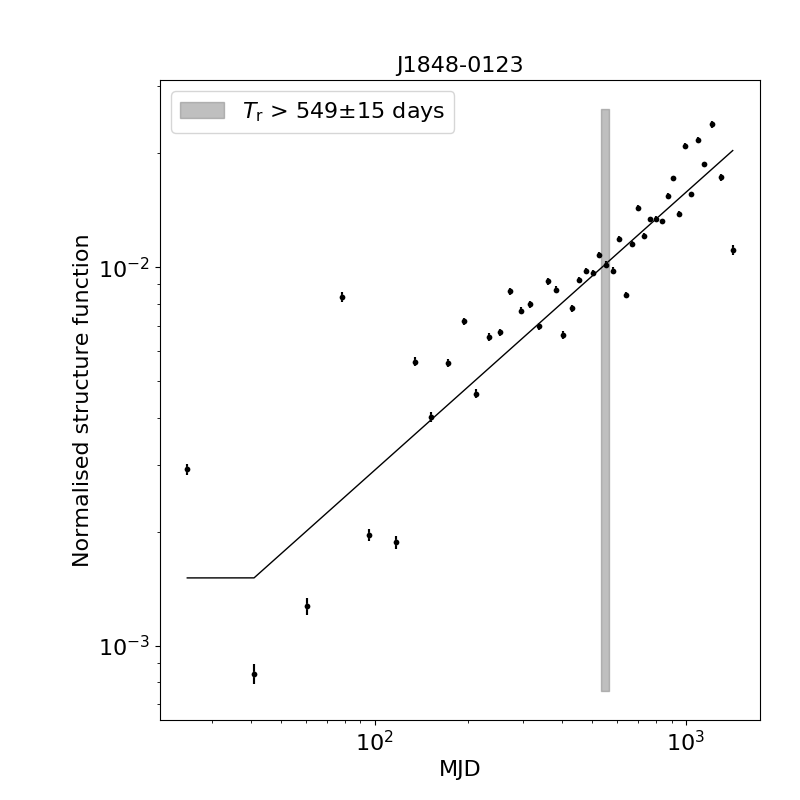}
\end{center}
\caption{Structure function of PSR J1848$-$0123.}
\end{figure*}

\bsp	
\label{lastpage}
\end{document}